\documentclass[11pt]{article}
\usepackage[letterpaper, margin=1in]{geometry}
\usepackage{enumitem}
\usepackage{mathtools}
\usepackage{amsthm}
\usepackage{amssymb}
\usepackage{mathrsfs}
\usepackage{xspace}
\usepackage{bm}
\usepackage[breakable]{tcolorbox}
\usepackage{longtable}
\usepackage{booktabs}
\usepackage{graphicx}
\usepackage{relsize}
\usepackage{tikz}
  \usetikzlibrary{positioning,fit,calc}
  \tikzstyle{block} = [draw=black, ultra thin, text width=2cm, minimum height=1.3cm, font = {\footnotesize\itshape},align=center]  
  \tikzstyle{arrow} = [thick,->,>=stealth]
  
  \renewenvironment{quote}
  {\list{}{\leftmargin=0.5cm \rightmargin=0.5cm} 
   \item\relax}
  {\endlist}

\newcommand{\tree}{\mathrm{tree}}

  \usepackage[page,toc,titletoc,title]{appendix}
\usepackage[disable,textsize=footnotesize]{todonotes}
\usetikzlibrary{shapes,arrows,positioning,patterns}
\usepackage{comment}
\usepackage{hyperref}
\usepackage{cleveref}
\hypersetup{
    colorlinks=true,
    citecolor=blue,
    linkcolor=blue,
    filecolor=blue,      
    urlcolor=blue,
}
\usepackage{thm-restate}
\usepackage{placeins}
\usepackage[font=small,labelfont=bf,margin=0.06in]{caption}

\usepackage[style=alphabetic,maxnames=999]{biblatex}
\AtEveryBibitem{
  \clearfield{editor}
  \clearlist{editor}
  \clearname{editor}
  \clearfield{location}
  \clearlist{location}
  \clearname{location}
  \clearfield{isbn}
  \clearlist{isbn}
  \clearname{isbn}
  \ifentrytype{online}
    {}
    {\clearfield{url}}
}

\RequirePackage[osf,sc]{mathpazo}
\usepackage{microtype}
\usepackage{framed}
\usepackage{mdframed}
\usepackage{float}
\usepackage{xcolor}

\newcommand{\ind}[2][]{%
  \mathrel{
    \mathop{
      \vcenter{
        \hbox{\oalign{\noalign{\kern-.3ex}\hfil$\vert$\hfil\cr
              \noalign{\kern-.7ex}
              $\smile$\cr\noalign{\kern-.3ex}}}
      }
    }^{#2}\displaylimits_{#1}
  }
}
\newcommand{\wh}{\widehat}

\newcommand{\poly}{\mathrm{poly}}

\DeclareMathOperator{\Types}{Types}

\DeclareMathOperator{\Ball}{Ball}

\newlength{\leftbarwidth}
\newlength{\leftbarsep}
\renewenvironment{leftbar}[1][blue]
{%
\def\FrameCommand
{%
{\hspace{-8pt} \color{#1} \vrule width 3pt}%
\hspace{0pt}
\fboxsep=\FrameSep\colorbox{#1!5!}%
}%
\MakeFramed{\hsize\hsize\advance\hsize-\width\FrameRestore}%
}
{\endMakeFramed}
\makeatletter

  \if@todonotes@disabled 
  \excludecomment{szin}
  \else 
  
  \fi

  \if@todonotes@disabled
  \excludecomment{janin}
  \else
  
  \fi

  \if@todonotes@disabled 
  \excludecomment{new}
  \else 
  
  \fi

\makeatother

\newcommand\myitem[1]{\hyperref[item:#1]{\emph{\ref*{item:#1}.}{}}\xspace}

\newtheorem{theorem}{Theorem}[section]

\newtheorem*{theorem*}{Theorem}
\newtheorem{theorem?}[theorem]{?`Theorem?}

\newtheorem*{mcconjecture}{Model Checking Conjecture}

\newtheorem{corollary}[theorem]{Corollary}
\newtheorem*{corollary*}{Corollary}
\newtheorem{lemma}[theorem]{Lemma}
\newtheorem*{lemma*}{Lemma}
\newtheorem{lemma?}[theorem]{?`Lemma?}
\newtheorem{fact}[theorem]{Fact}
\newtheorem{observation}[theorem]{Observation}

\newtheorem*{proposition*}{Proposition}

\newtheorem*{wrong*}{False Statement}

\crefname{theorem?}{Theorem{\scriptsize{?}}}{Theorems{\scriptsize{?}}}
\crefname{lemma?}{Lemma{\scriptsize{?}}}{Lemmas{\scriptsize{?}}}
\crefname{Theorem?}{Theorem{\scriptsize{?}}}{Theorems{\scriptsize{?}}}
\crefname{Lemma?}{Lemma{\scriptsize{?}}}{Lemmas{\scriptsize{?}}}

\newtheorem{claim}{Claim}[section]

\crefname{claim}{claim}{Claims} 
\Crefname{claim}{Claim}{Claims} 

\newenvironment{claimproof}[1][\proofname]{%
  \begin{proof}[#1]%
}{%
  \end{proof}%
}

\theoremstyle{remark}
\newtheorem{remark}[theorem]{Remark}

\newtheorem{definition}[theorem]{Definition}

\def\marker{\bar}

\def\bw{{\marker w}}
\def\bx{{\marker x}}

\renewcommand{\preceq}{\preccurlyeq}

\def\Nesetril{Ne\v{s}et\v{r}il\xspace}

\newcommand{\from}{\colon}

\newcommand{\set}[1]{\{#1\}}

\newcommand{\setof}[2]{\set{#1 \,:\, #2 }}
\newcommand{\setoff}[2]{\set{#1 \,:\, #2 }}

\def\phi{\varphi}
\def\cal{\mathcal}
\def\N{\mathbb N}
\def\R{\mathbb R}
\def\epsilon{\varepsilon}
\def\eps{\varepsilon}
\renewcommand{\subset}{\subseteq}
\renewcommand{\setminus}{-}
\renewcommand{\le}{\leqslant}
\renewcommand{\leq}{\le}
\renewcommand{\ge}{\geqslant}
\renewcommand{\geq}{\ge}

\newcommand{\dist}{\mathrm{dist}}

\renewcommand{\AA}{\cal A}
\newcommand{\BB}{\cal B}
\newcommand{\CC}{\mathscr C}

\newcommand{\HH}{\cal H}

\newcommand{\PP}{\cal P}

\newcommand{\QQ}{\cal Q}

\newcommand{\ATypes}{\mathrm{ATypes}}

\newcommand{\tup}{\bar}
\newcommand{\tp}{\textnormal{tp}}

\newcommand{\atp}{\textnormal{atp}}

\newcommand{\stp}{\textnormal{stp}}

\newcommand{\ftree}{\textnormal{ftree}}
\newcommand{\distFO}{\textup{dist-FO}\xspace}
\newcommand{\merge}[2]{\mathrm{merge}_{#1,#2}}
\newcommand{\resolve}[3]{\mathrm{resolve}^{#1}_{#2,#3}}
\newcommand{\rhom}{\rho^-}
\newcommand{\rhop}{\rho^+}
\newcommand{\size}[1]{\lVert #1\rVert}
\newcommand{\crad}{R}  

\newcommand{\rep}{\textnormal{rep}}

\newcommand{\sample}{\mathrm{sample}}

\begin{document}
\pagenumbering{roman}
\pagestyle{empty}

\newcommand{\funding}{
JD: Funded by the Deutsche Forschungsgemeinschaft (DFG, German Research Foundation) -- project number 563448077.
NM and ST received funding from the European Research Council (ERC) with grant agreement No.\ 101126229 -- BUKA.
NM also received funding from the European Union through an ERA Fellowship with grant agreement No.\ 101334340 -- LoCoMoDe.
\\[0.5em]
\hspace*{\fill}\includegraphics[width=40px]{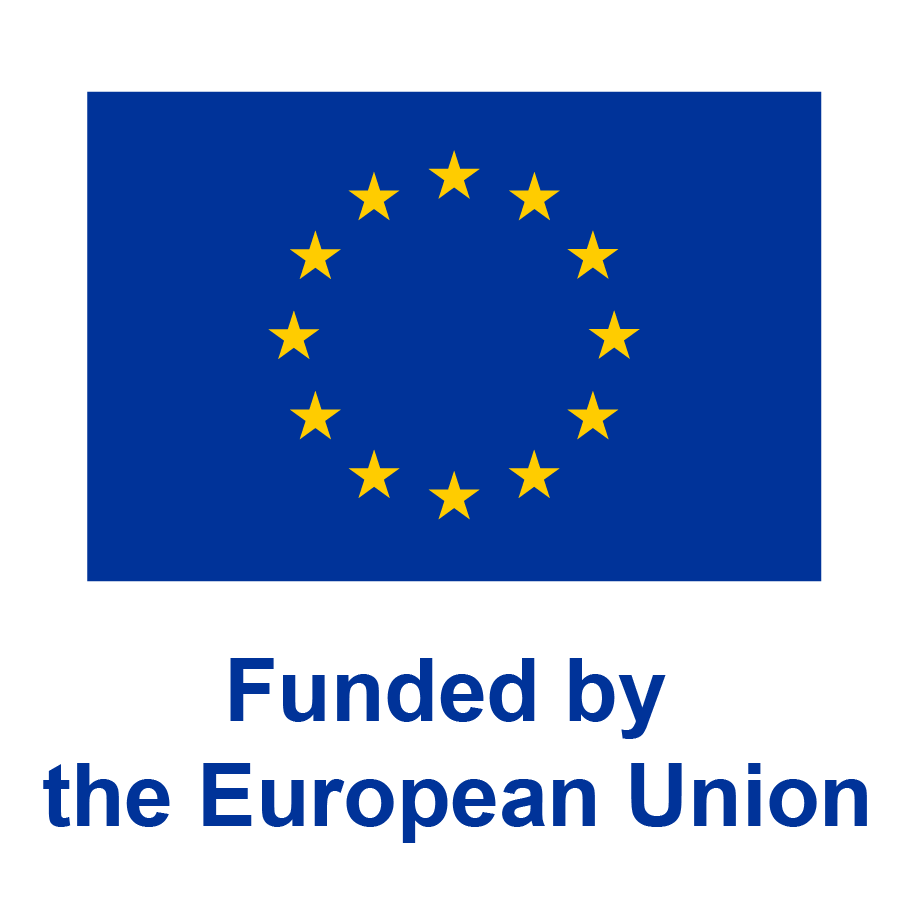}\hspace{1em}\includegraphics[width=40px]{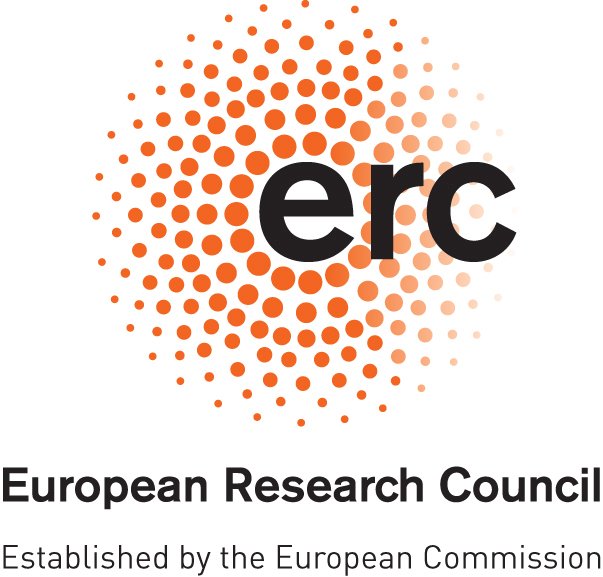}
}

\title{First-Order Model Checking\\on Monadically Dependent Graph Classes\thanks{\funding}}
\author{Jan Dreier\\{\small HPI Potsdam, Germany}
  \and Nikolas M{\"a}hlmann\\{\small University of Warsaw, Poland}
  \and Szymon Toru{\'n}czyk\\{\small University of Warsaw, Poland}}
\date{}
\maketitle
\thispagestyle{empty}

\begin{abstract}
Monadically dependent graph classes are those in which first-order logic cannot encode all graphs.
They include all nowhere dense classes and all classes of bounded twin-width.

We show that first-order model checking is fixed-parameter tractable on every monadically dependent graph class.
Together with the known hardness result for hereditary classes that are not monadically dependent, this gives a complete characterization of the hereditary graph classes with fixed-parameter tractable first-order model checking. 
This resolves a central open problem in algorithmic model theory.

Our proof combines ideas from 
structural graph theory (graph decompositions), 
finite model theory (locality of first-order logic), and model theory (counting types).
In particular, we show that in monadically dependent classes, tuples of fixed length have only $|U|^{O(1)}$ types of fixed quantifier rank over any set of parameters $U$.
\end{abstract}

\newpage
\setcounter{tocdepth}{1}
\tableofcontents

\clearpage
\pagestyle{plain}
\pagenumbering{arabic}
\newpage
\section{Introduction}
\newcommand{\AWstar}{$\mathrm{AW}[\star]$}

Instead of solving a single algorithmic problem, \emph{algorithmic meta-theorems} 
establish the tractability of whole classes of problems, in a uniform way.
The targeted problem class is commonly defined using logic, and hence many algorithmic meta-theorems fit the following template.

\newcommand{\placeholder}[1]{\textup{\textbf{$\boldsymbol{\langle}$#1$\boldsymbol{\rangle}$}}}
\begin{quote}
\emph{Every problem expressible in \placeholder{logic} can be solved in time \placeholder{runtime} on every input satisfying \placeholder{tameness condition}.}
\end{quote}
A prototypical example is Courcelle's Theorem, which states that every problem definable in \emph{monadic second-order logic} is solvable in \emph{linear time} on graphs with constant \emph{tree-width}~\cite{courcelle1990monadic}.
Similarly, a result of Frick and Grohe
implies that every graph problem definable in \emph{first-order logic} (FO) can be decided in \emph{linear time} on all \emph{planar graphs}~\cite{frick2001deciding}.
In particular, the \emph{(first-order) model checking problem} 
-- does a given graph $G$ satisfy a given FO sentence $\phi$?  -- is \emph{fixed-parameter tractable}~(fpt) on planar graphs. 
That is, it is solvable\footnote{Throughout this paper, we use the uniform -- but not strongly uniform -- version of fpt, and of fpt reductions. That means that the function $f$ need not be computable, but there is a single algorithm which solves the problem, independent of the parameter ($\phi$ in the case of model checking).} in time $f(|\phi|)\cdot |V(G)|^c$, for some function $f$ and constant~$c$ (with $c=1$ in this specific case).
Many natural graph problems are expressible in FO.
For example, the sentence
\[
    \exists x_1 \ldots \exists x_k  \forall y\ 
    \bigvee_i\big(E(y,x_i) \vee (y = x_i)\big)
\]
expresses $k$-\textsc{Dominating-Set}: the existence of $k$ vertices  such that every vertex is either adjacent or equal to at least one of them.
Further expressible problems include $k$-\textsc{Independent-Set}, $k$-\textsc{Clique}, and $H$-\textsc{Subgraph-Isomorphism}.
With no further restrictions made on the input graphs,
these problems appear to be highly intractable.
Indeed, model checking on general graphs is believed not to be fpt; this is equivalent to the $\mathrm{FPT}{\neq}\mathrm{AW}[\star]$ conjecture.
This raises the question -- first posed by Grohe~\cite[Section~8.2]{logic-graphs-algorithms} -- on which graph classes\footnote{A \emph{graph class} is a (usually infinite) set of graphs.} 
the problem becomes fpt.
After decades of research, \emph{monadic dependence} has crystallized as the conjectured tractability limit for \emph{hereditary}\footnote{A graph class is \emph{hereditary} if it is closed under taking induced subgraphs, i.e., deleting vertices.} classes (see, e.g.,~\cite{warwick-problems,new-perspective-journal,tww4,ms-mc1}).

\begin{mcconjecture}\label{conj:main2}
  A hereditary graph class admits fixed-parameter tractable first-order model checking if and only if it is monadically dependent (assuming $\mathrm{FPT}{\neq}\mathrm{AW}[\star]$).
\end{mcconjecture}

\begin{figure}[htb]
  \centering
  \begin{tikzpicture}[
        prop/.style={draw, line width=0.5pt, minimum width=\bw, minimum height=\bh,
            path picture={\def\pb{path picture bounding box}
                \fill[black!10] (\pb.north west) rectangle ([yshift=-\hz]\pb.north east);
                \fill[black!4] ([yshift=-\hz]\pb.north west) rectangle ([yshift=\fz]\pb.south east);
                \draw[head] ([yshift=-\hz]\pb.north west) -- ([yshift=-\hz]\pb.north east);
                \draw[foot] ([yshift=\fz]\pb.south west) -- ([yshift=\fz]\pb.south east);}},
        proplabel/.style={font=\small\bfseries, inner sep=0pt},
        imp/.style={line width=0.5pt, -latex},
        detail/.style={inner sep=0pt, anchor=base, font=\scriptsize},
        refs/.style={detail, text=black!60,
            execute at begin node={\hypersetup{citecolor=black!60}}},
        example/.style={detail, font=\scriptsize\itshape, text=black!60},
        head/.style={line width=0.5pt},
        foot/.style={line width=0.5pt, black!25},
    ]
    \def\bw{5.0cm}\def\bh{2.2cm}\def\cs{5.6}\def\rs{2.75}
    \def\tz{0.55cm}\def\rb{0.745cm}\def\fz{0.55cm}\def\eb{0.225cm}\def\hz{1.1cm}
    \def\prop#1#2#3{\node[prop] (#1) at (#2) {};
        \node[proplabel] at ([yshift=-\tz]#1.north) {\begin{tabular}{@{}c@{}}#3\end{tabular}};}
    \def\lbl#1{\textsc{#1}}
    \def\mcdec#1#2#3{\node[refs] at ([yshift=\rb]#1.south)
        {\lbl{mc:}\enspace#2\quad\lbl{dec:}\enspace#3};}
    \def\mcanddec#1#2{\node[refs] at ([yshift=\rb]#1.south) {\lbl{mc+dec:}\enspace#2};}
    \def\ex#1#2{\node[example] at ([yshift=\eb]#1.south) {#2};}
    \def\excite#1{{\hypersetup{citecolor=black!60}\textup{\cite{#1}}}}

    \prop{nd}{0,0}{nowhere denseness}
    \prop{be}{0,-\rs}{bounded expansion}
    \prop{sptw}{0,-2*\rs}{bounded sparse\\twin-width}
    \prop{trw}{0,-3*\rs}{bounded tree-width}

    \prop{ms}{\cs,0}{monadic stability}
    \prop{smw}{\cs,-\rs}{bounded stable\\merge-width}
    \prop{stw}{\cs,-2*\rs}{bounded stable\\twin-width}
    \prop{scw}{\cs,-3*\rs}{bounded stable\\clique-width}

    \prop{md}{2*\cs,0}{monadic dependence}
    \prop{mw}{2*\cs,-\rs}{bounded merge-width}
    \prop{tww}{2*\cs,-2*\rs}{bounded twin-width}
    \prop{cw}{2*\cs,-3*\rs}{bounded clique-width}

    \mcanddec{nd}{\cite{gks}}
    \mcdec{be}{\cite{DvorakKT13-journal}}{\cite{grad-and-bounded-expansion-Nesetril}}
    \mcdec{trw}{\cite{courcelle1990monadic}}{\cite{minorsXIII}}
    \mcdec{ms}{\cite{ms-mc1}}{\cite{flippers,ms-mc2}}
    \mcdec{md}{this paper}{\cite{md-abmw}}
    \mcdec{smw}{\cite{lsd-journal}}{\cite{dreier2026efficient}}
    \mcdec{mw}{\cite{mw-stoc}}{open}
    \mcdec{tww}{\cite{tww1}}{open}
    \mcdec{cw}{\cite{courcelle2000linear}}{\cite{OumSeymour-approximating}}

    \ex{sptw}{planar~\excite{frick2001deciding}, minor-free~\excite{flum2001fixed}}
    \ex{be}{bounded degree~\excite{seese_1996}}
    \ex{smw}{structurally bounded degree~\excite{new-perspective-journal}}
    \ex{stw}{map graphs~\excite{eickmeyer2017map}}
    \ex{tww}{unit interval graphs~\excite{ganian2015fo}}
    \ex{md}{struc.\ bd.\ loc.\ clique-width~\excite{DBLP:conf/lics/BonnetDGKMST22}}
    \ex{nd}{locally minor-free~\excite{dawar2007locally}}

    \foreach \a/\b in {trw/sptw, sptw/be, be/nd,
                       scw/stw, stw/smw, smw/ms,
                       cw/tww, tww/mw, mw/md}
        \draw[imp] (\a) -- (\b);

    \foreach \a/\b in {nd/ms, ms/md, be/smw, smw/mw, sptw/stw, stw/tww,
                       trw/scw, scw/cw}
        \draw[imp] ([yshift=-\tz]\a.north east) -- ([yshift=-\tz]\b.north west);
\end{tikzpicture}%
  \caption{A hierarchy of graph class properties that admit fpt first-order model checking.
    Arrows represent implications between properties.
    In particular, \emph{bounded tree-width} is the most restricted class property in this hierarchy and \emph{monadic dependence} is the most general one.
    In each box where applicable, \textsc{mc} cites the fpt model checking result which assumes a suitable decomposition is given, while \textsc{dec} cites the result providing such decompositions efficiently; \textsc{dec:}~open means that no algorithm for computing such a decomposition is known.
    Example graph classes are listed in \textcolor{black!60}{\emph{italics}}, each under the most restricted property that subsumes it, together with a citation to an fpt model checking algorithm for that class.}
  \label{fig:hierarchy}
\end{figure}
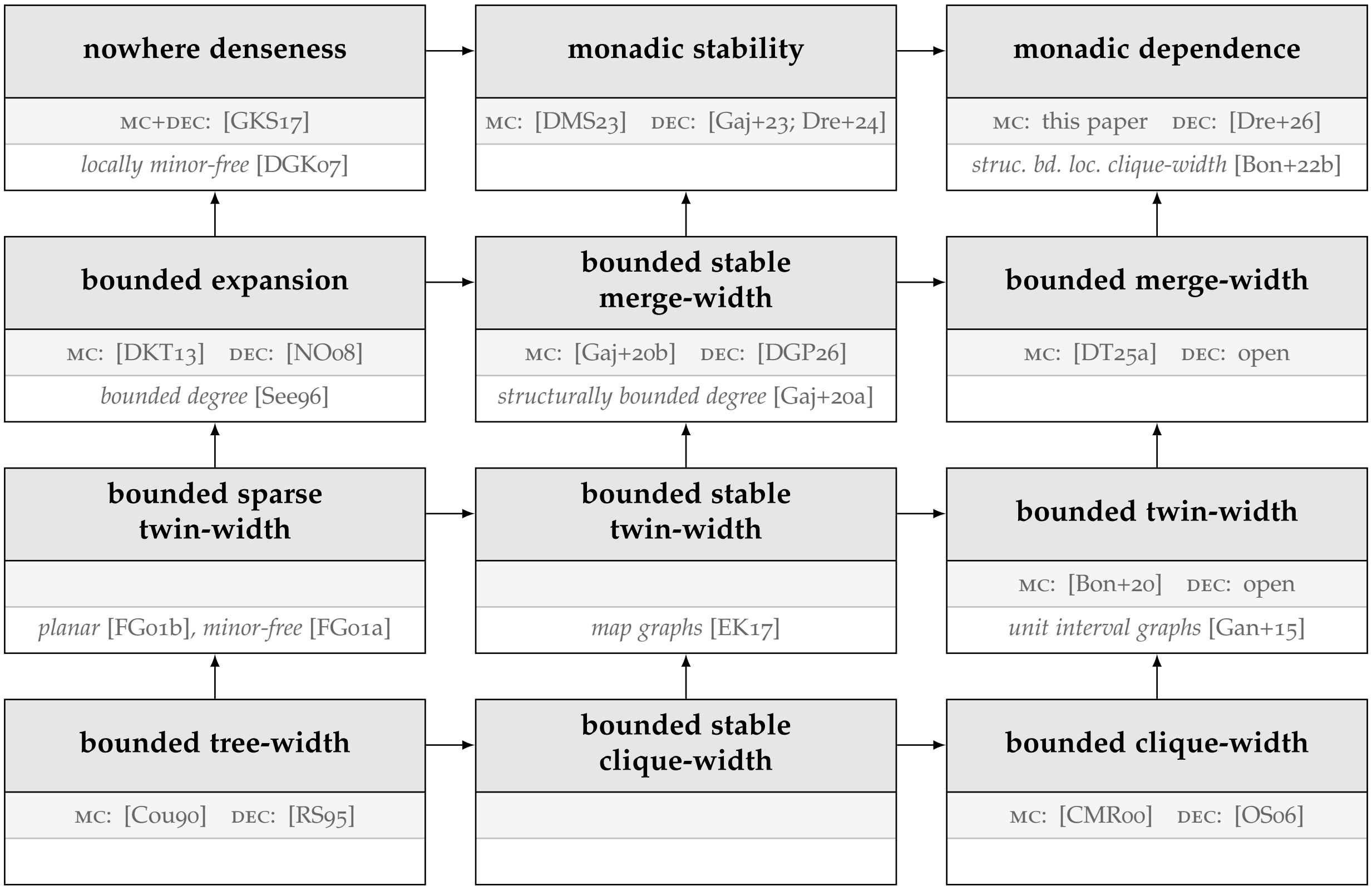

Monadic dependence is a model-theoretic tameness notion that originates in Shelah's classification theory~\cite{BS1985monadic}.
It subsumes many tameness notions from structural graph theory, such as planarity, excluding a minor, bounded degree, bounded expansion, nowhere denseness, and bounded clique-width, twin-width, and merge-width.
See \Cref{fig:hierarchy} for a hierarchy of included graph classes.
For hereditary graph classes, monadic dependence is characterized as follows~\cite{flip-breakability}.
A hereditary graph class $\CC$ is \emph{monadically independent} if there is some formula $\phi(x,y)$ with the following property: for every graph \(H\) there is some graph $G \in \CC$ 
such that $H$ is an induced subgraph of the graph with vertices $V(G)$ and edges \(\{ uv \,:\, \text{ $\phi(u,v)$ or $\phi(v,u)$ holds in $G$} \}\).
Intuitively, this means that every graph can be encoded by $\phi$ in $\CC$.
Then $\CC$ is \emph{monadically dependent} if it is not monadically independent.

\medskip
In a previous paper~\cite{flip-breakability} we resolved the hardness direction of the Model Checking Conjecture, by showing that model checking is 
\AWstar-hard on every hereditary graph class that is not monadically dependent.

\subsection{Our results}
In this paper, we confirm the Model Checking Conjecture by proving its tractability direction.

\begin{restatable}{theorem}{mainthm}\label{thm:main}
  There is an algorithm which
  solves the model checking problem for first-order logic on graphs --
  does a given graph $G$ satisfy a given first-order sentence $\phi$?  -- with the following guarantee.
  For every monadically dependent graph class $\CC$ and real $\eps>0$, the algorithm runs in time $O_{\CC,\phi,\eps}(n^{3+\eps})$ for $n$-vertex graphs $G\in\CC$.
\end{restatable}

In particular, this gives fpt first-order model checking on all classes of bounded twin-width and bounded merge-width, without requiring a decomposition as part of the input. 

We remark that our algorithm is uniform:
it takes both $G$ and $\phi$ as input, and does not depend on the graph class $\CC$, which only appears in the running-time guarantee.
Our result lifts from graphs to monadically dependent classes of binary structures -- structures over a fixed signature consisting of binary relation symbols -- using a standard encoding (see, e.g., Lemma~19.11 in the full version of~\cite{flip-breakability}).

\medskip
Our algorithm is based on dynamic programming on graph decompositions called \emph{construction sequences} of \emph{almost bounded merge-width}.
In a companion paper with Michał Pilipczuk~\cite{md-abmw}, we prove that they exist and can be computed efficiently on monadically dependent graph classes.

Existing model checking algorithms for classes of bounded twin- or merge-width have a tower-of-exponentials dependence on the width. This prevents their direct application to decompositions of \emph{almost} bounded merge-width, where the width is only bounded by a subpolynomial function $n^{o(1)}$ of the input size $n$, rather than a constant. 
To overcome this obstacle, we make the running time depend on the number of \emph{types} that are realized in the graph. We control this number using a type-counting result for monadically dependent classes, which we believe to be of independent interest.

This result provides polynomial bounds on the number of quantifier rank $q$ types $\tp_q^G(\tup v/U)$ of $k$-tuples $\tup v$ over a set $U$ of parameters.
The \emph{type} $\tp_q^G(\tup v/U)$ is the set of formulas of quantifier rank at most \(q\), with parameters from \(U\), which hold of the tuple $\tup v$ in $G$
(see \Cref{sec:prelims} for the precise definition).
For $q=0$, the type $\tp_0^G(v/U)$ of a vertex $v\notin U$ represents its neighborhood $N(v)\cap U$ in $U$. 
In general graphs, there could be $2^{|U|}$ many such neighborhoods, but the fundamental Sauer--Shelah lemma~\cite{Sauer72,She72} bounds their number by $(|U|+1)^d$, where $d$ is the VC-dimension of the graph $G$.
The VC-dimension $d$ is bounded uniformly in every monadically dependent class.
For $q=1$, the type $\tp_1^G(v/U)$ of a vertex $v$ 
determines $\setof{N(w)\cap U}{w\in N(v)}$, the set of neighborhoods in $U$ of the neighbors of $v$. In general graphs there could be 
$2^{2^{|U|}}$ many such sets, and so on for higher $q$: the number of $q$-types over a set $U$ can reach a tower of height $q+1$.
The following result, which is the cornerstone of our approach, can thus be seen as an extension of the Sauer--Shelah lemma to higher quantifier ranks, in monadically dependent graph classes.

\begin{theorem}[{Simplified form of \Cref{thm:type-bound}}]\label{thm:intro-type-bound}
Let $\CC$ be a monadically dependent graph class and $k,q\in\N$. There is some $d\in\N$ such that for every (possibly infinite) graph $G \in \CC$,  and finite subset $U \subseteq V(G)$ with $|U|>1$
we have
\[
  \left|
    \{\tp^{G}_q(\tup v/U) \,:\, \tup v \in V(G)^k \}\right|
    \leq  |U|^d.
\]
\end{theorem}

\Cref{thm:intro-type-bound} is a finitary analogue of a fundamental result in stability theory due to Shelah~\cite{Shelah1969}, which bounds the number of complete types (of unbounded quantifier rank) over an infinite set $U$, by $|U|^\kappa$ in every \emph{stable} class of structures, for some cardinal $\kappa$ depending on the class. 
Its proof blends ideas from \cite{number-of-types} and \cite{flippers}.

\medskip
A technical overview of the proofs of the main results is presented in \Cref{sec:tech-overview}. 
All our results are formalized in Lean. At \url{https://laxarchive.org/lax-185574/}, each definition, lemma, and theorem is annotated with its proven Lean counterpart.

\subsection{Related work}
\Cref{thm:main} completes a long line of research on first-order model checking on hereditary graph classes.
An overview of previous model checking results is given in \Cref{fig:hierarchy}. 
In particular, the Model Checking Conjecture was previously known to hold in the following three settings, which we discuss below:

\newcommand{\footnoteMonotone}{\footnote{A graph class is \emph{monotone} if it is closed under taking subgraphs, i.e., deleting vertices and edges.}}
\newcommand{\footnoteEdgeStable}{\footnote{\interlinepenalty=10000 
A graph class is \emph{edge-stable} if it excludes some semi-induced half-graph, i.e., there is $k\in\N$ such that no graph in the class contains vertices $a_1,\ldots,a_k,b_1,\ldots,b_k$ where $a_i$ and $b_j$ are adjacent if and only if $i\le j$. Equivalently, the edge relation is a stable formula on the class in the sense of model theory.}}
\newcommand{\footnoteOrdered}{\footnote{An \emph{ordered graph} is a graph equipped with a total order on its vertex set, which formulas can access via an additional binary relation symbol~$<$.}}

\begin{itemize}
  \item in \emph{monotone}\footnoteMonotone~graph classes,
  \hfill \cite{gks}

  where monadic dependence coincides with \emph{nowhere denseness};\hfill \cite{AdlerA14,podewski1978stable}

  \item in \emph{edge-stable}\footnoteEdgeStable~graph classes,
  \hfill \cite{ms-mc1,ms-mc2}

  where monadic dependence coincides with \emph{monadic stability};
  \hfill \cite{rankwidth-meets-stability}

  \item in classes of \emph{ordered graphs}\footnoteOrdered,
  \hfill \cite{tww4,tww1}

  where monadic dependence coincides with \emph{bounded twin-width}.
  \hfill \cite{tww4}
\end{itemize}

\paragraph*{Sparsity.}
The first model checking result in the sparse regime is the seminal Courcelle's Theorem, which established tractability for more expressive monadic second-order logic on all classes of bounded tree-width~\cite{courcelle1990monadic}.
Subsequently, Seese initiated the study of first-order model checking by showing tractability on all classes of bounded degree~\cite{seese_1996}.
This research direction gained momentum through the work of Grohe and his coauthors~\cite{frick2001deciding,flum2001fixed,dawar2007locally}; see Grohe's 2007 survey~\cite{logic-graphs-algorithms} for an account of these early developments.
This line of research (see also~\cite{dawar2009domination,DvorakKT13-journal,GroheKreutzer2011}) culminated in the model checking algorithm of Grohe, Kreutzer, and Siebertz~\cite{gks} for \emph{nowhere dense} classes.
This notion is one of the cornerstones of the \emph{sparsity theory} by \Nesetril and Ossona de Mendez~\cite{NesetrilM11a,sparsity-book} that shaped the field.
For monotone classes, this result is tight: first-order model checking is \AWstar-hard on every monotone class that is not nowhere dense~\cite{DvorakKT13-journal,dawar2009parameterized}.

\medskip
With the sparse classes resolved, the research focus has shifted towards possibly dense classes,
and two parallel research directions emerged.

\paragraph*{Stability.}
One research direction aims to lift the theory of nowhere dense classes to dense graphs.
Here a key notion is that of \emph{monadic stability}, a restriction of monadic dependence that also originates in model theory~\cite{BS1985monadic}.
The study of the model checking problem has revealed multiple new characterizations of monadic stability, such as \emph{flip-flatness}~\cite{dreier2023indiscernibles}, \emph{flipper games}~\cite{flippers}, and \emph{low shrub-depth covers}~\cite{horizons}.
They subsequently led to an fpt algorithm for first-order model checking for monadically stable classes~\cite{ms-mc1,ms-mc2}.
This result is tight in hereditary classes that are additionally edge-stable~\cite{ms-mc2,rankwidth-meets-stability}.
The ideas from this line of research have been further developed into combinatorial characterizations of monadically dependent graph classes such as \emph{flip-breakability}~\cite{flip-breakability} and \emph{flip-separability}~\cite{bonnet2025separability-icalp}.
This also includes the characterization by forbidden induced subgraphs, which settled the hardness part of the Model Checking Conjecture for all hereditary graph classes~\cite{flip-breakability}.

\paragraph*{Clique-width, twin-width, and merge-width.}
A second, parallel line of research has pushed the tractability frontier of first-order model checking through the development of new graph decompositions, often accompanied by new graph width parameters.
Mirroring tree-width in the sparse setting, it was shown early on by Courcelle, Makowsky, and Rotics that also classes of bounded \emph{clique-width} admit fpt monadic second-order model checking~\cite{courcelle2000linear}.

Twenty years later, inspired by the work of Guillemot and Marx on the permutation pattern problem~\cite{guillemot2014finding}, Bonnet, Kim, Thomass\'e, and Watrigant introduced the more general classes of bounded \emph{twin-width}, and showed that they admit fpt first-order model checking~\cite{tww1}.
Their algorithm is conditional in the sense that it requires as an additional input a decomposition witnessing the twin-width bound.
As of today it is not known how to efficiently compute or approximate these decompositions in general graphs. However, they can be efficiently approximated in many specific cases of interest~\cite{tww8}, in particular, in hereditary classes of ordered graphs~\cite{tww4}.
There, bounded twin-width forms the exact tractability limit for first-order model checking.

Inspired by twin-width and the related flip-width parameters~\cite{flip-width},
Dreier and Toru\'nczyk introduced \emph{merge-width}, a family of graph parameters based on decompositions called \emph{construction sequences}~\cite{mw-stoc}. 
A class has \emph{bounded merge-width} if for every radius $r$, every graph in the class has a construction sequence of radius-$r$ width bounded by a constant.
Again, those classes admit fpt first-order model checking under the condition that a construction sequence is given.
Crucial to our paper is the following relaxation of merge-width, called \emph{almost bounded merge-width}, which replaces the constant bound on the radius-$r$ width by $n^{o(1)}$, where $n$ is the vertex count of the graph. 
Dreier and Toru\'nczyk proved that hereditary classes of almost bounded merge-width are monadically dependent, and conjectured the converse~\cite{mw-stoc}.
The converse was first confirmed for radius one~\cite{neighborhood-complexity-mw}, and then in full~\cite{md-abmw}.
The proof of~\cite{md-abmw} is algorithmic: given a graph from a monadically dependent class, it computes in polynomial time a construction sequence of radius-$r$ width $n^{o(1)}$, for every fixed $r$.
Construction sequences of almost bounded merge-width form the basis for our model checking algorithm.

Note that it remains an open question how to approximate decompositions of constant width for classes with constant twin- and merge-width.
Thus, up until now and even in the light of \cite{md-abmw}, the previous model checking results for classes of bounded twin- and merge-width have remained conditional.
Therefore, our new algorithm marks the first time that unconditional model checking for all classes of bounded twin- and merge-width is established.

\section{Technical overview}
\label{sec:tech-overview}

In the related work section, we surveyed two approaches to the model checking problem in previous results.
Our algorithm combines ideas from both approaches:
\begin{itemize}
  \item[--] decomposition-based methods, to do dynamic programming along a construction sequence with almost bounded merge-width; and
  \item[--]  stability-theoretic bounds on the \emph{number of types}, allowing us to bound the runtime of the algorithm.
\end{itemize}
In this section, we give a high-level overview of the algorithm and the interplay between the two components.
We start by introducing the relevant definition.

\paragraph{Merge-width.}
The merge-width parameters are defined through decompositions called \emph{construction sequences}~\cite{mw-stoc}.
We can define them as follows.

\begin{itemize}
    \item A \emph{partial graph} $H$ consists of a finite vertex set $V=V(H)$, two disjoint sets $E(H),N(H)\subseteq\binom V2$ of \emph{edges} and \emph{non-edges}, and a partition of $V$ whose parts we treat as colors.
    The pairs in $R(H)\coloneqq E(H)\cup N(H)$ are \emph{resolved}, and all other pairs are \emph{unresolved}.
    We view $H$ as a logical structure with binary relations $E$ and $N$ and a unary predicate for every color.
    Distances $\dist^H$ and balls $\Ball^H_r(v)$ are measured in the Gaifman graph $(V,R(H))$, the graph of resolved pairs.
    \item For two distinct parts $P,Q$ of $H$, the \emph{merge} of $P$ and $Q$ replaces them by the single part $P\cup Q$.
    For two parts $P,Q$ of $H$ (possibly $P=Q$), the \emph{positive resolve} of $P$ and $Q$ adds all unresolved pairs $uv$ with $u\in P$ and $v\in Q$ to $E(H)$, while the \emph{negative resolve} adds them to $N(H)$.
    \item A \emph{construction sequence} of a graph $G=(V,E)$ is a sequence $G_1,\ldots,G_\lambda$ of partial graphs with vertex set $V$,
    such that $G_1$ has no resolved pairs and only singleton parts, every $G_{t+1}$ is obtained from $G_t$ by one merge or one resolve,
    and, finally, $G_\lambda$ has a single part, $E(G_\lambda)=E$, and $N(G_\lambda)=\binom V2\setminus E$.
    \item The \emph{radius-$r$ width} of the sequence is the maximum, over all $t\in[\lambda]$ and $v\in V$, of the number of parts of $G_t$ that intersect $\Ball^{G_t}_r(v)$.
    The \emph{radius-$r$ merge-width} of $G$ is the minimum radius-$r$ width over all construction sequences of $G$.
\end{itemize}

In a companion paper, together with Micha\l{} Pilipczuk, we prove that monadically dependent classes have \emph{almost bounded merge-width}~\cite{md-abmw}, and that witnessing construction sequences can be computed efficiently.
These construction sequences form the backbone of our model checking algorithm.

\begin{theorem}[\cite{md-abmw}]\label{thm:ov-abmw}
  Let $\CC$ be a monadically dependent graph class and $r\in\N$.
  Every $n$-vertex graph $G\in\CC$ has radius-$r$ merge-width $n^{o(1)}$, and a witnessing construction sequence can be computed in time $O_{\CC,r}(n^{3+o(1)})$.
\end{theorem}

\paragraph{Types along decompositions.}
Many model checking algorithms (including ours) process a decomposition of the input graph step by step, and maintain \emph{types}: formulas of bounded quantifier rank satisfied by the intermediate graphs processed so far.
For example, Courcelle's theorem~\cite{courcelle1990monadic} maintains, for every node $t$ of a tree decomposition with bag $\tup b=(b_1,\ldots,b_k)$, the type
\[
\setof{\psi(x_1,\ldots,x_k)}{\text{$\psi$ is an MSO formula of quantifier rank at most $q$, such that $\psi(\tup b)$ holds in $G_t$}},
\]
where $G_t$ is the subgraph induced by the bags in the subtree of $t$. Here $q$ is the quantifier rank of the input formula $\phi$.
The type above is computed from the types of the children of $t$.
Similarly, the algorithms for bounded twin-width maintain FO types along contraction sequences~\cite{tww1,tww-types-icalp}, and the algorithm for bounded merge-width along construction sequences~\cite{mw-arxiv}.

We call a first-order formula \emph{local} if, roughly speaking, every quantifier has the form 
\begin{center}
  $\exists y\,(\dist(\tup x,y)\le r\land\chi(\tup x y))$ \quad or \quad $\forall y\,(\dist(\tup x,y)\le r\to\chi(\tup x y))$  
\end{center}
ranging only over vertices close to some free or previously quantified variables $\tup x$.
The merge-width model checking algorithm~\cite{mw-arxiv} maintains, at every step $H$ of the construction sequence and for every vertex $v$, its \emph{local type} of quantifier rank $q$, for some constant $q$ depending on $\phi$:
\[
\setof{\psi(x)}{\text{$\psi$ is a local formula of quantifier rank at most $q$, such that $\psi(v)$ holds in $H$}}.
\]
In the first step of the construction sequence, nothing is resolved, and the local types are trivial.
In the last step, all pairs are resolved, and the local types determine whether the input sentence $\phi$ holds in $G$.
In between, the local types have to be updated after every merge and every resolve.
Merges are handled easily, by simply rewriting color names; the main difficulty in~\cite{mw-arxiv} is to lift the local types through resolves.

\paragraph{Combinatorial explosion.}
For all these type-based model checking algorithms, the number of possible types grows non-elementarily with the width of the decomposition.
In particular, in each step of a construction sequence of radius-$r$ width $w$, each vertex locally reaches at most $w$ colors, and the number of local formulas over $w$ colors is a tower of exponentials in $w$.
This is a constant if the width \(w\) is treated as a constant (the radius $r$ here is fixed and depends on the quantifier rank $q$).
However, monadically dependent classes may have unbounded merge-width, and we can only guarantee a merge-width bound of \(w = n^{o(1)}\). 
In this case the number of all possible local types explodes.

\paragraph{Our algorithm.}
The tower above counts the types that are \emph{possible} with $w$ colors.
Our starting point is the observation that in monadically dependent classes, only few of them are \emph{realized}, that is, actually occur as the type of some tuple of vertices of the input graph.
We keep the overall structure of the merge-width algorithm, but change it so that its running time depends only on the realized types.
This has three parts, which we describe in the rest of this section.
\begin{enumerate}
  \item We represent local types by \emph{type trees}, whose size depends only on the realized types.
  \item We show that only $n^{o(1)}$ types are realized in a ball, which bounds the size of type trees (\Cref{lem:ov-tree-size}).
  \item We lift type trees through merges and resolves in time polynomial in $n$ and the number of realized types (\Cref{lem:ov-resolve-lift}).
\end{enumerate}

\paragraph{Type trees.}
Type trees (defined in \Cref{sec:typetrees}) are a variant of Hintikka types~\cite{Scott1979}.
Simplifying for the sake of presentation, up to the precise choice of radii $r_1<r_2<\ldots$, the type tree of a tuple $\tup a$ is 
\[
\tree_q(\tup a)\coloneqq
\begin{cases}
\atp(\tup a)&\text{if $q=0$,}\\
\bigl(\atp(\tup a),\ \setof{(i,\tree_{q-1}(\tup a b))}{i\in[|\tup a|],\ b\in V,\ \dist(a_i,b)\le r_q}\bigr)&\text{if $q>0$,}
\end{cases}
\]
where $\atp(\tup a)$ is the set of atomic formulas satisfied by $\tup a$, including atoms that compare distances between the entries of $\tup a$ and the colors.
The index $i$ records which entry of $\tup a$ is close to $b$.
Two tuples have the same type tree if and only if they have the same local type (\Cref{cor:tp-tree}), so type trees are just another representation of local types.
Type trees resemble the morphism-trees of~\cite{tww1} and serve a similar purpose.
\Cref{fig:ov-type-tree} shows an example.

\begin{center}
  \captionsetup{type=figure}
  \begin{tikzpicture}[
      vx/.style={circle, fill, inner sep=1.6pt},
      vy/.style={circle, fill=blue!65!black, inner sep=1.4pt},
      vz/.style={rectangle, fill=orange!85!black, inner sep=1.8pt},
      e/.style={gray!70!black},
      tnode/.style={draw, rounded corners=2pt, font=\scriptsize, align=center, inner sep=3pt, fill=white},
      tedge/.style={gray!70!black},
      lbl/.style={font=\scriptsize},
      ilbl/.style={font=\scriptsize, inner sep=1pt}
    ]
    \fill[gray!10] (0,0) circle (1.4);
    \draw[gray, dashed] (0,0) circle (1.4);
    \node[lbl, gray!60!black] at (135:1.85) {$\Ball_{r_1}(a)$};
    \node[vx] (a) at (0,0) {};
    \node[lbl] at (270:0.3) {$a$};
    \foreach \ang/\v in {90/b,162/c,234/d} {
      \node[vy] (\v) at (\ang:0.95) {};
      \node[lbl] at (\ang:1.2) {$\v$};
      \draw[e] (a) -- (\v);
    }
    \foreach \ang/\v in {306/e,18/f} {
      \node[vz] (\v) at (\ang:0.95) {};
      \draw[e] (a) -- (\v);
    }
    \node[lbl] at (306:1.2) {$e$};
    \node[lbl, above=2pt of f] {$f$};
    \node[vy] (g) at (2.2,0.75) {};
    \node[lbl, right=2pt of g] {$g$};
    \draw[e] (f) -- (g);
    \node[vx] at (-0.9,-1.95) {};
    \node[lbl, right] at (-0.85,-1.95) {$X$};
    \node[vy] at (-0.05,-1.95) {};
    \node[lbl, right, text=blue!65!black] at (0,-1.95) {$Y$};
    \node[vz] at (0.8,-1.95) {};
    \node[lbl, right, text=orange!85!black] at (0.85,-1.95) {$Z$};
    \begin{scope}[yshift=-0.65cm]
      \node[tnode] (root) at (6.3,1.35) {$X(x_1)$};
      \node[lbl, left=6pt of root] {$\tree_1(a)$};
      \node[tnode] (c1) at (4.0,-0.55) {$x_1=x_2$\\ $X(x_2)$};
      \node[tnode] (c2) at (6.3,-0.55) {$x_1\neq x_2$\\$E(x_1,x_2)$\\ \textcolor{blue!65!black}{$Y(x_2)$}};
      \node[tnode] (c3) at (8.6,-0.55) {$x_1\neq x_2$\\$E(x_1,x_2)$\\ \textcolor{orange!85!black}{$Z(x_2)$}};
      \draw[tedge] (root) -- node[ilbl, pos=0.38, 
      above=1pt] {$1$} (c1);
      \draw[tedge] (root) -- node[ilbl, pos=0.38, 
      left=1pt] {$1$} (c2);
      \draw[tedge] (root) -- node[ilbl, pos=0.38, 
      above=1pt] {$1$} (c3);
      \node[ilbl, below=4pt of c1] {$a$};
      \node[ilbl, below=4pt of c2, text=blue!65!black] {$b,c,d$};
      \node[ilbl, below=4pt of c3, text=orange!85!black] {$e,f$};
    \end{scope}
  \end{tikzpicture}
  \caption{\emph{Left:} A partial graph with $7$ vertices and colors $X,Y,Z$. \emph{Right:} $\tree_q(a)$ for $q=1$ and $r_1=1$, drawn as a tree of depth $q$ whose nodes are the sets $\atp(a)$ and $\atp(a\,v)$. The edge label $1$ records that $v$ is close to the first entry of the tuple $(a)$. The lists of vertex labels below the boxes show which vertices of $\Ball_{r_1}(a)$ give each child of the root node. The vertex $g$ outside the locality radius \(r_1 = 1\) does not occur in the tree. Note that the size of the tree adapts to the number of atomic types $\atp(a\,v)$ realized in the ball.}
  \label{fig:ov-type-tree}
\end{center}

Type trees carry the same information as local types, but they may have a much smaller representation.
The type tree of a tuple $\tup a$ has one subtree for every type tree realized by an extension $\tup a b$ with $b$ close to some component $a_i$ of $\tup a$, so its size is governed by the nearby realized types rather than by the possible ones.
Roughly, this is like storing a small subset of a huge universe as the list of its elements, whereas naive type-based algorithms store it as its characteristic vector.
Showing that type trees have size $n^{o(1)}$ is a key step of our algorithm.
For this, we first leave the algorithm aside for a moment.

\paragraph{Counting types over parameters.}
We turn to a purely quantitative result about monadically dependent classes, which counts types over a set of parameters and may be of independent interest.
Let $H$ be a graph or partial graph, $U\subseteq V(H)$, and $k,q\in\N$.
The \emph{type over $U$} of a tuple $\tup a\in V(H)^k$ is the set of first-order formulas $\psi(x_1,\ldots,x_k)$ of quantifier rank at most $q$ which are satisfied by $\tup a$ in $H$,
where $\psi$ may use any number of vertices of $U$ as constants, called \emph{parameters}. We denote this type by $\tp^H_q(\tup a/U)$.
For $X\subseteq V(H)$, we write $\Types^H_{k,q}(X/U)$ for the set of types over $U$ of tuples from $X^k$,
$$\Types^H_{k,q}(X/U)=\setof{\tp^H_q(\tup a/U)}{\tup a\in X^k}.$$
This set has at most $|X|^k$ elements, but can be much smaller, as two tuples  have the same type if no formula distinguishes them.
We write $\Types^H_{k,q}(X)$ if $U=\emptyset$.
One of the main contributions of this paper is the following bound on the number of types over parameters in monadically dependent graph classes.

\begin{theorem}[Simplified form of \Cref{thm:type-bound}]\label{thm:ov-type-bound}
  Let $\CC$ be a monadically dependent graph class and $k,q\in\N$. There is some $d\in\N$ such that for every $G\in\CC$ and every set $U\subseteq V(G)$ with $|U|>1$,
  $$|\Types^G_{k,q}(V(G)/U)|\le |U|^d.$$
\end{theorem}

The proof of \Cref{thm:ov-type-bound} proceeds as follows. We fix some radius $r$ depending on the quantifier rank $q$.
We then show (in \Cref{thm:closures})  that there is a set $\wh U$ of size polynomial in $|U|$, 
such that for every vertex $v\in V(G)$,
there is some  \emph{separator} $S\subseteq \wh U$ of size bounded by some constant $s$,
such that $S$ \emph{separates} $v$ from $U$ \emph{at a distance} $r$.
The notion of separation is technical, 
but it ensures that the following holds
for all $v\in V(G)$ which are $r$-separated from $U$ by $S$:
\begin{quote}\centering
The type $\tp^G_q(v/U)$ is determined by the type $\tp^G_{q'}(v/S)$.
\end{quote}
Here $q'$ depends only on $q$.
Since $S$ has size bounded by $s$, and both $s$ and $q'$ are constants (depending only on $\CC$ and $q$),
the number of types $\tp^G_{q'}(v/S)$ is bounded by a constant $c$ (depending on $\CC$ and $q$). Since there are at most $|\wh U|^s$ considered separators $S$, it follows that altogether there are at most $c\cdot |\wh U|^s$  types over $U$ of single vertices $v\in V(G)$, of quantifier rank $q$. This gives the required polynomial bound on the number of types of single vertices $v$ over a set~$U$. The argument for $k$-tuples $\tup v$ proceeds similarly.

\paragraph{The size of type trees.}
We now return to type trees and bound their sizes using the bound on the number of types.
The following lemma makes precise that type trees store only the types realized nearby.

\begin{lemma}[informal version of \Cref{lem:ball-tree-bound}]\label{lem:ov-tree-size}
  Let $q\in\N$. There is $d\in\N$ depending only on $q$ such that for every partial graph $H$ and vertex $v$,
  the type tree $\tree_q(v)$ in $H$ has size polynomial in $K\coloneqq|\Types^H_{d,d}(\Ball^H_d(v))|$.
\end{lemma}

We bound $K$ in \Cref{lem:ov-tree-size} using the bound on the number of types (\Cref{thm:ov-type-bound}).
The two count different things.
$K$ counts types without parameters in a partial graph $G_t$, whose colors are not present in $G$, and resolved pairs a priori form an arbitrary subset of vertex pairs.
On the other hand, the type bound counts types in $G$ itself, over a set of parameters.
While \(G\) comes from a monadically dependent graph class, \(G_t\) a priori does not, making it difficult to enforce structural properties on \(G_t\).

We bridge this gap by computing construction sequences that are \emph{locally definable} in $G$:
in every local ball, the colors and resolved pairs of $G_t$ can be expressed by first-order formulas in $G$ that mention only $n^{o(1)}$ vertices as parameters.
A challenge in encoding $G_t$ into $G$ is that $G_t$ accumulates many resolves over time, which cannot be modeled nicely by parameters.
To overcome this, we turn to \emph{merge sequences}, which group many merges and resolves into a single step, and can therefore be much shorter than construction sequences.
For every $\eps>0$,~\cite{md-abmw} provides a chain of sets $V(G)=S_1\supseteq\dots\supseteq S_\ell=\emptyset$, a closely related decomposition, with only $\ell=O(1/\eps)$ steps, which we convert into a merge sequence of width $n^{O(\eps)}$ with the same number of steps.
Each step can be encoded in $G$ by $O(1)$ unary predicates, and when we refine the merge sequence into a construction sequence, every $G_t$ stays locally definable using these $O(1/\eps)$ predicates and few parameters.
The bound on the number of types (\Cref{thm:ov-type-bound}) extends to such expansions of $G$ by boundedly many unary predicates, so it bounds $K$ by $n^{O(\eps)}$, and by \Cref{lem:ov-tree-size}, every type tree has size $n^{O(\eps)}$ as well.

\paragraph{Lifting type trees through merges and resolves.}
Now that we know that type trees are small, it remains to compute them efficiently.
Our algorithm processes the construction sequence step by step and maintains the type tree of every vertex.
So after every merge and every resolve, it has to compute the new type trees from the old ones.
Merges are easy.
Merging two parts $P$ and $Q$ into $P\cup Q$ does not change the resolved pairs, and hence not the distances.
Given a type tree from before the merge, we can update it as follows:
in every node, remove all atoms that mention $P$ or $Q$, and add an atom mentioning $P\cup Q$ whenever the same atom with $P$ or with $Q$ in its place was present.
The difficulty lies in resolves, which create new short paths and thereby bring new vertices close to a vertex $v$ (see \Cref{fig:ov-resolve}).

\begin{center}
  \captionsetup{type=figure}
  \begin{tikzpicture}[
      vx/.style={circle, fill, inner sep=1.6pt},
      vp/.style={circle, fill=blue!65!black, inner sep=1.4pt},
      vq/.style={rectangle, fill=orange!85!black, inner sep=1.8pt},
      e/.style={gray!70!black},
      new/.style={red!75!black, thick, dashed},
      lbl/.style={font=\scriptsize}
    ]
    \fill[gray!5] (0,1.4) -- (2.7,1.4) arc[start angle=90, end angle=-90, radius=1.4] -- (0,-1.4) arc[start angle=270, end angle=90, radius=1.4] -- cycle;
    \draw[gray, dashed] (0,1.4) -- (2.7,1.4) arc[start angle=90, end angle=-90, radius=1.4] -- (0,-1.4) arc[start angle=270, end angle=90, radius=1.4] -- cycle;
    \node[lbl, gray!60!black] at (1.35,1.65) {$\Ball^{H'}_r(v)$};
    \fill[gray!15] (0,0) circle (1);
    \draw[gray, dashed] (0,0) circle (1);
    \node[lbl, gray!60!black] at (-0.2,-0.5) {$\Ball^H_r(v)$};
    \node[vx] (v) at (0,0) {};
    \node[lbl, left] at (v) {$v$};
    \node[vp] (p1) at (0.65,0.55) {};
    \node[vp] (p2) at (0.65,-0.55) {};
    \draw[e] (v) -- (p1);
    \draw[e] (v) -- (p2);
    \node[vq] (q1) at (2.15,0.75) {};
    \node[vq] (q2) at (2.15,-0.75) {};
    \node[vx] (c) at (2.95,0) {};
    \draw[e] (q1) -- (c);
    \draw[e] (q2) -- (c);
    \foreach \p in {p1,p2} {
      \foreach \q in {q1,q2} {
        \draw[new] (\p) -- (\q);
      }
    }
    \node[vp] at (0.7,-1.85) {};
    \node[lbl, right, text=blue!65!black] at (0.75,-1.85) {$P$};
    \node[vq] at (1.6,-1.85) {};
    \node[lbl, right, text=orange!85!black] at (1.65,-1.85) {$Q$};
  \end{tikzpicture}
  \caption{A resolve of the parts $P$ and $Q$, positive or negative, resolves all pairs between $P$ and $Q$ (red). This creates short paths from $v$ to vertices that were far from $v$ before, so the ball of $v$ grows from $\Ball^H_r(v)$ to $\Ball^{H'}_r(v)$. The old type tree of $v$ contains no information about the new vertices in the ball. The new type tree of $v$ therefore has to be assembled from the old type trees of $v$ and of these vertices.}
  \label{fig:ov-resolve}
\end{center}

In~\cite{mw-arxiv}, resolves are handled by a rank-preserving locality theorem~\cite{locality-note}.
Fix a vertex $v$.
For every formula that could \emph{potentially} belong to the local type of $v$ after the resolve, the theorem expresses whether $v$ satisfies it in terms of the local type of $v$ before the resolve (and some global information about the partial graph).
So the new local type is assembled backwards, one potential formula at a time.
For width $n^{o(1)}$, there are far too many of them, even if only few types are realized.

We instead work forwards.
We use the same ideas as in the proof of the locality theorem~\cite{locality-note}, but turned around:
we construct the new type tree directly from the old type trees, and spend work only on type trees that are actually realized.
This is summarized by the following lemma.

\begin{lemma}[informal version of \Cref{lem:resolve-lift}]\label{lem:ov-resolve-lift}
  Let $q\in\N$. There is $d\in\N$ depending only on $q$ such that the following holds.
  Given a partial graph $H$, a resolve producing $H'$, and the type trees of all vertices of $H$,
  one can compute the type trees of all vertices of $H'$ in time polynomial in $n$ and
  $K\coloneqq\max_{v}|\Types^H_{d,d}(\Ball^{H'}_d(v))|$.
\end{lemma}

\paragraph{A route not taken: games and covers.}
The algorithms for nowhere dense and monadically stable classes face the same problem of an unbounded number of local types.
They sidestep it by a bounded-depth recursion into local neighborhoods, guided by the splitter game~\cite{gks} or the flipper game~\cite{flippers,ms-mc1,ms-mc2}, with neighborhood covers keeping the recursion tree small.
We tried and failed to adapt the approach to the monadically dependent setting, before settling on our current approach.

\paragraph{Organization.}
After the preliminaries in \Cref{sec:prelims}, \Cref{sec:main-proof} states the ingredients formally and derives \Cref{thm:main} from them.
The remaining parts prove the ingredients:
\Cref{part:types} proves the bound on the number of types (\Cref{thm:ov-type-bound}),
\Cref{part:definable-construction-sequences} computes the locally definable construction sequences,
and \Cref{part:mc} gives the model checking algorithm based on type trees (\Cref{lem:ov-tree-size,lem:ov-resolve-lift}).

\section{Preliminaries}\label{sec:prelims}

$\N$ denotes the set of non-negative integers and $[n]\coloneqq\set{1,\ldots,n}$ for $n\in\N$.
For a set $V$, we write $\binom V2$ for the set of unordered pairs $uv$ of distinct elements of $V$, and for $X,Y\subseteq V$ we put $XY\coloneqq\setof{xy}{x\in X,\ y\in Y,\ x\ne y}$. A graph $G$ consists of a finite set $V$ of \emph{vertices} and a set $E\subseteq \binom{V}{2}$ of \emph{edges}, denoted by $V(G)$ and $E(G)$.
The (open) \emph{neighborhood} of a vertex $v\in V(G)$ is the set $N^G(v)\coloneqq\setof{w\in V(G)}{vw\in E(G)}$.
For $r\in \N$, denote $\Ball^G_r(v)\coloneqq\setof{w\in V(G)}{\dist^G(v,w)\le r}$, where $\dist^G(\cdot,\cdot)$ is the shortest-path distance in~$G$.
These notations extend to sets $A,B\subseteq V(G)$ with $\Ball_r^G(A) \coloneqq \bigcup_{a\in A} \Ball_r^G(a)$ and $\dist^G(A,B)\coloneqq\min_{a\in A,\,b\in B}\dist^G(a,b)$, and to tuples of vertices as expected.

Let $G$ be a graph with vertex set $V$ and edge set $E$. By $\overline G$ we denote the edge-complement of $G$, which has  vertex set $V$ and edge set $\binom{V}{2}-E$.
The subgraph of $G$ \emph{induced} by a set $X\subseteq V$ is the graph $G[X]$ with vertex set $X$ and edge set $E\cap \binom{X}{2}$. For two disjoint sets $X,Y\subseteq V$,
by $G[X,Y]$ we denote the bipartite graph with parts $X$ and $Y$ and edge set $XY\cap E$.

A pair of sets $X,Y\subseteq V(G)$ is \emph{complete} if $XY\subseteq E(G)$ and \emph{anticomplete} if $XY\subseteq E(\overline G)$. The pair is \emph{homogeneous} if it is complete or anticomplete.

A \emph{graph class} is a set $\CC$ of graphs.

\paragraph{Partitions.}
For two partitions $\PP,\QQ$ of the same set we write $\PP\preceq \QQ$ to denote that $\PP$ refines $\QQ$, i.e., each part of $\PP$ is contained in some part of $\QQ$.
If $\PP$ is a partition of $V$ and $X\subseteq V$, denote
$X/\PP\coloneqq\{P\in\PP:P\cap X\ne\emptyset\}$.
The part of $\PP$ containing an element $v$ is denoted by $\PP(v)$.
Further, $\equiv_\PP$ is the equivalence relation on $V$ whose classes are the parts of $\PP$.

\paragraph{Asymptotic notation.}
For a possibly empty list of parameters $k$, 
 by $O_k(\cdot)$ we denote
 an anonymous function $f\from \R_{\ge 0}\to \R_{\ge 0}$ (depending only on the parameters~$k$) which is upper bounded by $x\mapsto a\cdot x+b$, for some $a,b\in\N$.


Similarly, we write $\poly_k(\cdot)$ for an anonymous function $f\from \R_{\ge 0}\to \R_{\ge 0}$ which is upper bounded by some polynomial
;
in other words, $g(n)\le \poly_k(n)$ has the same implications as $g(n)\le (n+2)^{O_k(1)}$.
We use $\widehat O_k(n^c)$ for a bound $O_{k,\eps}(n^{c+\eps})$ for every $\eps>0$; thus it hides a subpolynomial factor that may depend on the parameters~$k$.

\paragraph{Monadic dependence.} We view graphs as relational structures in the signature with one (symmetric, irreflexive) binary relation $E$.
For $k\in\N$, a \emph{$k$-colored graph} is a graph equipped with $k$ unary predicates $U_1,\ldots,U_k$.
A \emph{$k$-coloring} $G^+$ of a graph $G$ assigns interpretations of these predicates on $V(G)$; the predicates need not be disjoint.
A \emph{coloring} of $G$ is a $k$-coloring for some $k\in\N$.

A graph class $\CC$ is \emph{monadically independent} if there are $k\in\N$ and a first-order formula $\varphi(x,y)$ in the signature of $k$-colored graphs such that  every finite graph $H$ can be obtained from some graph $G\in\CC$ as follows: first choose a $k$-coloring $G^+$ of $G$, then consider the graph $\varphi(G^+)$ with vertex set $V(G)$ and edges $uv\in\binom{V(G)}2$ such that $G^+\models\varphi(u,v)\vee\varphi(v,u)$; finally take some induced subgraph of $\varphi(G^+)$, obtaining a graph isomorphic to $H$.
A graph class $\CC$ is \emph{monadically dependent} if it is not monadically independent.

\paragraph{Neighborhood complexity.}
The \emph{neighborhood complexity} of a graph $G$ is the function $\pi_G\from\N\to\N$ defined by:
\[
    \pi_G(k) \coloneqq
    \max_{\substack{S\subseteq V(G)\\|S|\le k}}
    \left|\{N^G(v)\cap S:v\in V(G)\}\right|,
\]
 which counts the maximum number of distinct neighborhoods of vertices in $G$ restricted to a set of size at most~$k$.

The \emph{VC-dimension} of a graph $G$ is the maximal size of a set $A\subseteq V(G)$ such that $\setof{N(v)\cap A}{v\in V(G)}=2^A$. The Sauer--Shelah lemma~\cite{Sauer72,She72} implies the following.
\begin{lemma}\label{lem:ss}
  If $G$ has VC-dimension $d$ then 
  $$\pi_G(k)\le (k+1)^d\quad\text{for all }k\in\N.$$
\end{lemma}
The following is easy to see from the definition of monadic dependence; see~\cite[Observation~3.2]{md-abmw} for a proof.
\begin{observation}\label{obs:md-vc}
For every monadically dependent graph class $\CC$ there is some $d\in\N$ 
such that every graph $G\in\CC$ has VC-dimension at most $d$.
\end{observation}


\paragraph{Types and parameters.}
For a structure $A$ and a set $W$ of its elements, 
let $A_W$ denote the structure obtained by expanding $A$  with constants representing each element of $W$. 
A formula $\phi(x_1,\ldots,x_k)$ \emph{with parameters} from $W$ is a formula in the signature of the structure $A_W$.
For $\bar v\in V(A)^k$ we write $A\models\phi(\bar v)$ to denote $A_W\models\phi(\bar v)$.

For $k,q\in\N$
and a tuple $\bar v\in V(A)^k$, 
define
\emph{the type of $\bar v$ of quantifier rank $q$ over $W$}, denoted
$$\tp^A_q(\bar v/W),$$ as the set of all formulas $\phi(x_1,\ldots,x_k)$ with parameters from $W$ and of quantifier rank at most $q$ such that 
$A\models \phi(\bar v)$.

For two sets $X,Y\subseteq V(A)$,
denote $$\Types^A_{k,q}(X/Y)\coloneqq \setof{\tp^A_q(\bar v/Y)}{\bar v\in X^k}.$$ 
We omit $/Y$ from the notation if $Y=\emptyset$.
The \emph{atomic type} of $\bar v$ over $Y$ is defined as $\atp^A(\bar v/Y) \coloneqq \tp_0^A(\bar v/Y)$, and we set $\ATypes^A(X/Y)\coloneqq \Types^A_{1,0}(X/Y)$.

Note that for a graph $G$ and vertex set $S\subseteq V(G)$, two vertices $u,v\in V(G)$ have the same atomic type over $S$ if and only if either $u=v$, or $u,v\notin S$ 
and $N(u)\cap S=N(v)\cap S$. 
The \emph{partition into atomic types}, denoted $\PP_S$, is the corresponding partition of 
$V(G)$, which partitions $S$ into singletons, and vertices in $V(G)-S$ according to their neighborhood in $S$.

\paragraph{Gaifman graph, distances, balls.}
For a structure $A$ with domain $V$ equipped with unary and binary relations, the \emph{Gaifman graph} of $A$ is the graph 
with vertex set $V$ and edge set consisting of all pairs $uv$ such that 
$(u,v)$ or $(v,u)$ is in some binary relation of $A$.
We write $\dist^A(u,v)$ to denote the distance between two vertices in the Gaifman graph of $A$, and $\Ball^A_r(v)$  for the set of vertices at distance at most $r$ from $v$ in the Gaifman graph of $A$.

\paragraph{Flips and flip metrics.}
For a graph $G$ and partition $\PP$ of $V\coloneqq V(G)$, a \emph{$\PP$-flip} of $G$ is a graph $H$ with $V(H)=V$ such that for every pair $X,Y\in \PP$ 
(possibly $X=Y$), either 
$H[X,Y]=G[X,Y]$ or $H[X,Y]=\overline G[X,Y]$.
If $|\PP| \leq s$, then we call $H$ an \emph{$s$-flip of $G$}. 

Define the associated \emph{flip metric} and its balls by:
\begin{align*}
  \dist_\PP(u,v)&\coloneqq
  \max\{\dist^H(u,v):H\text{ is a $\PP$-flip of }G\},
  \\
  \Ball^\PP_r(v)&\coloneqq\{u:\dist_\PP(u,v)\le r\}=\bigcap\setof{\Ball^H_r(v)}{\text{$H$ is a $\PP$-flip of $G$}}.
\end{align*}

As a shorthand, for a graph $G$ and a set $S\subseteq V(G)$, a $\PP_S$-flip of $G$ is called an \emph{$S$-flip}, and we set $\dist_S^G \coloneqq \dist_{\PP_S}$ and $\Ball_r^S \coloneqq \Ball_r^{\PP_S}$.

\paragraph{Partial graphs and merge-width.}
Partial graphs, construction sequences, their radius-$r$ width, and the radius-$r$ merge-width of a graph are defined in \Cref{sec:tech-overview} (see also~\cite{mw-stoc}).
Moreover, for a partial graph $H$ with partition $\PP$ and a set $W\subseteq V(H)$, the induced partial graph $H[W]$ is the partial graph with vertex set $W$,
edges $E(H)\cap \binom{W}{2}$, non-edges 
$N(H)\cap \binom{W}{2}$, and partition $\setof{P\cap W}{P\in \PP}-\set{\emptyset}$.

\section{Proof of the main theorem}\label{sec:main-proof}

The upcoming Parts~\ref{part:types}, \ref{part:definable-construction-sequences}, and~\ref{part:mc} prove three independent results.
This section combines them into the main result, \Cref{thm:main}. It does so via two theorems that are connected by the following type-based width notion.

\begin{restatable}{definition}{defrqwidth}\label{res:width}
For $r,q\in\N$,
the \emph{radius-$r$ depth-$q$ width}
of a construction sequence $G_1,\ldots,G_\lambda$
is defined as:
   $$\max_{t\in[\lambda-1]}\max_{v\in V(G)}|\Types^{G_t}_{q,q}({\Ball^{G_{t+1}}_r(v)})|.$$
  \end{restatable}

The offset between $G_t$ and $G_{t+1}$ mirrors the offset between $\PP_{i-1}$ and $R_i$ in the width definition of chains and of merge sequences (see \Cref{def:chain-width,def:merge-sequence}).

\begin{theorem}\label{thm:skel-few-types}
  Let $\CC$ be a monadically dependent graph class, $r,q\in\N$, and let $\eps>0$ be a rational.
  There is a constant $\alpha\ge1$ depending only on $\CC$ and $q$ such that every $n$-vertex graph $G\in\CC$
  has a construction sequence of \text{radius-$r$} depth-$q$ width $O_{\CC,r,q,\eps}(n^{\alpha\eps})$ and length $O_{\CC,r,q,\eps}(n^{1+\alpha\eps})$. 
  Moreover, there is an algorithm (independent of~$\CC$) which, given $r,q\in\N$, a rational $\eps>0$, and an $n$-vertex graph $G\in\CC$,
  computes such a construction sequence in time $\widehat O_{\CC,r,q,\eps}(n^3)$.
\end{theorem}
\noindent 
It is important that $\alpha$ does not depend on $\eps$, as this allows us to make the exponent $\alpha\eps$ of the width as small as we like.
The upcoming \Cref{sec:skel-few-types} assembles \Cref{thm:skel-few-types} from the key results of 
\Cref{part:types,part:definable-construction-sequences}.

\newcommand{\mcdepth}{q}
\begin{restatable}{theorem}{mcthm}\label{thm:mc}
  There is an algorithm which, given
  a sentence $\phi$ and an $n$-vertex graph~$G$ together with its construction sequence of length $\lambda$, determines whether $G\models\phi$ in time
  $\lambda\cdot n^2\cdot\poly_{\phi}(K)$,
  where $K$ is the radius-$r$ depth-$\mcdepth$ width of the construction sequence,
  for some $r$ and $\mcdepth$ which are computable from $\phi$.
\end{restatable}


\begin{proof}
  See \Cref{part:mc}.
\end{proof}

From these two theorems, we can derive
\Cref{thm:main} without much effort.
The only subtlety is that the algorithm does not know $\CC$, so it cannot choose $\eps$ itself.

\mainthm*

\begin{proof}[Proof of \Cref{thm:main}]
  Let $\phi$ be the input sentence, and let $r,q$ be the numbers from \Cref{thm:mc}.
  For $j\ge1$, let $\mathcal A_j$ be the algorithm that computes a construction sequence of $G$ by \Cref{thm:skel-few-types} with $r,q$, and $\eps\coloneqq1/j$,
  and then decides whether $G\models\phi$ by \Cref{thm:mc}.

  Fix a monadically dependent graph class $\CC$, and let $\alpha\ge1$ be the constant from \Cref{thm:skel-few-types} for $\CC$ and $q$.
  Let $c\ge1$ be such that the polynomial in \Cref{thm:mc} is $O_\phi(K^c)$.
  On $n$-vertex graphs $G\in\CC$, the sequence computed by $\mathcal A_j$ has radius-$r$ depth-$q$ width $K\le O_{\CC,\phi,j}(n^{\alpha/j})$ and length $\lambda\le O_{\CC,\phi,j}(n^{1+\alpha/j})$.
  So $\mathcal A_j$ runs in time
  \[
    \widehat O_{\CC,\phi,j}(n^3)+O_\phi(\lambda\cdot n^2\cdot K^c)
    \le O_{\CC,\phi,j}(n^{3+\beta/j}),
  \]
  where $\beta\coloneqq(c+1)\alpha$.
  Since $\beta/j>0$, the subpolynomial factor hidden by $\widehat O$ is absorbed by $n^{\beta/j}$.
  The number $\beta$ depends on $\CC$ and $\phi$, but not on~$j$.
  
  The algorithm does not know $\CC$, so it cannot choose $j$.
  Instead, it runs all algorithms $\mathcal A_j$ in parallel, giving $\mathcal A_j$ a $2^{-j}$ share of the computation, and returns the first answer.
  Its answer is correct, as every $\mathcal A_j$ correctly solves the model checking problem on every input graph.

  For the running time, let $\eps>0$ and $j\coloneqq\lceil2\beta/\eps\rceil$.
  If $\mathcal A_j$ halts within time $T$, then it completes within the first $2^jT$ simulated steps, so the algorithm halts within time
  \[
    O\bigl(2^jT\log(2^jT+2)\bigr)=O_j\bigl(T\log(T+2)\bigr).
  \]
  Hence, on $n$-vertex graphs in $\CC$, the algorithm runs in time
  \[
    O_{\CC,\phi,\eps}\bigl(n^{3+\beta/j}\log(n+2)\bigr)
    \le O_{\CC,\phi,\eps}(n^{3+\eps}),
  \]
  since $\beta/j\le\eps/2$ and $\log(n+2)=O_\eps(n^{\eps/2})$.
\end{proof}

\subsection{Proving \texorpdfstring{\Cref{thm:skel-few-types} from \Cref{part:types,part:definable-construction-sequences}}{Theorem \getrefnumber{thm:skel-few-types} from Parts I and II}}\label{sec:skel-few-types}

We first state the key results of the two parts, and derive \Cref{thm:skel-few-types} afterwards.

\subsubsection{The key result of \texorpdfstring{\Cref{part:types}}{Part I}}\label{sec:skel-part-types}

The key ingredient towards proving \Cref{thm:skel-few-types} is the following type bound derived in \Cref{part:types}.

\newif\iftypeboundgeneral
\begin{restatable}{theorem}{thmtypes}\label{thm:type-bound}

Let $\CC$ be a monadically dependent graph class and $k,q,c\in\N$.
For every \iftypeboundgeneral(possibly infinite) \fi graph $G \in \CC$, expansion $G^+$ of $G$ by $c$ unary predicates, and \iftypeboundgeneral finite \fi subset $U \subseteq V(G)$,
\[
    \left|\Types^{G^+}_{k,q}(V(G)/U)\right| \leq O_{\CC,q,k,c}(1)\cdot \poly_{\CC,q,k}(|U|).
\]

\end{restatable}

The number $c$ of unary predicates affects only the constant factor, not the degree of the polynomial.
This matters, as our construction sequences use $O(r/\eps)$ unary predicates, while the constant $\alpha$ in \Cref{thm:skel-few-types} must not depend on $\eps$.

\subsubsection{The key result of \texorpdfstring{\Cref{part:definable-construction-sequences}}{Part II}}\label{sec:skel-part-defseq}

The type bound counts types in $G$, while the width (\Cref{res:width}) counts types in the partial graphs $G_t$.
To bridge this gap, \Cref{part:definable-construction-sequences} produces a construction sequence in which local balls of $G_t$ are defined by formulas in $G$.

\begin{restatable}{definition}{defpartialdef}\label{def:partial-definable}
Let $G$ be a graph with vertex set $V$ and let $H$ be a partial graph with vertex set $W\subseteq V$.
For $q,c,w\in \N$,
we say that $H$ is \emph{definable} in $G$  by formulas of quantifier rank $q$ using $c$ unary predicates and $w$ parameters, if
there is an expansion $G^+$ of $G$ by $c$ unary predicates
and a set $S\subseteq V(G)$ of parameters with $|S|\le w$,
such that for every relation symbol $Q$ of $H$ (when viewed as a relational structure) of arity $k$ there is some formula $\phi_Q(x_1,\ldots,x_k)$ of quantifier rank $q$ with parameters from $S$
such that
$$G^+\models\phi_Q(\bar v)\quad\iff\quad \bar v\in W^k\text{ and }H\models Q(\bar v)\qquad\quad\text{for all $\bar v\in V^k$.}$$
\end{restatable}

The vertex set $W$ of $H$ is the union of its parts, so it is defined by the disjunction of the formulas defining the parts.

\begin{restatable}{definition}{defskellocdef}\label{def:skel-loc-def}
  Let $G$ be a graph and $r,c,s,q_0\in\N$.
  A construction sequence $G_1,\ldots,G_\lambda$ of $G$ is \emph{radius-$r$ definable} with $c$ unary predicates, $s$ parameters, and quantifier rank $q_0$
  if for all $t\in[\lambda-1]$ and $v\in V(G)$, the induced partial graph
  \[
      G_t[\Ball^{G_{t+1}}_r(v)]
  \]
  is definable in $G$ by formulas of quantifier rank $q_0$ using $c$ unary predicates and $s$ parameters.
\end{restatable}

\noindent Note that the expansion and the parameters may depend on $t$ and $v$.

\medskip

The key result of \Cref{part:definable-construction-sequences} is the following theorem.
\begin{restatable}{theorem}{thmskeldefseq}\label{thm:skel-def-seq}
  Let $r\ge 1$ be an integer, $\eps>0$ be a rational, and $\CC$ be a monadically dependent graph class. There is a constant $\delta\ge1$ depending only on $\CC$ such that every $n$-vertex graph $G\in\CC$
  has a construction sequence of $G$ which is radius-$r$ definable with $O(r/\eps)$ unary predicates, $O_{\CC,r,\eps}(n^{\delta\eps})$ parameters, and quantifier rank $7$, and has length $O_{\CC,r,\eps}(n^{1+\delta\eps})$. Moreover, there is an algorithm (independent of $\CC$)
  which, given $r$, $\eps$, and $G$, computes such a sequence, and runs in time $\widehat O_{\CC,r,\eps}(n^3)$ whenever $G\in\CC$.
\end{restatable}

For nowhere dense classes, the following stronger bounds follow easily from known results. The proof is given in \Cref{sec:nd}.

\begin{restatable}{theorem}{thmnddefseq}\label{thm:nd-def-seq}
  Let $\CC$ be a nowhere dense graph class, $r\in\N$, and $\eps>0$.
  Every $n$-vertex graph $G\in\CC$ has a radius-$r$ definable construction sequence with one unary predicate, $O_{\CC,r,\eps}(n^\eps)$ parameters, quantifier rank $0$, and length $O_{\CC,r,\eps}(n^{1+\eps})$.
  Such a sequence can be computed in time $O_{\CC,r,\eps}(n^{1+\eps})$.
\end{restatable}

In particular, \Cref{thm:nd-def-seq}, together with the results of \Cref{part:types,part:mc}, gives an alternative proof of the result of Grohe, Kreutzer, and Siebertz~\cite{gks} that first-order model checking is fixed-parameter tractable on nowhere dense classes.
(Alternatively, in nowhere dense classes one can replace the type bound from \Cref{part:types}  by a stronger bound deducible from \cite{number-of-types}).

\subsubsection{Assembling the parts}\label{sec:skel-assembly}

We use the following consequence of Gaifman locality, proved in \Cref{app:gaifman}.

\begin{restatable}[Corollary of Gaifman locality]{lemma}{lemLocalToGlobal}\label{lem:local-to-global}
For every $k,q,\alpha \in\N$, there are numbers $h=h(q)$ and $q'=q'(k,q,\alpha)$ computable from $k,q,\alpha$,
 with the following property.
Fix a finite relational signature $\sigma$ of arity at most $\alpha$ and a $\sigma$-structure $A$.
Then, for all $\bar a\in V(A)^k$
and $B\supseteq \Ball_h^A(\bar a)$,
$\tp_q^A(\bar a)$ depends only on
 $\tp_{q'}^{A[B]}(\bar a)$.
\end{restatable}

\begin{lemma}\label{lem:skel-def-to-types}
  For every $q\in\N$, there is $h\in\N$, computable from $q$, such that the following holds for every monadically dependent graph class $\CC$ and all $r,c,s,q_0\in\N$.
  If a construction sequence of a graph $G\in\CC$ is radius-$(r+h)$ definable with $c$ unary predicates, $s$ parameters, and rank $q_0$,
  then its radius-$r$ depth-$q$ width is at most $O_{\CC,q,q_0,c}(1)\cdot\poly_{\CC,q,q_0}(s)$.
\end{lemma}
\begin{proof}
  We will apply \Cref{lem:local-to-global} to partial graphs from a construction sequence of $G$.
  These partial graphs will have a finite signature $\sigma$ with up to $|V(G)|$ unary predicates marking the partition, but the arity is fixed to $\alpha=2$.
  Hence, we can choose $h\coloneqq h(q)$ and $q'\coloneqq q'(q,q,2)$ from \Cref{lem:local-to-global}.
  Let $G_1,\ldots,G_\lambda$ be a construction sequence of $G\in\CC$ which is radius-$(r+h)$ definable with $c$ unary predicates, $s$ parameters, and rank $q_0$.
  Fix $t\in[\lambda-1]$ and $v\in V(G)$, and let $B\coloneqq\Ball^{G_{t+1}}_r(v)$ and $W\coloneqq\Ball^{G_{t+1}}_{r+h}(v)$.
  By \Cref{res:width}, we have to bound $\left|\Types^{G_t}_{q,q}(B)\right|$.

  First, we move from $G_t$ to the partial graph $G_t[W]$.
  As $G_{t+1}$ has all resolved pairs of $G_t$, every tuple $\bar u\in B^q$ satisfies $\Ball^{G_t}_h(\bar u)\subseteq W$.
  So by \Cref{lem:local-to-global}, $\tp^{G_t}_q(\bar u)$ is determined by $\tp^{G_t[W]}_{q'}(\bar u)$, and hence
  \[
    \left|\Types^{G_t}_{q,q}(B)\right|
    \le\left|\Types^{G_t[W]}_{q,q'}(W)\right|.
  \]

  Next, we move from $G_t[W]$ to $G$.
  By assumption, $G_t[W]$ is definable in an expansion $G^+$ of $G$ by $c$ unary predicates, by formulas of quantifier rank $q_0$ with parameters from a set $S$ of at most $s$ vertices.
  Replace every atom of a formula over $G_t[W]$ by its defining formula, and restrict every quantifier to the vertex set of $G_t[W]$.
  This turns a formula of quantifier rank $q'$ over $G_t[W]$ into a formula of quantifier rank at most $q'+q_0$ over $G^+$ with parameters from $S$,
  which holds for the same tuples from $W$.
  So for every $\bar u\in W^q$, $\tp^{G_t[W]}_{q'}(\bar u)$ is determined by $\tp^{G^+}_{q'+q_0}(\bar u/S)$.

  Together with \Cref{thm:type-bound}, we obtain
  \[
    \left|\Types^{G_t}_{q,q}(B)\right|
    \le\left|\Types^{G_t[W]}_{q,q'}(W)\right|
    \le\left|\Types^{G^+}_{q,q'+q_0}(V(G)/S)\right|
    \le O_{\CC,q,q'+q_0,c}(1)\cdot\poly_{\CC,q,q'+q_0}(s).
  \]
  As $q'$ depends only on $q$, this is the claimed bound $O_{\CC,q,q_0,c}(1)\cdot\poly_{\CC,q,q_0}(s)$.
\end{proof}

\begin{proof}[Proof of \Cref{thm:skel-few-types}]
  Let $\CC$ be a monadically dependent graph class containing $G$.
  Let $h$ be the number given by \Cref{lem:skel-def-to-types} for depth $q$.
  We run the algorithm of \Cref{thm:skel-def-seq} with radius $r+h$ and $\eps$, which takes time $\widehat O_{\CC,r,q,\eps}(n^3)$.
  
  By \Cref{thm:skel-def-seq}, there is a constant $\delta\ge1$ depending only on $\CC$ such that the sequence is radius-$(r+h)$ definable with $c\le O((r+h)/\eps)\le O_{r,q,\eps}(1)$ unary predicates, $s\le O_{\CC,r,q,\eps}(n^{\delta\eps})$ parameters, and rank $q_0\coloneqq7$,
  and it has length $O_{\CC,r,q,\eps}(n^{1+\delta\eps})$.
  By \Cref{lem:skel-def-to-types}, there is an integer $D\ge1$ such that its radius-$r$ depth-$q$ width is at most
  \[
    O_{\CC,q,q_0,c}(1)\cdot\poly_{\CC,q,q_0}(s)\le O_{\CC,r,q,\eps}(n^{D\delta\eps}),
  \]
  where we used $q_0=7$ and $c\le O_{r,q,\eps}(1)$. As $D$ depends only on $\CC$ and $q$, we can choose $\alpha\coloneqq D\delta$ to satisfy the statement.
  Since $D\ge1$ and $\delta\ge1$, we have $\alpha\ge \delta\ge1$, so this choice also bounds the sequence length by $O_{\CC,r,q,\eps}(n^{1+\alpha\eps})$.
\end{proof}

\clearpage
\part{Number of types}\label{part:types}
In this part we prove the following result.
As we believe it to be of independent interest also in the context of infinite graphs,
all considered graphs may be finite or infinite in \Cref{part:types},
and we track the necessary finiteness assumptions in the arguments below.
Everywhere else in the paper, graphs are assumed to be finite.

{\typeboundgeneraltrue\thmtypes*}

We make two remarks, which are of independent interest in the context of model theory.
\begin{remark}\label{rem:converse}
Conversely, the conclusion of \Cref{thm:type-bound} implies monadic dependence. 
Indeed, let $c\in\N$, and let $\phi(x,y)$ be any formula in the signature of $c$-colored graphs, and let $q$ denote its quantifier rank. Let $N$ be a number and $H$ be the bipartite graph with parts $A=\{a_1,\ldots,a_N\}$ and $B=\setof{b_I}{I\subset [N]}$, where $a_i$ is adjacent to $b_I$ if and only if $i\in I$. Suppose there is some $G\in\CC$ and $c$-coloring $G^+$ of $G$, such that $H$ is an induced subgraph of $\phi(G^+)$, the graph with vertex set $V(G)$ and edge set $\setof{uv\in \binom{V(G)}{2}}{G^+\models \phi(u,v)\lor \phi(v,u)}$. Then $\tp^q(b_I/A)$ is different for every $I\subset [N]$,
so the number of types of quantifier rank $q$ over $A$ is at least $2^N$.
On the other hand, the bound in \Cref{thm:type-bound} is at most polynomial in $|A|=N$, which implies that $N$ is bounded by a constant depending on $\CC$ and $\phi$.
Hence, not arbitrary graphs can be encoded in~$\CC$ by~$\phi$. 
\end{remark}

\begin{remark}
     An inspection of our proof of \Cref{thm:type-bound} shows that it only uses the assumption that $\CC$ is a \emph{dp-minimal} class of graphs (see \Cref{app:patterns} for a definition).
    Thus, the conclusion of \Cref{thm:type-bound} holds under the assumption of dp-minimality.

    Monadic dependence implies dp-minimality, as shown in \Cref{lem:mondep-dpmin}.
    For general classes of structures, however, dp-minimality is a strictly weaker assumption than monadic dependence\footnote{For example, $(\mathbb Z,+)$ is dp-minimal \cite[paragraph before Proposition 6.6]{ADHMS2016} but not monadically dependent:
    For $B=\setof{2^i}{i\in\mathbb N}$, the function $\set{u,v}\mapsto u+v$ defines an injection $f\from {B\choose 2} \to \mathbb Z$.
    Let $E\subset {B\choose 2}$ be arbitrary.
    In the unary expansion of $(\mathbb Z,+)$ with unary predicate $U$ marking $f(E)$,
    the formula $\phi(x,y):=\exists z. U(z)\land (x+y=z)$, restricted to $B$, defines the graph $(B,E)$. As $E$ is arbitrary and $\phi$ is fixed, $(\mathbb Z,+)$ is not monadically dependent.
    }.
     With the previous remark, we conclude that for classes $\CC$ of graphs, the notions of dp-minimality and monadic dependence coincide. This extends to classes of structures over a finite  signature consisting of binary relations, via the standard encoding of such structures as graphs (see \cite[full version, Section 19]{flip-breakability}).
\end{remark}

\section{Closures}\label{sec:closures}
The goal of this section is to prove the following theorem, which can be seen as a finitary analogue of \cite[Theorem~7.1 in the full version]{flippers}.

\begin{restatable}[Closure Theorem]{theorem}{thmClosures}\label{thm:closures}
For every monadically dependent graph class $\CC$ and radius $r\in\N$ there are some $c,s\in\N$ with the following property.

\smallskip\noindent
For every $G \in \CC$ and finite set $A \subseteq V(G)$ there is a set $A^*\subseteq V(G)$ of size $O_{\CC,r}(|A|^c)$ such that for every vertex $v \notin A$ there is a set $S(v)\subseteq A^*$ of size at most $s$
and an $S(v)$-flip $H$ with
\[
    \dist^H(v,A) > r.
\]
\end{restatable}

We also extend the statement to tuples.
For simplicity, and in contrast to \Cref{thm:closures}, the conclusion uses non-definable flips. 
Definability can easily be restored using the methods from~\cite{DBLP:conf/lics/BonnetDGKMST22} (see also~\cite{bonnet2025separability-icalp}).

\begin{restatable}[Closure Theorem for Tuples]{theorem}{thmClosuresTuples}\label{thm:closuresTuples}
For every monadically dependent graph class $\CC$, radius $r\in\N$, and tuple size $k \in \N$, there are some $c,s\in\N$ with the following property.

\smallskip\noindent
For every $G \in \CC$ and finite set $A \subseteq V(G)$ there is a set $\HH$ of $O_{\CC,r,k}(|A|^c)$ many $s$-flips of $G$ such that for every tuple $\bar v \in (V(G) \setminus A)^k$ there is a flip $H \in \HH$ with
\[
    \dist^H(\bar v,A) > r.
\]
\end{restatable}

\subsection{Preliminaries}

\paragraph*{Definable flips.}
For $s \in \N$, let $\Pi=\set{(x=y), (x\neq y)\land E(x,y),(x\neq y)\land \neg E(x,y)}$ denote the set of three possible atomic types of a pair of vertices in a graph.
For an $s$-tuple $\bar p=(p_1,\ldots,p_s)$ of vertices in a graph $G$ and a vertex $v$,
we identify $\atp(v/\bar p)$ with an $s$-tuple in $\Pi^s$, whose $i$th component is the atomic type of the pair $(v,p_i)$.
For a symmetric relation $R \subseteq \Pi^s \times \Pi^s$, we denote by $G \oplus (\bar p, R)$ the $\set{p_1,\ldots,p_s}$-flip $H$ of $G$ obtained by partitioning $V(G)$ into atomic types over $\bar p$ and then flipping according to $R$.
More precisely, for $uv\in \binom{V}{2}$,
we have that $uv\in E(H)\triangle E(G)$ (where $\triangle$ denotes the symmetric difference) if and only if $(\atp^G(u/\bar p), \atp^G(v/ \bar p)) \in R$.
We call such a flip \emph{$s$-definable}.
Equivalently, an $s$-definable flip is an $S$-flip for some $S\subseteq V(G)$ of size at most $s$.
Note that an $s$-definable flip is also $s'$-definable for every $s'\ge s$.

\begin{observation}\label{obs:sim-def-flip}
    We can define the edges of a definable flip in the original graph.
    More precisely, for every $s\in\N$ and symmetric relation $R \subseteq \Pi^s \times \Pi^s$ there exists a quantifier-free formula $E_{\oplus R}(x,y,\bar z)$ such that for every graph $G$, tuple $\bar p \in V(G)^s$, and vertices $u,v\in V(G)$ we have 
    \[
        G\oplus(\bar p, R) \models E(u,v)
        \iff
        G \models E_{\oplus R}(u,v,\bar p).
    \]
\end{observation}

The following is an easy consequence of the characterization of monadic dependence through transductions~(see, e.g., \cite{flip-breakability}). It also holds in the greater generality of $s$-flips (rather than $s$-definable flips).
\begin{fact}\label{obs:flips-mon-dep}
    Let $\CC$ be a monadically dependent graph class and $s\in\N$.
    The class 
    consisting of all $s$-definable flips of graphs from $\CC$ is monadically dependent.
\end{fact}

The following is well-known, see, e.g., Lemma~2.6 in the full version of~\cite{flippers}.

\begin{fact}[Transitivity of definable flips]\label{lem:definable-flip-transitive}
    Let $G_1$ be an $s_1$-definable flip of $G_0$ and $G_2$ be an $s_2$-definable flip of $G_1$. Then $G_2$ is an $(s_1 + s_2)$-definable flip of $G_0$.
\end{fact}

\subsection{The main argument}\label{sec:closures-main}

The argument is a finitary analogue of the reasoning in \cite[Section 7]{flippers}.
We use two lemmas. The first enlarges a given set $A$ to a set $A^*$ such that balls avoiding $A^*$ have few types over $A$. The second uses this bound to increase the separation radius by one.

\begin{restatable}{lemma}{lemBallTypes}\label{lem:ball-types}
    For every monadically dependent graph class $\CC$ and $r,s\in\N$, there are some $k,t\in\N$ with the following property.
    For every $G\in\CC$ and finite set $A\subseteq V(G)$, there is $A^*\supseteq A$ of size $O_{\CC,r,s}(|A|^k)$ such that, for every $s$-definable flip $H$ of $G$ and every vertex $v$,
    \[
        \Ball_r^H(v)\cap A^*=\emptyset
        \quad\Longrightarrow\quad
        |\ATypes^H(\Ball_r^H(v)/A)|\leq t.
    \]
\end{restatable}

The inequality above
can be equivalently stated as follows:
 $\Ball_r^H(v)$ intersects at most $t$ parts of 
the partition of $V(G)$ into atomic types over $A$ in $H$.
The proof of \Cref{lem:ball-types} is deferred to \Cref{sec:few-types-in-balls}.

\begin{restatable}{lemma}{lemBallFlip}\label{lem:ball-flip}
For all $d,t\in \N$, there is some $z\in\N$ with the following property.

\smallskip\noindent
For every graph $G$ of VC-dimension at most $d$, radius $r\in\N$, finite set $A\subseteq V(G)$, and arbitrary set $C\subseteq V(G)$ satisfying
\begin{center}
    $\Ball_r^G(C) \cap A = \emptyset$
    \qquad and \qquad
    $|\ATypes^G(\Ball_r^G(C)/A)|\leq t$,
\end{center}
there exists a $z$-definable flip $H$ of $G$ in which $\Ball_{r+1}^H(C) \cap A = \emptyset$.
\end{restatable}

The proof is deferred to \Cref{sec:definable-flips-for-balls}. We now conclude \Cref{thm:closures} from the two lemmas.

\thmClosures*

For brevity, we say a flip $H$ of $G$ \emph{$r$-separates} two sets $A, B\subseteq V(G)$ if $\dist^H(A,B)>r$.
This terminology extends from sets to vertices as expected. 

\begin{proof}[Proof of \Cref{thm:closures}]
    We induct on $r$, with $A^*\coloneqq A$, $S(v)\coloneqq \emptyset$, and $H\coloneqq G$ proving the base case $r=0$. 
    Suppose the theorem holds at radius $r$, with parameter bound $s$. 
    By \Cref{obs:flips-mon-dep}, and since monadically dependent classes have bounded VC-dimension (\Cref{obs:md-vc}), it follows that
    every $s$-definable flip of a graph in $\CC$ has VC-dimension at most some $d\leq O_{\CC,r}(1)$.

    Fix $G\in\CC$ and a finite set $A\subseteq V(G)$. Assume $A$ is nonempty, as otherwise the statement is trivial.
    Applying \Cref{lem:ball-types} to $r,s,G,A$ gives a polynomial-size superset $A^*\supseteq A$ and a constant $t \leq O_{\CC,r}(1)$ such that every $r$-ball in any $s$-definable flip $H$ of $G$ that avoids $A^*$ realizes at most $t$ types over $A$ in $H$.
    (Crucially, to avoid cyclic dependencies, note that $s$ is an input, not an output, of \Cref{lem:ball-types}.)

    Apply the induction hypothesis to $A^*$, obtaining a vertex set $B$ of size $\poly_{\CC,r}(|A^*|)\leq\poly_{\CC,r}(|A|)$ such that every vertex outside $A^*$ is $r$-separated from $A^*$ by a flip defined over a subset of $B$ of size at most $s$. We may assume that $B$ is nonempty.

    For each $\bar p\in B^s$ and each symmetric relation $R\subseteq\Pi^{s}\times\Pi^{s}$, put $H\coloneqq G\oplus(\bar p,R)$.
    Let $V_H$ be the set of vertices $v$ that are $r$-separated from $A^*$ by $H$, that is, $\Ball_r^H(v)\cap A^*=\emptyset$.

    We assign to each $v \in V_H$ its \emph{profile}
    \[
        \mathrm{profile}_H(v) \coloneqq \ATypes^H(\Ball_r^H(v)/A).
    \]
Equivalently,
if  $\PP_A$ denotes the partition of $V(G)$ into atomic types over $A$ in $H$,
then $\mathrm{profile}_H(v)=
\Ball_r^H(v)/\PP_A$
is the set of equivalence classes that are intersected by $\Ball_r^H(v)$.

Then $\mathrm{profile}_H(v)$ 
is a subset of $\ATypes^H(V(G)/ A)$ of size at most $t$, 
by \Cref{lem:ball-types}.
By the Sauer--Shelah lemma (\Cref{lem:ss}), we have that
\[
    |\ATypes^H(V(G)/ A)|\leq (|A|+1)^d+|A|.
\]
It follows that 

    \[
        |\{\mathrm{profile}_H(v) \,:\, v \in V_H\}|
        \leq
        |\ATypes^H(V(G)/ A)|^t \leq \big((|A|+1)^d+|A|\big)^t 
        \leq 
        \poly_{\CC,r}(|A|).
    \]
    
    Group the vertices of $V_H$ by their profiles.
    For each group $C$, the ball $\Ball_r^H(C)$:
    \begin{itemize}
        \item avoids $A$, because each $\Ball_r^H(v)$ with $v \in V_H$ does, and
        \item realizes at most $t$ types over $A$ in $H$, because all $v \in C$ have the same profile of size ${\leq}t$.    
    \end{itemize}
    Applying \Cref{lem:ball-flip} yields a 
    number $z\le O_{d,t}(1)\le O_{\CC,r}(1)$ and a
    further $z$-definable flip $H_C$ of $H$ that $(r+1)$-separates $C$ from $A$.
    By transitivity of definable flips (\Cref{lem:definable-flip-transitive}), $H_C$ is an $(s+z)$-definable flip of $G$.
    Choose a set $U_{\bar p,R,C}\subseteq V(G)$ of size at most $s+z$ such that $H_C$ is a $U_{\bar p,R,C}$-flip of $G$.

    We finish the proof by setting
    \[
        A^{**}\coloneqq A^*\cup\bigcup_{\bar p,R,C}U_{\bar p,R,C}.
    \]
    There are at most $|B|^s$ choices of $\bar p$, $O_s(1)$ choices of $R$ for each $\bar p$, and $\poly_{\CC,r}(|A|)$ groups $C$ for each choice of $\bar p,R$.
    Since each $U_{\bar p,R,C}$ has size at most $s+z$, we have $|A^{**}|\leq\poly_{\CC,r}(|A|)$.

    For each $v\notin A^*$, the induction hypothesis gives a tuple $\bar p \in B^s$ and a $\bar p$-flip $H$ that $r$-separates $v$ from $A^*$.
    Let $R$ be its flip relation, so that $H=G\oplus (\bar p,R)$, and let $C$ be the group of $V_H$ containing $v$.
    Then we may take $S(v)\coloneqq U_{\bar p,R,C}\subseteq A^{**}$, since $H_C$ $(r+1)$-separates $v$ from $A$.
    For $v\in A^*\setminus A$, take $S(v)\coloneqq \{v\}\subseteq A^{**}$ and the flip that isolates $v$.
    Hence, the theorem holds at radius $r+1$ with parameter set $A^{**}$ and parameter bound $s+z$.
\end{proof}

\subsection{Few types in balls}\label{sec:few-types-in-balls}
In this subsection we prove \Cref{lem:ball-types}.
We use the following Ramsey-type result for set systems due to Ding, Oporowski, Oxley, and Vertigan~\cite{DBLP:journals/jct/DingOOV96}.

\begin{definition}
    For a graph $G$ and tuples $\bar a, \bar b \in V(G)^t$, we say $\bar a = a_1 \ldots a_t$ and $\bar b = b_1 \ldots b_t$ are \emph{atomically matched} 
    if there is a relation ${\sim} \in \{=,\neq,\geq\}$ such that $G \models E(a_i, b_j)$ if and only if $i \sim j$ for all $i,j \in [t]$.
    In particular, the bipartite graph semi-induced between the tuples $\bar a$ and $\bar b$ is a matching, a co-matching, or a half-graph.
\end{definition}

\begin{theorem}[{\cite[Corollary~2.4]{DBLP:journals/jct/DingOOV96}, see also~\cite[Theorem~2]{gravier2004}}]\label{thm:matching}
    There exists a computable function $Q:\N\rightarrow\N$ such that for every $\ell\in\N$ and for every bipartite graph $G = (L,R,E)$ with
    $|\ATypes(L/R)| > Q(\ell)$, there are two tuples $\bar a \in L^\ell$ and $\bar b \in R^\ell$ that are atomically matched in $G$.
\end{theorem}

We use the above result to construct the following obstructions to monadic dependence.

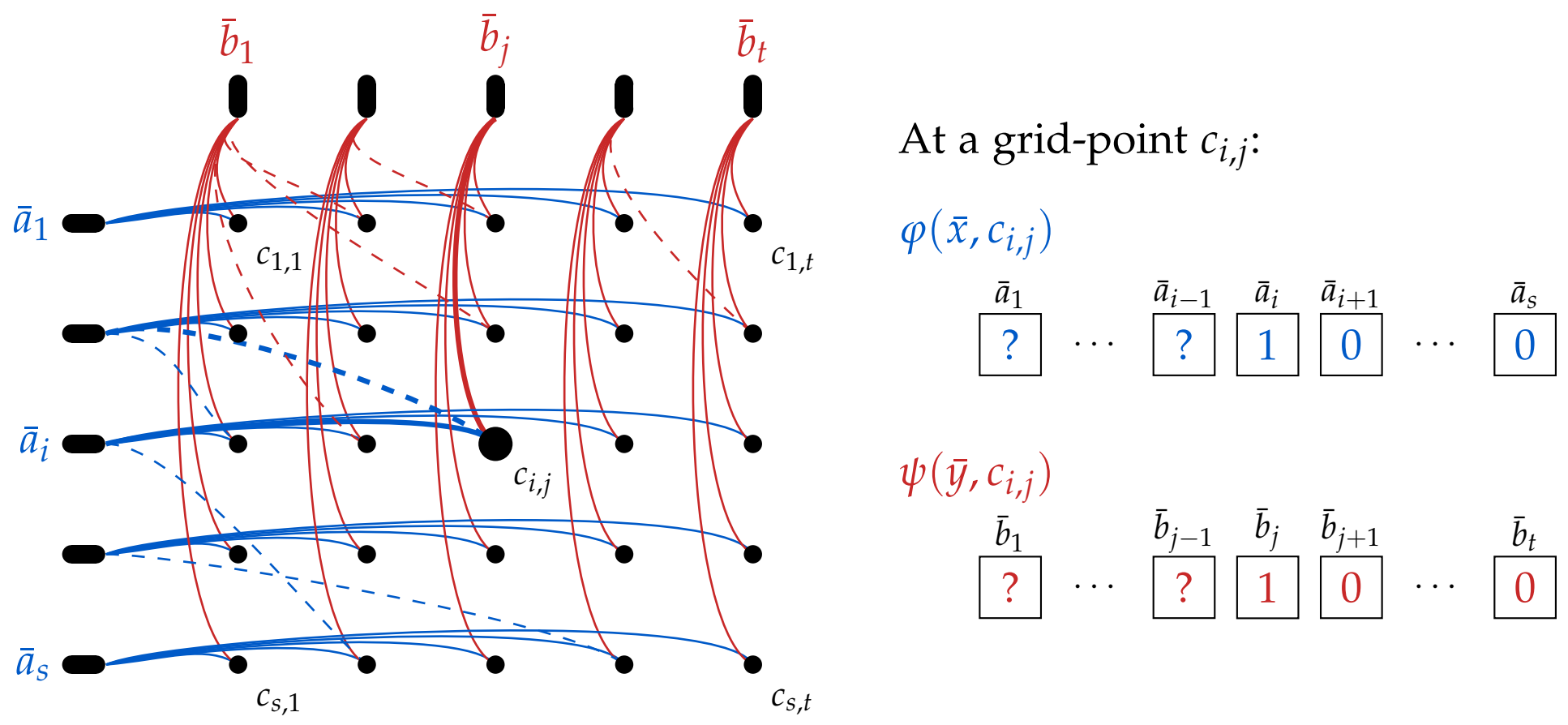
\begin{figure}[htbp]
    \definecolor{patternphi}{RGB}{0,90,200}
    \definecolor{patternpsi}{RGB}{200,40,40}
    \centering
    \begin{tikzpicture}[
        x=1.05cm,y=0.90cm,font=\small,
        grid point/.style={circle,fill=black,inner sep=1.5pt},
        a tuple/.style={rectangle,inner sep=0pt,
            minimum width=10pt,minimum height=5pt,
            path picture={
                \fill ([xshift=2.12pt]path picture bounding box.west) circle (2.12pt);
                \fill ([xshift=-2.12pt]path picture bounding box.east) circle (2.12pt);
                \fill ([xshift=2.12pt,yshift=-2.12pt]path picture bounding box.west)
                    rectangle ([xshift=-2.12pt,yshift=2.12pt]path picture bounding box.east);
            }},
        b tuple/.style={rectangle,inner sep=0pt,
            minimum width=5pt,minimum height=10pt,
            path picture={
                \fill ([yshift=-2.12pt]path picture bounding box.north) circle (2.12pt);
                \fill ([yshift=2.12pt]path picture bounding box.south) circle (2.12pt);
                \fill ([xshift=-2.12pt,yshift=-2.12pt]path picture bounding box.north)
                    rectangle ([xshift=2.12pt,yshift=2.12pt]path picture bounding box.south);
            }},
        phi incidence/.style={draw=patternphi,line width=0.5pt},
        psi incidence/.style={draw=patternpsi,line width=0.5pt}
    ]
        \foreach \i in {1,...,5} {
            \node[a tuple] (a-\i) at (-0.2,-\i) {};
        }
        \foreach \i/\aindex in {1/1,5/s} {
            \node[anchor=east,text=patternphi,inner sep=1pt] at ([xshift=-1pt]a-\i.west) {$\bar a_{\aindex}$};
        }
        \node[anchor=east,text=patternphi,inner sep=1pt] at ([xshift=-1pt]a-3.west) {$\bar a_i$};
        \foreach \j in {1,...,5} {
            \node[b tuple] (b-\j) at (\j,0.15) {};
        }
        \foreach \j/\bindex in {1/1,5/t} {
            \node[text=patternpsi] at (\j,0.65) {$\bar b_{\bindex}$};
        }
        \node[text=patternpsi] at (3,0.65) {$\bar b_j$};
        \foreach \i in {1,...,5} {
            \foreach \j in {1,...,5} {
                \coordinate (c-\i-\j) at (\j,-\i);
            }
        }
        \foreach \i in {1,...,5} {
            \foreach \j in {1,...,5} {
                \pgfmathsetmacro{\incidencewidth}{0.5+0.7*(\i==3)*(\j==3)}
                \draw[phi incidence,line width=\incidencewidth pt] (a-\i.east)
                    .. controls (0.45,-\i+0.05*\j)
                    and (\j-0.25,-\i+0.10+0.09*\j) .. (c-\i-\j);
                \draw[psi incidence,line width=\incidencewidth pt] (b-\j.south)
                    .. controls (\j-0.12-0.10*\i,-0.35)
                    and (\j-0.12-0.10*\i,-\i+0.35) .. (c-\i-\j);
            }
        }
        \foreach \earlier/\i/\j in {2/3/1,3/5/2,4/5/4} {
            \pgfmathsetmacro{\incidencewidth}{0.5+0.7*(\i==3)*(\j==3)}
            \draw[phi incidence,dashed,line width=\incidencewidth pt] (a-\earlier.east)
                .. controls (0.55,-\earlier)
                and (\j-0.40,-\i+0.40) .. (c-\i-\j);
        }
        \draw[phi incidence,dashed,line width=1.2pt] (a-2.east)
            .. controls (0.55,-1.75) and (2.60,-2.60) .. (c-3-3);
        \foreach \earlier/\i/\j in {1/1/2,1/2/3,2/1/3,4/2/5} {
            \pgfmathsetmacro{\incidencewidth}{0.5+0.7*(\i==3)*(\j==3)}
            \draw[psi incidence,dashed,line width=\incidencewidth pt] (b-\earlier.south)
                .. controls (\earlier-0.45,-0.40)
                and (\j-0.35,-\i+0.35) .. (c-\i-\j);
        }
        \draw[psi incidence,dashed] (b-1.south)
            .. controls (0.35,-0.65) and (1.65,-2.90) .. (c-3-2);
        \foreach \i in {1,...,5} {
            \foreach \j in {1,...,5} {
                \node[grid point] at (c-\i-\j) {};
            }
        }
        \foreach \i/\j/\rowindex/\colindex in {1/1/1/1,1/5/1/t,5/1/s/1,5/5/s/t} {
            \node[anchor=north west,fill=white,inner sep=1pt,font=\scriptsize]
                at (\j+0.10,-\i-0.17) {$c_{\rowindex,\colindex}$};
        }
        \node[anchor=north west,fill=white,inner sep=1pt,font=\scriptsize]
            at (3.10,-3.17) {$c_{i,j}$};
        \node[grid point,inner sep=2.8pt] at (c-3-3) {};

        \node[anchor=west] at (6.0,-0.3) {At a grid-point $c_{i,j}$:};
        \node[anchor=west,text=patternphi] at (6.0,-1.1)
            {$\phi(\bar x,c_{i,j})$};
        \node[anchor=west,text=patternpsi] at (6.0,-3.3)
            {$\psi(\bar y,c_{i,j})$};
        \foreach \k/\profilex/\p/\q in {1/7/?/?,2/8.35/?/?,3/9/1/1,4/9.65/0/0,5/11/0/0} {
            \node[draw,minimum size=5mm,text=patternphi] (phi-box-\k)
                at (\profilex,-2.1) {$\p$};
            \node[draw,minimum size=5mm,text=patternpsi] (psi-box-\k)
                at (\profilex,-4.3) {$\q$};
        }
        \foreach \k/\aindex/\bindex in {1/1/1,2/i-1/j-1,3/i/j,4/i+1/j+1,5/s/t} {
            \node[font=\scriptsize] (phi-label-\k)
                at ($(phi-box-\k.center)+(0,0.45)$) {$\bar a_{\aindex}$};
            \node[font=\scriptsize,anchor=south] (psi-label-\k)
                at ($(phi-label-\k.south)+(0,-2.2)$) {$\bar b_{\bindex}$};
        }
        \foreach \leftindex/\rightindex in {1/2,4/5} {
            \node[font=\scriptsize]
                at ($(phi-box-\leftindex.center)!0.5!(phi-box-\rightindex.center)$) {$\cdots$};
            \node[font=\scriptsize]
                at ($(psi-box-\leftindex.center)!0.5!(psi-box-\rightindex.center)$) {$\cdots$};
        }
    \end{tikzpicture}
    \caption{A $(\phi,\psi)$-pattern of order $s\times t$ for $s=t=5$.
        Edges indicate $\phi(\bar a_{i'},c_{i,j})$ or $\psi(\bar b_{j'},c_{i,j})$.
        Each grid-point $c_{i,j}$ is related via $\phi$ to the tuple $\bar a_i$ in its row and possibly (indicated by dashed edges) to some earlier row tuples, and analogously for $\psi$ and the column tuples.
        The profiles on the right illustrate these conditions at a grid-point $c_{i,j}$.}
    \label{fig:phi-psi-pattern}
\end{figure}

\begin{restatable}{definition}{defphipsi}\label{def:phi-psi-patterns}
    Fix first-order formulas $\phi(\bar x,z)$ and $\psi(\bar y,z)$. A \emph{$(\phi,\psi)$-pattern} of order $s \times t$ in a graph~$G$ consists of tuples $\bar a_1, \ldots, \bar a_s \in V(G)^{|\bar x|}$, $\bar b_1, \ldots, \bar b_t \in V(G)^{|\bar y|}$, and vertices ${(c_{i,j}: i\in [s],j\in [t])}$ such that the following holds
    for all $i,i'\in [s]$ and $j,j'\in[t]$:
    \begin{enumerate}
        \item $G\models \phi(\bar a_i,c_{i,j})$ 
        and $G\not\models \phi(\bar a_{i'},c_{i,j})$ if $i'>i$,
        \item $G\models \psi(\bar b_j,c_{i,j})$ 
        and $G\not\models \psi(\bar b_{j'},c_{i,j})$ if $j'>j$.
    \end{enumerate}
\end{restatable}
\noindent We call $\setof{c_{i,j}}{i\in[s],j\in[t]}$ the \emph{grid-points} of the pattern, and $s$ its \emph{height}. See \Cref{fig:phi-psi-pattern}.

\medskip
The following lemma is a corollary of~\cite[Lemma~C.1]{ordered-tww}. We include a self-contained proof in \Cref{app:patterns} for completeness.

\begin{restatable}{lemma}{lemphipsi}\label{lem:phi-psi-patterns}
    For every monadically dependent graph class $\CC$ and formulas $\phi(\bar x,z),\psi(\bar y,z)$, there is some $t\in\N$ such that no graph $G\in \CC$ contains a $(\phi,\psi)$-pattern of order $t \times t$.
\end{restatable}

\begin{definition}
    For a first-order formula $\phi(\bar x,z)$, a \emph{$\phi$-pattern} is a $(\phi,\psi)$-pattern
    with $\psi(y,z)=E(y,z)$ or $\psi(y,z)=\neg E(y,z)$.
    We call those a
    \emph{$(\phi,E)$-pattern} 
    and a \emph{$(\phi,\neg E)$-pattern}, respectively.
    Such a pattern consists of tuples $\bar a_1,\ldots,\bar a_s\in V(G)^{|\bar x|}$,
    vertices $b_1,\ldots,b_t\in V(G)$ and $(c_{i,j}:i\in[s],j\in [t])$.
    We call $\bar b=(b_1,\ldots,b_t)$
    the \emph{base} of the pattern. 
\end{definition}

\begin{corollary}\label{lem:forbidden-patterns}
        For every monadically dependent graph class $\CC$ and formula $\phi(\bar x,z)$, there is some $t\in \N$ such that no graph $G\in \CC$ contains a $\phi$-pattern of order $t \times t$.
\end{corollary}

We now prove \Cref{lem:ball-types}, which we repeat for convenience.

\lemBallTypes*
\begin{proof}
    For each symmetric relation $R\subseteq\Pi^s \times \Pi^s$, let $\phi_R(\bar z x,y)$ be a first-order formula expressing that $x$ and $y$ have distance at most $r$ in the flip $G\oplus(\bar z,R)$, where $\bar z$ is an $s$-tuple.
Its existence follows from \Cref{obs:sim-def-flip}.

Let $k\ge 1$ be the bound from 
     \Cref{lem:forbidden-patterns}, such that no $G\in\CC$ contains a $\phi_R$-pattern of order $k\times k$, for any $R$.
    Let $Q\from\N\to\N$ be as in \Cref{thm:matching}, and set
    \[
        t\coloneqq Q\big(k \cdot |\Pi^s|^2\big)=Q(k\cdot 9^s).
    \]

    Fix $G\in\CC$ and a finite set $A\subseteq V(G)$.
    For every $R$ as above and $N\in \set{E,\neg E}$, and every $\bar b\in A^k$, choose a maximum-height $(\phi_R,N)$-pattern with base $\bar b$ in $G$. Each of them is an $\ell\times k$ pattern for some $\ell<k$ (possibly $\ell=0$).
    
    Let $A^*$ consist of $A$ and all grid-points of the chosen patterns.
    There are $O_s(1)$ many choices of $R$ and $|A|^k$ many choices of $\bar b$, and each chosen pattern contributes at most $k^2$ vertices, so
    \[
        |A^*|\leq |A|+O_s(1)\cdot|A|^k\cdot k^2
        =O_{\CC,r,s}(|A|^k).
    \]

    Fix a symmetric relation $R\subseteq\Pi^s \times \Pi^s$, tuple $\bar p \in V(G)^s$, and vertex $v \in V(G)$. Let
    $H=G\oplus(\bar p,R)$ and suppose $B\coloneqq \Ball_r^H(v)$ satisfies $B\cap A^*=\emptyset$.
    We show that $B$ realizes at most $t$ atomic types over $A$ in $H$, thus proving the lemma.
    
    Suppose the contrary.
    By \Cref{thm:matching}, there are tuples $\bar u'$ in $B$ and $\bar w'$ in $A$, each of length $k \cdot |\Pi^s|^2$, that are atomically matched in $H$.

    Let $\PP$ be the partition of $V(G)$ into atomic types over $\bar p$; then $H$ is a $\PP$-flip of $G$ and $|\PP|\le |\Pi^s|$.
    By the pigeonhole principle, we can pick corresponding subsequences $\bar u$ of $\bar u'$ and $\bar w$ of $\bar w'$ of length $k$,
    such that $\bar u$ 
    is contained in a single part $X\in \PP$
    and $\bar w$ is contained in a single part $Y\in\PP$.
    Write $\bar u=(u_1,\ldots,u_k)$ and $\bar w=(w_1,\ldots,w_k)$.
    Since $H$ is a $\PP$-flip of $G$, adjacency between $X$ and $Y$ is either preserved or complemented.
    Together with the matching condition in $H$, this gives a choice of $N\in\set{E,\neg E}$ such that
    \[
        \begin{aligned}
            G&\models N(w_j,u_j)&\text{for every }j\in[k],\\
            G&\not\models N(w_{j'},u_j)&\text{for }1\le j<j'\le k.
        \end{aligned}
    \]
    
    Consider the maximum-height $(\phi_R,N)$-pattern chosen for $R$ and $N$ and for  base $\bar b\coloneqq\bar w=(b_1,\ldots,b_k)$.
    Write $\bar a_1,\ldots,\bar a_\ell$ for its witnessing tuples and $c_{i,j}$ for its grid-points, where $i\in[\ell]$ and $j\in[k]$ (possibly $\ell=0$).
    All these grid-points belong to $A^*$.
    Append a new row by setting
    \[
        \bar a_{\ell+1}\coloneqq\bar p v,
        \qquad
        c_{\ell+1,j}\coloneqq u_j\quad\text{for every }j\in[k].
    \]
    We verify that this yields a $(\phi_R,N)$-pattern with the same base $\bar b$ and height $\ell+1$, contradicting the maximality of $\ell$.        
    The column condition holds by the choice of $N$, $b_j=w_j$ and $c_{\ell+1,j}=u_j$:
    \[
        \begin{aligned}
            G&\models N(b_j,c_{\ell+1,j})&\text{for every }j\in[k],\\
            G&\not\models N(b_{j'},c_{\ell+1,j})&\text{for }1\le j<j'\le k.
        \end{aligned}
    \]    
    For every $j\in[k]$, we have $G\models\phi_R(\bar a_{\ell+1},c_{\ell+1,j})$, since $c_{\ell+1,j}\in B=\Ball_r^H(v)$.
    Moreover, for every $i\in[\ell]$, we have $G\not\models\phi_R(\bar a_{\ell+1},c_{i,j})$, since $c_{i,j}\in A^*$ and $B\cap A^*=\emptyset$. This proves the lemma.
\end{proof}

\subsection{Definable flips for balls}\label{sec:definable-flips-for-balls}

The following lemma is a corollary of the $(p,q)$-Theorem for bounded VC-dimension from~\cite{Matousek:p-q-theorem}.
See~\cite[Corollary~6.13]{simon_book} or~\cite[full version, Theorem~E.2]{flip-width} for self-contained proofs.

For a graph $G$ and sets $S,X\subseteq V(G)$, we say that $S$ \emph{dominates} $X$ if every vertex of $X \setminus S$ has a neighbor in $S$, and that $S$ \emph{antidominates} $X$ if every vertex of $X \setminus S$ has a nonneighbor in $S$.

\begin{lemma}[Bounded-VC duality]\label{lem:vc-duality}
    For every $d\in\N$ there is some $k\in\N$ with the following property.
    Let $G$ be a graph of VC-dimension at most $d$, and let $X,Y\subseteq V(G)$ be two (possibly overlapping) sets with at least one of $X,Y$ finite.
    Then there exists a set $\sample(X,Y)$ of size at most $k$ such that at least one of the following holds:
    \begin{enumerate}
        \item $\sample(X,Y)$ is contained in $Y$ and dominates $X$;
        \item $\sample(X,Y)$ is  contained in $X$ and antidominates $Y$.
    \end{enumerate}
\end{lemma}
We remark that in the sources we have given, the above lemma is stated either for finite set systems or finite binary relations.
The translation to finite graphs is straightforward, by either viewing the graph as a set system of neighborhoods or working with the adjacency relation.
Moreover, the statement lifts to infinite graphs if at least one of the sets $X$ and $Y$ is finite, as the other one can be replaced by its finite subset representing all neighborhoods in the  finite one.
(However, the statement may fail if both $X$ and $Y$ are infinite, as witnessed by the bi-infinite half-graph with vertices $\mathbb{Z} \times \{1,2\}$ and edges between $(x,1)$ and $(y,2)$ if $x\leq y$.)

It was observed already in~\cite{DBLP:conf/lics/BonnetDGKMST22} that \Cref{lem:vc-duality} is useful to obtain definable flips.
We will use it in a similar way to deduce the remaining result of \Cref{sec:closures-main}.

\lemBallFlip*

Note that, given the assumptions of the lemma, it is easy to find a (non-definable) $(2^t+t+1)$-flip $H'$ of $G$ in which $\Ball_{r+1}^{H'}(C) \cap A = \emptyset$: group the vertices of $A$ and of $\Ball_r^G(C)$ by their neighborhoods in the other set, which gives at most $2^t$ and $t$ groups, respectively, and flip every group of $A$ with every group of $\Ball_r^G(C)$ that it is complete to.
So the challenge is really to get such a flip in a definable way.

We also remark that \Cref{lem:ball-flip} can be seen as a finitization of Lemma~7.3 in the full version of~\cite{flippers}.
That lemma works in a more model-theoretic setting, where $A$ is a (possibly infinite) model instead of a set.
It uses a version of \Cref{lem:vc-duality} which also works for infinite $X,Y$, under the stronger 
assumption that $G$ excludes a half-graph of order $d$.

\begin{proof}[Proof of \Cref{lem:ball-flip}]
    Put $B\coloneqq \Ball_r^G(C)$.
    Let $\AA$ be the partition of $A$ that groups vertices by their neighborhood in $B$, and let $\BB$ be the partition of $B$ that groups vertices by their neighborhood in $A$.
    By assumption $|\BB|\leq t$.
    The neighborhood in $B$ of a vertex of $A$ is a union of parts of $\BB$, so $|\AA|\leq2^t$.
    Moreover, every pair $P\in\AA$, $Q\in\BB$ is homogeneously connected: is either complete or anticomplete.

    Let $k \leq O_d(1)$ be given by \Cref{lem:vc-duality}.
    Pick any set $S$ which contains
    \begin{itemize}
        \item an arbitrary representative $\rep(P)\in P$ for each part $P\in\AA\cup\BB$, and
        \item
        the set $\sample(X,Y)$ supplied by \Cref{lem:vc-duality}, for every $X\in\AA$ and $Y\in\AA\cup\{V(G)\setminus B\}$         
        (the lemma is applicable because $X \subseteq A$ is finite).
    \end{itemize}
    We can pick such a set with $|S|\le z$ for some $z\le O_{k,t}(1)\le O_{d,t}(1)$.
    We will construct $H$ as an $S$-flip of $G$; in particular, $H$ is $z$-definable.
    
    Let $\PP_S$ be the partition of $V(G)$ into atomic types over $S$.
    Write $x\sim y$ when $x$ and $y$ are in the same $\PP_S$-part.
    Observe the following two properties of $\PP_S$.

    \begin{claim}
    $\PP_S$ refines $\AA$ on $A$ and $\BB$ on $B$.
    \end{claim}
    \begin{claimproof}
        If two vertices $a$ and $a'$ of $A$ are in different $\AA$-parts, then their adjacencies differ on some vertex $b \in B$.
        Let $P$ be the $\BB$-part containing $b$.
        The vertex $\rep(P) \in S$ has the same adjacencies to $A$ as $b$ and hence distinguishes $a$ and $a'$ in $\PP_S$.
        Thus, $\PP_S$ refines $\AA$ on $A$ and symmetrically also $\BB$ on $B$.
    \end{claimproof}

    \begin{claim}\label{clm:dagger}
    Let $X\in \AA$ and $Y\in \AA\cup\set{V(G)-B}$. 
    There do not exist vertices $x,y,x',y'$ with:
    \begin{equation*}\tag{$\dagger$}\label{eq:ball-flip-crossing}
        x\in X,\quad y\in Y,\quad x\sim x',\quad y\sim y',
        \quad Y\subseteq N^{\overline G}(x'),\quad X\subseteq N^G(y').
    \end{equation*}    
    \end{claim}
    \begin{claimproof}
    Suppose that $\sample(X,Y)$ lies in $Y$ and dominates $X$. 
    Then there is some  $s\in\sample(X,Y)\cap Y$ with $sx\in E(G)$.
    If
    $Y\subseteq N^{\overline G}(x')$,
then $sx'\notin E(G)$.
Thus, $x\nsim x'$.

The case when $\sample(X,Y)$ lies in $X$ and antidominates $Y$, is symmetric, by swapping $X$ with $Y$ and $G$ with $\overline G$.
    Both cases are illustrated in \Cref{fig:ball-flip-crossing}.
    \end{claimproof}
    \begin{center}
        \captionsetup{type=figure}
%
%
%
\definecolor{bfAnti}{RGB}{200,40,40}%
\definecolor{bfHl}{RGB}{0,90,200}
{\setlength{\fboxsep}{6pt}\makebox[\linewidth]{\colorbox{white}{%
\begin{tikzpicture}[
        vertex/.style={circle, fill=black, inner sep=0pt, minimum size=3.2pt},
        blob/.style={draw=black!70, line width=0.6pt, fill=black, fill opacity=0.04},
        complete/.style={draw=black, line width=0.6pt},
        anticomplete/.style={draw=bfAnti, dash pattern=on 2.5pt off 2pt, line width=0.6pt,
            shorten <=2pt},
        sameType/.style={black!60, densely dotted, line width=0.7pt},
        lbl/.style={inner sep=1pt},
        sampleSet/.style={draw=bfHl, line width=0.6pt, fill=bfHl, fill opacity=0.12},
        hlEdge/.style={draw=bfHl, line width=1.2pt},
        hlNonedge/.style={draw=bfHl, dash pattern=on 3pt off 2pt, line width=1.2pt},
    ]
    \def\bx{1.7}   
    \def\rx{1.1}   
    \def\ry{0.8}   
    \def\off{0.7}  
    \def\h{2.2}    
    \def\panelsep{6.3} 

    \pgfmathsetmacro{\bfunit}{\linewidth / (2*\panelsep + 2*(\bx+\rx))}
    \begin{scope}[x=\bfunit pt, y=\bfunit pt]

    \pgfmathsetmacro{\tphi}{atan2(\h/\ry, (2*\bx+\off)/\rx)}
    \pgfmathsetmacro{\tdelta}{acos(1/sqrt(((2*\bx+\off)/\rx)^2 + (\h/\ry)^2))}

    \def\bfSkipComplete{-1}
    \def\bfSkipAnti{-1}
    \def\bfSimLabel{$\sim$}

    \def\bfCrossingBase{%
        \coordinate (X) at (-\bx, 0);
        \coordinate (Y) at ( \bx, 0);
        \coordinate (x) at (-\bx-\off, 0.15);
        \coordinate (y) at ( \bx+\off, 0.15);
        \coordinate (x') at (-\bx-\off, \h);
        \coordinate (y') at ( \bx+\off, \h);

        \pgfmathsetmacro{\px}{2*\bx+\off} 
        \pgfmathsetmacro{\fanfrom}{atan2(\ry*sin(\tphi+\tdelta) - \h, \rx*cos(\tphi+\tdelta) - \px)}
        \pgfmathsetmacro{\fanto}{atan2(\ry*sin(\tphi-\tdelta) - \h, \rx*cos(\tphi-\tdelta) - \px)}
        \foreach \i in {0,...,5} {
            \pgfmathsetmacro{\fanang}{\fanfrom + \i*(\fanto - \fanfrom)/5}
            \pgfmathsetmacro{\fanA}{(cos(\fanang))^2/\rx^2 + (sin(\fanang))^2/\ry^2}
            \pgfmathsetmacro{\fanB}{2*(\px*cos(\fanang)/\rx^2 + \h*sin(\fanang)/\ry^2)}
            \pgfmathsetmacro{\fanC}{\px^2/\rx^2 + \h^2/\ry^2 - 1}
            \pgfmathsetmacro{\fanlen}{(-(\fanB) - sqrt(max(0, \fanB*\fanB - 4*\fanA*\fanC)))/(2*\fanA)}
            \ifnum\i=\bfSkipComplete\relax\else
                \draw[complete] (y') -- ++(\fanang:\fanlen);
            \fi
            \ifnum\i=\bfSkipAnti\relax\else
                \draw[anticomplete, dash phase={mod(\i,2)*2.25pt}] (x') -- ++(180-\fanang:\fanlen);
            \fi
        }

        \draw[blob] (X) ellipse [x radius=\rx, y radius=\ry];
        \draw[blob] (Y) ellipse [x radius=\rx, y radius=\ry];
        \node[lbl, below=2pt] at ($(X) + (0, -\ry)$) {$X$};
        \node[lbl, below=2pt] at ($(Y) + (0, -\ry)$) {$Y$};

        \draw[sameType] (x') -- node[lbl, left=1pt, black] {\bfSimLabel} (x);
        \draw[sameType] (y') -- node[lbl, right=1pt, black] {\bfSimLabel} (y);

        \foreach \v in {x, y, x', y'}
            \node[vertex] at (\v) {};
        \node[lbl, below=2pt] at (x) {$x$};
        \node[lbl, below=2pt] at (y) {$y$};
        \node[lbl, above=2pt] at (x') {$x'$};
        \node[lbl, above=2pt] at (y') {$y'$};
    }

    \begin{scope}[name prefix=a-]
        \bfCrossingBase
    \end{scope}

    \begin{scope}[shift={(\panelsep, 0)}, name prefix=b-]
        \def\bfSkipAnti{3}
        \def\bfSimLabel{}
        \bfCrossingBase
        \pgfmathsetmacro{\bfsang}{180 - (\fanfrom + \bfSkipAnti*(\fanto - \fanfrom)/5)}
        \pgfmathsetmacro{\bfslen}{(2*\bx + \off - 0.4)/cos(\bfsang)}
        \coordinate (s) at ($(x') + (\bfsang:\bfslen)$);
        \draw[sampleSet] (s) ellipse [x radius=0.45, y radius=0.3];
        \draw[hlEdge] (s) -- (x);
        \draw[hlNonedge] (s) -- (x');
        \foreach \v in {x, x'}
            \node[vertex] at (\v) {};
        \node[vertex, fill=bfHl] at (s) {};
    \end{scope}

    \begin{scope}[shift={(2*\panelsep, 0)}, name prefix=c-]
        \def\bfSimLabel{}
        \bfCrossingBase
        \pgfmathsetmacro{\bfsang}{\fanfrom + 3*(\fanto - \fanfrom)/5}
        \pgfmathsetmacro{\bfslen}{-(2*\bx + \off - 0.4)/cos(\bfsang)}
        \coordinate (s) at ($(y') + (\bfsang:\bfslen)$);
        \draw[sampleSet] (s) ellipse [x radius=0.45, y radius=0.3];
        \draw[hlNonedge] (s) -- (y);
        \draw[hlEdge] (s) -- (y');
        \foreach \v in {y, y'}
            \node[vertex] at (\v) {};
        \node[vertex, fill=bfHl] at (s) {};
    \end{scope}

    \coordinate (bbtop) at (current bounding box.north);
    \coordinate (bbbot) at (current bounding box.south);
    \pgfresetboundingbox
    \path[use as bounding box]
        (-\bx-\rx,0 |- bbbot) rectangle (2*\panelsep+\bx+\rx,0 |- bbtop);
    \end{scope}
\end{tikzpicture}%
}}}

        \caption{%
            Left: the forbidden configuration~\eqref{eq:ball-flip-crossing}.
            Middle: the case of the proof of \Cref{clm:dagger} where $\sample(X,Y)$ (marked in blue) lies in $Y$. 
            Right: the case where $\sample(X,Y)$ lies in $X$.}
        \label{fig:ball-flip-crossing}
    \end{center}

    Define the $S$-flip $H$ of $G$ where a pair $(P,Q)$ of parts of $\PP_S$ is flipped if and only if there is an edge in $G$ with endpoints in $P$ and in $Q$ and also in $A$ and in $B$. More formally, $H[P,Q]=G[P,Q]$ whenever
    $E(G) \cap PQ \cap AB = 
    \emptyset,$
    and otherwise, $H[P,Q]=\overline G[P,Q]$.

    The following three properties will yield the conclusion of the lemma:
    \begin{enumerate}[label=(P\arabic*), ref=P\arabic*]
        \item\label{itm:ball-flip-P1} $A$ and $\Ball_{r}^G(C) $ are nonadjacent in $H$.
        \hfill (proven in \Cref{clm:ball-flip-P1})
        \item\label{itm:ball-flip-P2} $\Ball_{r-1}^H(C)=\Ball_{r-1}^G(C)$.
        \hfill (proven in \Cref{clm:ball-flip-P2})
        \item\label{itm:ball-flip-P3}$\Ball_{r-1}^G(C)$ and $V(G) \setminus \Ball_{r}^G(C)$ are nonadjacent in $H$.
        \hfill (proven in \Cref{clm:ball-flip-P3})
    \end{enumerate}
    If $r = 0$, then already \ref{itm:ball-flip-P1} alone yields the conclusion. This justifies the implicit assumption $r \geq 1$ made in  \ref{itm:ball-flip-P2} and \ref{itm:ball-flip-P3}.
    By combining \ref{itm:ball-flip-P2} and \ref{itm:ball-flip-P3}, we derive:
    \begin{enumerate}[label=(P\arabic*), ref=P\arabic*,start = 4]
        \item\label{itm:ball-flip-P4}  $\Ball_{r}^H(C) \subseteq \Ball_{r}^G(C)$.
    \end{enumerate}
    The conclusion of the lemma thus follows:
    \[
       A \cap \Ball_{r+1}^H(C)
       \  
    =
    \ 
       A \,\cap\, \Ball_1^H\big(\Ball_r^H(C)\big)
       \  \underset{\rule{0pt}{8pt}\text{\ref{itm:ball-flip-P4}}}{\subseteq} \ 
       A \cap \Ball_1^H\big(\Ball_r^G(C)\big)
       \  \underset{\rule{0pt}{8pt}\text{\ref{itm:ball-flip-P1}}}{=\vphantom{\subseteq}} \ 
       A \cap \Ball_r^G(C)
       \  \underset{\rule{0pt}{8pt}\text{ass.}}{=\vphantom{\subseteq}} \ 
       \emptyset.
    \]
    It only remains to establish \ref{itm:ball-flip-P1}--\ref{itm:ball-flip-P3}.

    \begin{claim}[\ref{itm:ball-flip-P1}]\label{clm:ball-flip-P1}
        $A$ and $B$ are nonadjacent in $H$.
    \end{claim}
    \begin{claimproof}
By contrapositive, suppose $ab\in E(H)$ for some $a \in A$ and $b \in B$.
    By definition of $H$,  $ab\in E(G)$ implies $ab\notin E(H)$, so
     $ab\notin E(G)$.
    Since $ab\in E(H)$, the parts of $\PP_S$ containing $a$ and $b$ are flipped, that is, there is some $a'b'\in E(G)$ with $a' \in A$ and $b' \in B$,
    and 
    \begin{center}
        $(a\sim a'$ and $b\sim b')$
        \qquad
        or
        \qquad
        $(a\sim b'$ and $b\sim a')$.
    \end{center}
    The first case cannot happen 
    as $\PP_S$ refines $\AA$ on $A$ and $\BB$ on $B$, but the two sets $\{a,a'\}$ and $\{b,b'\}$ are inhomogeneous as $ab\notin E(G)$ and $a'b'\in E(G)$.

    Thus, assume $a\sim b'$ and $b\sim a'$.
    Let $X,Y\in\AA$ contain $a',a$, respectively.
    Then $a'b'\in E(G)$ implies that $b'$ is complete to $X$, whereas  $ab\notin E(G)$ implies that $b$ is anticomplete to $Y$ (both by homogeneity). This contradicts \eqref{eq:ball-flip-crossing} with $(x,y,x',y')=(a',a,b,b')$.
\end{claimproof}

    For brevity, we define $D\coloneqq \Ball_{r-1}^G(C)$. Recall that $\PP_S(v)$ is the part of $\PP_S$ containing $v$.

    \begin{claim}\label{clm:disjoint-from-d}
        If $b$ is adjacent to some $a\in A$ in $G$ then $\PP_S(b)\cap D=\emptyset$.
    \end{claim}
    \begin{claimproof}
            The vertices in $B \cap \PP_S(b)$ have the same neighborhood on $A$ as~$b$. In particular, they are all adjacent to $a$, whereas $D \subseteq B$ is anticomplete to $A$ since $\Ball^G_1(D)=\Ball^G_r(C)$ is disjoint from $A$ by assumption.
    \end{claimproof}

    \begin{claim}[\ref{itm:ball-flip-P2}]\label{clm:ball-flip-P2}
        $\Ball_{r-1}^H(C)=D$.
    \end{claim}
    \begin{claimproof}
    The claim is trivially true for $r = 1$, hence assume $r\geq 2$.
    Consider any  $ab\in E(G)$ with $a \in A$ and $b \in B$ witnessing a flipped pair of $\PP_S$-parts $\PP_S(a)$ and $\PP_S(b)$.

    \begin{enumerate}
    \item \emph{$\PP_S(b)$ is disjoint from $\Ball_{r-2}^G(C)$:}
        
    This is implied by \Cref{clm:disjoint-from-d}, because $\Ball_{r-2}^G(C) \subseteq \Ball_{r-1}^G(C) =D$.

    \item \emph{$\PP_S(a)$ is disjoint from $\Ball_{r-2}^G(C)$:}

    Let $Q\in\BB$ contain $b$.
    Every vertex of $Q$ is adjacent to $a$ and hence lies in $B\setminus D$.
    Thus, $\rep(Q)\in S$ is adjacent to $a$ and nonadjacent to every vertex of $\Ball_{r-2}^G(C)$.
    Hence, $\rep(Q)$ witnesses that $a$ is in a different $\PP_S$-part than all the vertices in $\Ball_{r-2}^G(C)$. 
    \end{enumerate}

    This means that whenever two parts $\PP_S(a)$ and $\PP_S(b)$ are flipped, then neither of them intersects~$\Ball_{r-2}^G(C)$.
    Hence, the vertices in $\Ball_{r-2}^G(C)$ are incident to the same edges in $H$ as in $G$, which gives the desired conclusion 
    $\Ball_{r-1}^H(C) = \Ball_{r-1}^G(C)$.
    \end{claimproof}

    For brevity, we define $\overline B\coloneqq V(G)\setminus B$.

    \begin{claim}[\ref{itm:ball-flip-P3}]\label{clm:ball-flip-P3}
        $D$ and $\overline B$ are nonadjacent in $H$.
    \end{claim}

    \begin{claimproof}
            Otherwise, an edge $dz\in E(H)$ with $d \in D$ and $z \in \overline B$ is witnessed by some $ab\in E(G)$ with $a \in A$ and $b \in B$ (because $D$ and $\overline B$ are nonadjacent in $G$).
    By \Cref{clm:disjoint-from-d}, the $\PP_S$-part of $b$ is disjoint from $D$, so necessarily $d\sim a$ and $z\sim b$.
    Let $X\in\AA$ contain $a$.
    The vertex $b$ is complete to $X$, whereas $d$ is anticomplete to $\overline B$, contradicting \eqref{eq:ball-flip-crossing} for $(X,\overline B)$ with $(x,y,x',y')=(a,z,d,b)$.
    \end{claimproof}
    As described above, establishing \ref{itm:ball-flip-P1}--\ref{itm:ball-flip-P3} finishes the proof.
\end{proof}

\subsection{Extension to tuples}\label{sec:closures-tuples}
In this subsection we prove \Cref{thm:closuresTuples}.
Recall that for $s\in\N$, an \emph{$s$-flip} of a graph $G$ is a $\PP$-flip of $G$ for some partition $\PP$ of $V(G)$ into at most $s$ parts.
For $S\subseteq V(G)$, the partition $\PP_S$ into atomic types over $S$ has at most $3^{|S|}$ parts, so every $s$-definable flip is a $3^s$-flip.
The flip metric $\dist_\PP$ associated with a partition $\PP$ of $V(G)$ is the maximum of $\dist^H$ over all $\PP$-flips $H$ of $G$, and $\Ball^\PP_r(v)\coloneqq\{u\in V(G):\dist_\PP(u,v)\leq r\}$ denotes its balls.
We use the same notation for the minimum distance between two tuples or sets.

\begin{observation}\label{obs:partition-distance-monotone}
If $\QQ$ refines $\PP$, then every $\PP$-flip is also a $\QQ$-flip, and hence
\[
    \dist_\QQ(u,v)\geq \dist_\PP(u,v).
\]
\end{observation}

We use the following metric-conversion result.    

\begin{lemma}[{\cite[Lemma 8]{bonnet2025separability-icalp}}]\label{lem:partition-metric-conversion}
    Let $G$ be a graph and let $\PP$ be a partition of $V(G)$. There is a refinement $\QQ$ of $\PP$ with $|\QQ|\leq |\PP|\cdot 2^{|\PP|}$ and a $\QQ$-flip $H$ of $G$ such that, for every $v\in V(G)$ and $r\in\N$,
    \[
        \Ball^H_r(v)\subseteq\Ball^\PP_{6r}(v).
    \]
\end{lemma}

We now combine the flips from  the Closure Theorem (\Cref{thm:closures}) for the individual entries of a tuple into a single flip.

\thmClosuresTuples*

\begin{proof}
    Let $c_0,s_0\in\N$ be given by \Cref{thm:closures} for $\CC$ and radius $6r$.
    Fix $G\in\CC$ and a finite set $A\subseteq V(G)$, and let $S$ and the sets $S(v)$ for $v\notin A$ be as in \Cref{thm:closures}.
    For a tuple $\bar v=(v_1,\ldots,v_k)\in(V(G)\setminus A)^k$, put
    \[
        S(\bar v)\coloneqq S(v_1)\cup\cdots\cup S(v_k).
    \]
    Then $S(\bar v)\subseteq S$ and $|S(\bar v)|\leq k\,s_0$.
    As $\PP_{S(\bar v)}$ refines every $\PP_{S(v_i)}$, \Cref{obs:partition-distance-monotone} gives
    \begin{equation}\label{eq:closures-tuples-far}
        \dist_{S(\bar v)}^G(v_i,A)\geq\dist_{S(v_i)}^G(v_i,A)>6r
        \qquad\text{for every }i\in[k].
    \end{equation}

    For every $T\subseteq S$ with $|T|\leq ks_0$, the partition $\PP_T$ has at most $3^{ks_0}$ parts.
    So \Cref{lem:partition-metric-conversion} yields an $s$-flip $H_T$ of $G$, for some $s\leq O_{\CC,r,k}(1)$ independent of $T$, such that
    \[
        \Ball^{H_T}_r(x)\subseteq\Ball^{T}_{6r}(x)
        \qquad\text{for every }x\in V(G).
    \]
    Let $\HH\coloneqq\{H_T \,:\, T\subseteq S,\ |T|\leq ks_0\}$.
    Then $|\HH|\leq(|S|+1)^{ks_0}\leq O_{\CC,r,k}(|A|^{c})$ for $c\coloneqq c_0\, k\, s_0$.

    Finally, let $\bar v=(v_1,\ldots,v_k)\in(V(G)\setminus A)^k$ and $H\coloneqq H_{S(\bar v)}\in\HH$.
    For every $i\in[k]$, the ball $\Ball^H_r(v_i)\subseteq\Ball^{S(\bar v)}_{6r}(v_i)$ is disjoint from $A$ by \eqref{eq:closures-tuples-far}.
    Thus $\dist^H(\bar v,A)>r$.
\end{proof}

\section{Number of types over a set of parameters}\label{sec:types-over-sets}
In this section we prove \Cref{thm:type-bound}, which bounds the number of types of $k$-tuples over a set $U$ of parameters polynomially in $|U|$.

We use a consequence of locality of first-order logic, which bounds the number of types of tuples that are far from $U$. 
The following result is an adaptation of~\cite[Corollary~16]{number-of-types}. We provide a self-contained proof in \Cref{app:gaifman}.

\begin{restatable}{lemma}{septypestuples}\label{lem:sep-types-tuple}
    For every $q,k,c\in\N$ there are numbers $r=r(q)$ and $T=T(q,k,c)$ with the following property. Let $H$ be a graph expanded by at most $c$ unary predicates, and let $U\subseteq V(H)$ be finite. Fix a tuple $\bar u$ of vertices of $U$ of finite length.
    Let $F \coloneqq \{a\in V(H) \,:\, \dist^H(a,U) > r\}$. 
    Then
    \[
      \left|
        \{
        \tp_q^H(\bar a\bar u/U) \,:\, \bar a \in F^k  
        \}
      \right|
      \leq
      T(q,k,c).      
    \]
\end{restatable}


We now prove \Cref{thm:type-bound} by combining \Cref{lem:sep-types-tuple} with the Closure Theorem for Tuples (\Cref{thm:closuresTuples}).

{\typeboundgeneraltrue\thmtypes*}

\begin{proof}
    Fix $G\in\CC$, its expansion $G^+$ by $c$ unary predicates, and a finite set $U\subseteq V(G)$.
    Let $r=r(q)$ and $T$ be given by \Cref{lem:sep-types-tuple}, and put $\overline{U}\coloneqq V(G)\setminus U$.
    By the Closure Theorem for Tuples (\Cref{thm:closuresTuples}) for $\CC$, radius $r$, and $k$, there is some $s\leq O_{\CC,q,k}(1)$ and some set $\HH$ of $\poly_{\CC,q,k}(|U|)$ many $s$-flips of $G$ such that every $\bar v\in\overline{U}^k$ has some $H(\bar v)\in\HH$ with $\dist^{H(\bar v)}(\bar v,U)>r$.
    Note that $s$ and $\HH$ do not depend on the unary predicates of $G^+$.
    For each $H\in\HH$, fix a partition $\QQ_H$ of $V(G)$ into at most $s$ parts such that $H$ is a $\QQ_H$-flip of $G$, and let $H^+$ be the expansion of $H$ by the $c$ unary predicates of $G^+$ and by at most $s$ unary predicates marking the parts of $\QQ_H$.
    The edge relation of $G$ is quantifier-free definable in $H^+$, by flipping back the pairs of parts that are flipped in $H$.
    Substituting this definition for every edge atom turns each formula over $G^+$ with parameters from $U$ into a formula over $H^+$ with the same parameters and quantifier rank that is satisfied by the same tuples.
    Hence, for every $H$ and tuple $\bar w$ of vertices,
    \begin{center}
        $\tp_q^{G^+}(\bar w/U)$ depends only on $\tp_q^{H^+}(\bar w/U)$.
    \end{center}

    We want to bound the number of types of tuples $\bar v \in V(G)^k$ over $U$.
    By padding and permuting $2k$-tuples, up to a multiplicative factor of $O_k(1)$, it suffices to bound the number of types for the slightly longer tuples of the form $\bar a\bar u$ with $\bar a \in \overline{U}^k$ and $\bar u \in U^k$.
    
    For $H\in\HH$, let $F_H\coloneqq\{w\in V(G) \,:\, \dist^H(w,U)>r\} \subseteq \overline{U}$.
    As $\bar a\in\overline{U}^k$ lies in $F_{H(\bar a)}^k$ for some $H(\bar a) \in \HH$, we have
    \[
        \{\tp_q^{G^+}(\bar a\bar u/U) \,:\, \bar a\in\overline{U}^k,\ \bar u\in U^k\}
        =
        \bigcup_{H\in\HH}\ \bigcup_{\bar u\in U^k}
        \{\tp_q^{G^+}(\bar a\bar u/U) \,:\, \bar a\in F_H^k\}.
    \]
    It remains to bound each of the $|\HH|\cdot|U|^k$ terms on the right.
    So fix $H\in\HH$ and $\bar u\in U^k$.
    As $H^+$ has at most $c+s$ unary predicates, \Cref{lem:sep-types-tuple}, applied to $H^+$, $U$, and $\bar u$, gives
    \[
        \left|\{\tp_q^{H^+}(\bar a\bar u/U) \,:\, \bar a\in F_H^k\}\right|
        \leq T(q,k,c+s).
    \]
    Since $\tp_q^{G^+}(\bar a\bar u/U)$ depends only on $\tp_q^{H^+}(\bar a\bar u/U)$, the same bound holds with $G^+$ in place of $H^+$.
    Hence,
    \[
        \left|\{\tp_q^{G^+}(\bar a\bar u/U) \,:\, \bar a\in\overline{U}^k,\ \bar u\in U^k\}\right|
        \leq|\HH|\cdot|U|^k\cdot T(q,k,c+s)
        \leq O_{\CC,q,k,c}(1)\cdot\poly_{\CC,q,k}(|U|),
    \]
    where we use $T(q,k,c+s)\leq O_{\CC,q,k,c}(1)$, as $s\leq O_{\CC,q,k}(1)$, and $|\HH|\leq\poly_{\CC,q,k}(|U|)$.
\end{proof}

\clearpage
\part{Definable construction sequences}\label{part:definable-construction-sequences}

Recall the following definitions from \Cref{sec:skel-part-defseq}.
\defpartialdef*

\defskellocdef*

In this part, we establish the following main result.
\thmskeldefseq*

We will use the following variation on the definition of merge-width, originating in~\cite{drabik2026variants}.

\begin{definition}\label{def:chain-width}
  A \emph{chain} in a graph $G$ is a decreasing sequence $V(G)=S_1\supseteq\dots\supseteq S_\ell=\emptyset$ of vertex sets. Its \emph{radius-$r$ width}, for $r\in\N$, is defined as
$$\max_{2\le i\le \ell} \max_{v\in V(G)}\left|\ATypes^G(\Ball_r^{S_{i}}(v)/S_{i-1})\right|=\max_{2\le i\le \ell} \max_{v\in V(G)}\left|\Ball_r^{S_{i}}(v)/\PP_{S_{i-1}}\right|,$$
where $\Ball_r^S(v)$ is defined above via $S$-flips and $\PP_{S}$ is the partition into atomic types over $S$.
\end{definition}


The offset between $S_{i-1}$ and $S_i$ in the definition of the radius-$r$ width ensures that the definition is robust. For instance, a chain can be completed into one in which two consecutive sets differ by one element, 
by removing the elements of $S_{i-1}$ one by one in any order, until $S_i$ is reached; this transformation
preserves the width parameters.

Chains can be converted to construction sequences and vice versa, with a loss in the radius-$r$ width.
Intuitively, a chain 
gives rise to a sequence of ever coarser partitions $\PP_{S_1}\preceq \PP_{S_2}\preceq\dots \preceq \PP_{S_\ell}$, which can then be converted 
into a sequence of partial graphs
(see~\cite{drabik2026variants} for details).

We use the following key result from~\cite{md-abmw}. It says that in monadically dependent classes, chains of constant length and small width can be computed efficiently.

\begin{theorem}[{\cite[Theorem~4.4]{md-abmw}}]\label{thm:monadic-dependence-implies-definable-mw}  
  There is an algorithm which, given 
  an $n$-vertex graph $G$, radius $r\in\N$, and rational $\eps>0$, computes a chain of $G$ of length $O(r/\eps)$ and radius-$r$ width $O_{\CC,r,\eps}(n^\eps)$, 
  and runs in time  $\widehat O_{\CC,r}(n^{3})$, 
  where $\CC$ is any monadically dependent graph class containing $G$.
\end{theorem}

To prove \Cref{thm:skel-def-seq}, we turn these chains into construction sequences using the following main technical contribution of this part.

\begin{restatable}{lemma}{lembetterdefinablemwfromsets}\label{lem:better-definable-mw-from-sets}
  For every $n$-vertex graph $G$ and its chain of length $\ell$ there is
  a construction sequence $G_1,\ldots,G_\lambda$ of $G$ such that the following properties hold for all $r\ge 1$.
  Let $k_r$ denote the radius-$(5r+1)$ width of the chain and $w_r \coloneqq k_r\cdot \pi_G(k_r)$.
  \begin{enumerate}
    \item The resulting construction sequence has radius-$r$ width at most $w_r$ and length $\lambda\le O(nw_1)$.
    \item For every $t\in[\lambda-1]$ and $v\in V(G)$, the induced partial graph
    \[
        G_t[\Ball^{G_{t+1}}_r(v)]
    \]
    is definable in some expansion of $G$
    with $2\ell+1$ unary predicates, using formulas of quantifier rank~$7$ with $w_r$ parameters.
    \item The construction sequence can be computed in time $O(\ell n^3)$, given $G$ and its chain.
  \end{enumerate}
\end{restatable}

\begin{proof}[Proof of \Cref{thm:skel-def-seq}]
  Let $\CC$ be a monadically dependent graph class containing $G$, and let $d$ be an upper bound on the VC-dimension of graphs in $\CC$, which exists by \Cref{obs:md-vc}.
  We apply \Cref{thm:monadic-dependence-implies-definable-mw} with radius $5r+1$ and $\eps$.
  This computes a chain of length $\ell\le O((5r+1)/\eps)=O(r/\eps)$ in time $\widehat O_{\CC,r}(n^3)$.
  We then apply \Cref{lem:better-definable-mw-from-sets} to $G$ and this chain, which takes time $O(\ell n^3)$.
  So the total running time is
  \[
    \widehat O_{\CC,r}(n^3)+O(\ell n^3)
    =\widehat O_{\CC,r,\eps}(n^3).
  \]

  The chain has radius-$(5r+1)$ width \[k_r\le O_{\CC,r,\eps}(n^\eps),\] and $\pi_G(k_r)\le(k_r+1)^d$
  by the Sauer--Shelah lemma (\Cref{lem:ss}), so
  \[
    w_r=k_r\cdot\pi_G(k_r)\le(k_r+1)^{d+1}\le O_{\CC,r,\eps}(n^{(d+1)\eps}).
  \]
  We set $\delta\coloneqq d+1$.
  The sequence is radius-$r$ definable with $2\ell+1\le O(r/\eps)$ unary predicates, $w_r$ parameters, and rank $7$.
  As $k_1\le k_r$, it has length $O(nw_1)\le O(nw_r)\le O_{\CC,r,\eps}(n^{1+\delta\eps})$.
\end{proof}

\paragraph{Proof strategy for \texorpdfstring{\Cref{lem:better-definable-mw-from-sets}}{the technical lemma}}\label{sec:defseq-lemma-strategy}

The proof of \Cref{lem:better-definable-mw-from-sets} 
occupies the remainder of \Cref{part:definable-construction-sequences}.
It combines constructions from~\cite{drabik2026variants} and~\cite{mw-stoc}, together with a careful analysis of definability.
We revisit several variants of merge-width studied in those papers.
In particular, we use the following variation of the definition of merge-width.

\begin{definition}\label{def:merge-sequence}
A \emph{merge sequence} of a graph $G$ is a sequence
$(\PP_1,R_1),\ldots,(\PP_\ell,R_\ell)$ such that:
\begin{enumerate}
  \item $\PP_1$ is the partition into
singletons and $\PP_\ell=\{V(G)\}$,  
\item $\PP_1\preceq \dots\preceq \PP_\ell$, 
\item $R_1\subseteq \dots\subseteq R_\ell\subseteq \binom{V(G)}{2}$,
\item for every $i\in[\ell]$ and $X,Y\in \PP_i$,
the set $XY-R_i$ is contained in either $E(G)$ or in $E(\overline G)$,
 that is, the pair $X,Y$ is homogeneous once pairs in $R_i$ are ignored.
\end{enumerate}

  The \emph{radius-$r$ width} of a merge sequence as above is
\begin{equation}\label{eq:merge-seq-width}
  \max_{2\le i\le \ell}\max_{v\in V(G)}
  \left|\Ball^{R_i}_r(v)/\PP_{i-1}\right|.
\end{equation}
\end{definition}
\noindent 
Here the offset in the indices of $R_i$ and $\PP_{i-1}$ is intentional.
Every construction sequence naturally induces a merge sequence of the same radius-$r$ width for all $r\in\N$, and vice versa (see~\cite{mw-arxiv,mw-stoc} for details).


\section{Merge-width with definable stages}\label{sec:better-definable-mw}
\label{sec:definable-mw}

In this section, we prove the following lemma.

\begin{restatable}{lemma}{lembettermw}\label{lem:better-mw}
For every graph $G$ and chain of length $\ell$ there is a merge sequence $(\QQ_1, R_1), \ldots, (\QQ_\ell, R_\ell)$ with the following properties.
    \begin{enumerate}
        \item For all $r\ge 1$,  the radius-$r$ width of the merge sequence is bounded by $k\cdot \pi_G(k)$, where $k$ is the radius-$(5r+1)$ width of the given chain.
        \item  For each $i\in [\ell]$, the relations $\equiv_{\QQ_i}$ and $R_i$ are definable in some expansion $G^i$ of $G$ with two unary predicates, using two fixed first-order formulas $\phi_\equiv(x,y)$ and $\phi_R(x,y)$ of quantifier rank at most~$5$.
        \item The merge sequence can be computed in time $O(\ell n^3)$, given $G$ and its chain.
    \end{enumerate}    
\end{restatable}

\subsection{Preliminaries}

The proof of \Cref{lem:better-mw} uses ideas developed in~\cite{drabik2026variants}. We first recall the required notions and lemmas from that paper.

\begin{lemma}[folklore]\label{lem:diamComplement}
    For every graph $G$, at least one of $G$ and $\overline{G}$ has diameter at most $3$.
\end{lemma}
\begin{proof}
Suppose $G$ has diameter greater than $2$. Then there are non-adjacent vertices $u,v$ with no common neighbor in $G$.
In $\overline G$, $uv$ is an edge and all vertices are neighbors of either $u$ or $v$.
Thus $\overline G$ has diameter at most $3$.
\end{proof}

If $B$
is a bipartite graph with parts $X,Y$ and edge set $E$,
its \emph{bipartite complement}, denoted $\overline B$, 
is the bipartite graph with parts $X,Y$ and edge set $XY\setminus E$.

\begin{lemma}[{\cite[Lemma~A.3]{drabik2026variants}}]\label{lem:bipartite-informal-lemma}
  Let $B = (X,Y,E)$ be a bipartite graph with bipartite complement~$\overline{B}$, and pick arbitrary vertices $x \in X$, $y \in Y$.
  Partition~$X$ according to adjacency to $y$,
  i.e.\ $X = X_+ \uplus X_-$ with $X_+ = N(y)$ and $X_- = X \setminus N(y)$,
  and similarly $Y = Y_+ \uplus Y_-$ according to~$x$.
  Then there is a bipartite graph $B' = (X,Y,E')$ obtained from $B$ by some $\{X_+,X_-,Y_+,Y_-\}$-flip satisfying:
  \[ \text{for all $uv \in E'$,} \quad \dist^B(u,v) \le 5 \quad \text{and} \quad \dist^{\overline{B}}(u,v) \le 5. \]
\end{lemma}
Note that~\cite[Lemma~A.3]{drabik2026variants} states the above lemma with $6$ instead of $5$; however, the two statements are equivalent, since 
if $uv\in E'$ then $u$ and $v$ lie in opposite parts of the bipartite graphs $B'$, $B$, and $\overline{B}$, and thus $\dist(u,v)\le 6$ is equivalent to $\dist(u,v)\le 5$ for parity reasons.

\begin{definition}
    Fix a graph $G$ and a partition $\PP$ of $V(G)$. We define $\PP \wedge T$ to be the refinement of~$\PP$ where each part $P \in \PP$ is split into atomic types over $T \setminus P$
    (i.e.\ into sets of the form $P\cap Q$, where $Q$ is the set of all vertices of $G$ with a given neighborhood in $T-P$).
\end{definition}

\noindent In this definition, we choose not to divide a part $P$ into types over $T \cap P$. This is for technical convenience only, to ensure that $\PP \wedge T = \PP$ whenever $\PP$ is the partition with only one part.

\begin{lemma}[{\cite[Corollary~5.4]{drabik2026variants}}]\label{lem:transversalRefinement}
	Let~$G $ be a graph, $\PP\preceq\QQ$ partitions of~$V(G)$, and~$T$ a transversal of~$\PP$.
  Suppose $|\Ball^\QQ_{r+1}(v)/\PP| \le k$ for every $v\in V(G)$.
  Then for every $v\in V(G)$, we have
  \[
    \left|\Ball^{\QQ}_r(v)/(\PP \wedge T)\right|\le k \cdot \pi_G(k).
  \]
\end{lemma}

\subsection{Constructing the merge sequence}


Recall that for a graph 
$G$ and disjoint vertex sets $X,Y\subseteq V(G)$,
by $G[X,Y]$ we denote the bipartite graph with parts $X$ and $Y$, and edges connecting pairs of vertices which are adjacent in $G$.
Furthermore, $G[X,X]$
denotes the bipartite graph whose parts are two copies of $X$,
and edges connect two vertices in opposite parts if the corresponding vertices in $G$ are either adjacent or equal.
We call the edges of $G[X,X]$ connecting the two copies of each vertex in $X$ \emph{diagonal edges}.
Thus,
${\overline G[X,X]}$ is the bipartite complement  $\overline {G[X,X]}$ of $G[X,X]$ extended additionally with the diagonal edges.

\begin{definition}
    Let $G$ be a graph and $\PP$ a partition of $V = V(G)$.
    Define $R = R(G,\PP) \subseteq \binom{V}{2}$ so that for all parts $X,Y \in \PP$ (possibly $X = Y$) and pairs $uv$ with $u \in X$, $v \in Y$, and $u \ne v$, we have
    $uv \in R$ if and only if
    \[
    \dist^{G[X,Y]}(u,v) \le 5 \quad \text{and} \quad \dist^{\overline{G}[X,Y]}(u,v) \le 5. \]
\end{definition}

We first describe the construction of the merge sequence that underlies the proof of \Cref{lem:better-mw}.
Consider a graph $G$ and its chain 
    $V(G)=S_1\supseteq\cdots\supseteq S_\ell=\emptyset$.
    Let $\PP_i$ be the partition into atomic types over $S_i$.
    Fix an arbitrary linear order $<$ on $V(G)$ 
    giving rise to representative sets $T_i \coloneqq \{\min_< P \,:\, P \in \PP_i\}$. 
    Then $T_1\supseteq T_2\supseteq \dots \supseteq T_\ell$.
    For $i=1,\ldots,\ell$ define $$R_i \coloneqq R(G,\PP_i)\qquad\text{ and }\qquad\QQ_i \coloneqq \PP_i \wedge T_i.$$
    The following lemmas establish that 
    $$(\QQ_1,R_1),\ldots,(\QQ_\ell,R_\ell)$$ is a merge sequence satisfying the conditions stated in \Cref{lem:better-mw}.

\begin{lemma}\label{lem:def-R}
    Suppose $G$ is a graph and $\PP$ is the partition of $V(G)$ into atomic types over some set $S \subseteq V(G)$. Then the relation $R \coloneqq R(G,\PP)$ is definable in $G$ expanded with one unary predicate marking~$S$, using a fixed first-order formula $\phi_R(x,y)$ of quantifier rank at most~$5$.
\end{lemma}

\begin{proof}
    The following formula $\theta_S(x,y)$, using a unary predicate marking $S$ and with one quantifier,  defines the equivalence relation $\equiv_\PP$:
    \begin{align}\label{eq:definableEquiv}
        \theta_S(x,y) \coloneqq
        x = y \lor \bigl(\neg S(x) \land \neg S(y) \land
        \forall z\,(S(z) \to (E(x,z) \leftrightarrow E(y,z)))\bigr).    
    \end{align}

  The following formula $\eta_S(x,y;a,b)$ says that $x$ and $y$ are in opposite parts of the bipartite graph $G[X,Y]$, where $X$ and $Y$ are the parts of $\PP$ containing $a$ and $b$:
\begin{align*}
  \eta_S(x,y;a,b)\coloneqq{}&
  (\theta_S(x,a)\land\theta_S(y,b))
  \lor(\theta_S(y,a)\land\theta_S(x,b)).
\end{align*}
The following formulas $e_S^+(x,y;a,b)$ and $e_S^-(x,y;a,b)$ 
 say that $x$ and $y$ are adjacent in the bipartite graph $G[X,Y]$ or $\overline G[X,Y]$, respectively, where $X$ and $Y$ are the parts of $\PP$ containing $a$ and $b$:
\begin{align*}
  e_S^+(x,y;a,b)&\coloneqq \eta_S(x,y;a,b)\land(x=y\lor E(x,y)),\\
  e_S^-(x,y;a,b)&\coloneqq \eta_S(x,y;a,b)\land(x=y\lor \neg E(x,y)).
\end{align*}
For $\sigma\in\{+,-\}$, define
\[
  \rho_S^\sigma(x,y)\coloneqq
  \bigvee_{p\in\{1,3,5\}}
  \exists z_1\cdots\exists z_{p-1}\,
  \bigwedge_{i=0}^{p-1}e_S^\sigma(z_i,z_{i+1};x,y),
\]
where $z_0\coloneqq x$ and $z_p\coloneqq y$.
This formula expresses the existence of a walk of length $1$, $3$, or $5$
between opposite copies of $x$ and $y$ in $G[X,Y]$ for $\sigma=+$, and
in $\overline G[X,Y]$ for $\sigma=-$, where $X\coloneqq \PP(x)$ and $Y\coloneqq \PP(y)$.
Each step switches sides, also when $X=Y$; an equality step uses a diagonal edge.
Since every path between opposite sides of a bipartite graph has odd length,
these formulas express precisely the two distance bounds defining $R$.
Consequently,
\[
  \phi_R(x,y)\coloneqq
  x\ne y\land\rho_S^+(x,y)\land\rho_S^-(x,y)
\]
defines $R(G,\PP)$.
Each $e_S^\sigma$ has quantifier rank~$1$, and each disjunct in
$\rho_S^\sigma$ quantifies at most four intermediate vertices.
Thus $\phi_R$ is a fixed first-order formula of quantifier rank at most~$5$.
\end{proof}

\begin{lemma}\label{lem:monotoneR}
    Fix a graph $G$ and partitions $\PP$ and $\PP'$ of $V(G)$ such that $\PP$ refines $\PP'$. Then $$R(G,\PP)\subseteq R(G, \PP').$$
\end{lemma}
\begin{proof}
    Choose any $uv \in R(G,\PP)$. Let
    \[
        X \coloneqq \PP(u), \quad
        X' \coloneqq \PP'(u), \quad
        Y \coloneqq \PP(v), \quad
        Y' \coloneqq \PP'(v)
    \]
    be the parts of $\PP$ and $\PP'$ that contain $u$ and $v$.
    As $\PP$ refines $\PP'$, we have $X \subseteq X'$ and $Y \subseteq Y'$.
    Hence $G[X,Y]$ and $\overline G[X,Y]$ are induced subgraphs of $G[X',Y']$ and $\overline G[X',Y']$, respectively.
    Thus, by the definition of $R$,
    \begin{align*}
        \dist^{G[X',Y']}(u,v) &\le \dist^{G[X,Y]}(u,v) \le 5,\\
        \dist^{\overline G[X',Y']}(u,v) &\le \dist^{\overline G[X,Y]}(u,v) \le 5.
    \end{align*}
    This proves $uv \in R(G, \PP')$, as required.
\end{proof}

\begin{lemma}\label{lem:ballContainment}
    Fix a graph $G$ and a partition $\PP$ of $V(G)$ and let $R \coloneqq R(G,\PP)$. For every radius $r \in \N$ and vertex $v \in V(G)$ we have
    \[
        \Ball^R_r(v) \subseteq \Ball^\PP_{5r}(v).
    \]
\end{lemma}

\begin{proof}
    It suffices to show that for every $uv \in R$ we have $\dist_\PP(u,v) \leq 5$.
    Let $X,Y\in\PP$ be the parts containing $u$ and $v$, respectively. Fix an arbitrary $\PP$-flip $H$ of $G$.
    Then the bipartite graph $H[X,Y]$ is either $G[X,Y]$ or $\overline G[X,Y]$.
    By the definition of $R$, this graph contains a $u$--$v$ path of length at most $5$.
    If $X \neq Y$, then this path is also a path in $H$.
    If $X = Y$, we can project this path to a length ${\leq}5$ path in $H$ by contracting diagonal edges. Hence, in both cases,
    $\dist^H(u,v) \leq 5$, which finishes the proof.
\end{proof}

\begin{lemma}\label{lem:default}
    Fix a graph $G$ and a partition $\PP$ of $V(G)$ and let $T$ be a transversal of $\PP$ and $R \coloneqq R(G,\PP)$.
    Then $\QQ \coloneqq \PP \wedge T$ is \emph{homogeneous modulo $R$}, i.e., for all $X,Y \in \QQ$ either
    \[
        XY \setminus R \subseteq E(G)
        \quad
        \text{or}
        \quad
        XY \setminus R \subseteq E(\overline{G}).
    \]
\end{lemma}

\begin{proof}
    We construct a $\QQ$-flip $H$ of $G$ with $E(H)\subseteq R$. This will yield the conclusion.
    Indeed, let $X,Y\in\QQ$. Since $H$ is a $\QQ$-flip of $G$, either
    \begin{center}
    $E(H)\cap XY \,=\, E(G)\cap XY$ \quad or \quad  $E(H)\cap XY \,=\, E(\overline G)\cap XY$.    
    \end{center}
    Since $E(H)\subseteq R$, no pair in $XY\setminus R$ is an edge of $H$.
    Hence $XY\setminus R\subseteq E(\overline G)$ in the first case, and
    $XY\setminus R\subseteq E(G)$ in the second.
    It remains to define $H$. We separately describe how to flip a part $X$ with itself, and how to flip two distinct parts $X$ and $Y$ with each other.

    \begin{itemize}
        \item     For every part $X\in\PP$, by \Cref{lem:diamComplement} we have that 
    either $G[X]$ or $\overline{G}[X]$ has diameter at most~$3$.
    If $G[X]$ has diameter at most $3$, set 
 $H[X]\coloneqq \overline G[X]$, otherwise set $H[X]\coloneqq G[X]$.

 Suppose now that $uv\in E(H[X])$ for some $u,v\in X$. 
 We consider the case when $G[X]$ has diameter at most $3$ and 
  $H[X] = \overline G[X]$, the other case being symmetric. 
Then $\dist^{G[X]}(u,v)\le 3$ and $\dist^{\overline G[X]}(u,v)= 1$.
The path of length at most $3$ in $G[X]$ lifts to a path of length at most $3$ between opposite copies in $G[X,X]$ after adding at most one diagonal step to make its length odd.
Similarly, the edge $uv$ in $\overline G[X]$ lifts to a path of length $1$ between opposite copies in $\overline G[X,X]$. This shows that $uv\in R$, and thus $E(H[X])\subseteq R$.

\item
For every pair of distinct parts $X,Y\in\PP$, let $x,y$ be their representatives in $T$ and apply \Cref{lem:bipartite-informal-lemma} to $G[X,Y]$.
    Use the resulting bipartite graph as $H[X,Y]$. The statement of the lemma implies that $E(H[X,Y])\subseteq R$.

    The partition $\QQ$ restricted to $X\cup Y$, denote it by $\QQ[X,Y]$, refines the neighborhood partitions of $X$ over $y$ and of $Y$ over $x$, so $H[X,Y]$ is a $\QQ[X,Y]$-flip of $G[X,Y]$.
    Altogether, $H$ is a     
    $\QQ$-flip of $G$ with $E(H)\subseteq R$.\qedhere
    \end{itemize}
\end{proof}

We now prove \Cref{lem:better-mw}.

\lembettermw*

\begin{proof}
    Let 
    $V(G)=S_1\supseteq\cdots\supseteq S_\ell=\emptyset$ be the given chain.
    Let $\PP_i$ be the partition into atomic types over $S_i$.
    Fix an arbitrary linear order $<$ on $V(G)$ 
    giving rise to representative sets $T_i \coloneqq \{\min_< P \,:\, P \in \PP_i\}$. 
    Then $T_1\supseteq T_2\supseteq \dots \supseteq T_\ell$.
    Define $$R_i \coloneqq R(G,\PP_i)\qquad\text{ and }\qquad\QQ_i \coloneqq \PP_i \wedge T_i.$$

    We claim that 
    \[
        (\QQ_1, R_1), \ldots, (\QQ_\ell, R_\ell)
    \]
    is a merge sequence.  Indeed:
    \begin{enumerate}
        \item Since $\PP_1$ is the partition into singletons, so is
        $\QQ_1$.  
        It follows from $\PP_\ell=\{V(G)\}$ and the definition of $\PP_\ell \wedge T_\ell$ that also
        $\QQ_\ell=\{V(G)\}$.

        \item For $i<j$ we have that $T_i\supseteq T_j$ and $\PP_i\preceq \PP_j$, which implies 
that  $\PP_i\wedge T_i\preceq \PP_j\wedge T_j$. Hence,

    \[
        \QQ_1 \preceq \QQ_2 \preceq \dots \preceq \QQ_\ell    
    \]
    is a sequence of ever coarser partitions.

    \item By \Cref{lem:monotoneR}, we have that 
    \[
        R_1 \subseteq R_2 \subseteq \dots \subseteq R_\ell.
    \]

    \item By \Cref{lem:default}, we have that for all $P,Q \in \QQ_i$ either
    \[
        PQ \setminus R_i \subseteq E(G)
        \quad
        \text{or}
        \quad
        PQ \setminus R_i \subseteq E(\overline{G}).
    \]
    \end{enumerate}
    Fix $r\in\N$, and let $k$ be the radius-$(5r+1)$ width of the given chain.
    For every $i\in\{2,\ldots,\ell\}$ and $v\in V(G)$, the definition of chain width gives
    \[
        |\Ball_{5r + 1}^{\PP_i}(v) / \PP_{i-1}| \leq k.
    \]
    Applying \Cref{lem:transversalRefinement} to
$\PP_{i-1}\preceq\PP_i$ and its transversal $T_{i-1}$ yields
    \begin{equation}\label{eq:ball-small}
         |\Ball_{5r}^{\PP_i}(v) / \QQ_{i-1} | = |\Ball_{5r}^{\PP_i}(v) / (\PP_{i-1} \wedge T_{i-1})| \leq k \cdot \pi_G(k).
    \end{equation}
    By \Cref{lem:ballContainment}, we have
    \begin{equation}\label{eq:ball-contained}
        \Ball_{r}^{R_i}(v) 
        \subseteq
        \Ball_{5r}^{\PP_i}(v).
    \end{equation}
    Combining \eqref{eq:ball-small} and \eqref{eq:ball-contained} yields the desired bound on the radius-$r$ width of the merge sequence:
    \begin{equation*}
        |\Ball_{r}^{R_i}(v)/\QQ_{i-1}| \leq k \cdot \pi_G(k).
    \end{equation*}

    Finally, consider the graph $G$ equipped with two unary predicates marking $S_i$ and $T_i$.    
    The relation $R_i$ is definable in this expansion by \Cref{lem:def-R} by a formula of quantifier rank $5$.
    The relation $\equiv_{\QQ_i}$ is
    defined easily using $S_i$ and $T_i$ by a formula of quantifier rank $2$.

    We now analyze the running time of a straightforward algorithm computing the resulting merge sequence. Put $n\coloneqq |V(G)|$.
    For each $i\in[\ell]$, keep every vertex of $S_i$ in a singleton part and group the remaining vertices by their adjacency vectors over $S_i$.
    This computes $\PP_i$ in $O(n(|S_i|+1))\le O(n^2)$ time.
    Scanning the resulting partition computes $T_i$ in $O(n)$ time.
    Similarly, compute the partition $\QQ_i$ in time $O(n(|T_i|+1))\le O(n^2)$.
    Thus computing and outputting $\PP_i$, $\QQ_i$, $S_i$, and $T_i$ takes $O(n^2)$ time for a fixed $i$.

    Compute $R_i$ by running breadth-first searches to depth five from every vertex in each of $G[X,Y]$ and $\overline G[X,Y]$, for all pairs $X,Y\in\PP_i$, allowing $X=Y$.
    The cost per pair is $O((|X|+|Y|)|X||Y|)$. We have:
    \[
        \sum_{X,Y\in\PP_i} (|X|+|Y|)|X||Y|
        =2\sum_{X,Y\in\PP_i} |X||Y|^2
        =2\sum_{X\in\PP_i} |X|\cdot \sum_{Y\in\PP_i}|Y|^2
        \le 2n(\sum_{Y\in\PP_i}|Y|)^2=2n^3,
    \]
     thus $R_i$ can be computed in time $O(n^3)$, and the whole merge sequence in time $O(\ell n^3)$.
\end{proof}

\section{Construction sequences with locally definable partial graphs}\label{sec:construction-sequences}

For the purpose of model checking, 
we need to work with construction sequences rather than merge sequences. 
The construction of~\cite[Lemma~3.3]{mw-arxiv} shows how to convert a merge sequence to a construction sequence. However, a single step in a merge sequence is translated into a potentially long sequence of steps in a construction sequence. When aiming at defining each step of the construction sequence in the original graph $G$, we need to remember the result of the entire sequence of all the steps performed so far. 
We cannot hope to obtain construction sequences in which each part is definable by a formula with fewer than $n/3$ parameters, even for the edgeless graph on $n$ vertices.
Therefore, in the following lemma, we only aim at proving that \emph{locally} -- in each ball of radius $r$ in the graph formed by the resolved vertex pairs -- the current set of resolved edges and each part of the current partition are definable in $G$ by formulas of bounded quantifier rank, involving a number of parameters which is bounded by the radius-$r$ merge-width of the input merge sequence. Recall the precise definition 
of definability stated in \Cref{def:partial-definable}, and used in the statement below.
It is crucial that when considering merge sequences of bounded length, we
can retain the unary predicates describing every stage, and thus obtain formulas using only boundedly many unary predicates.

\begin{lemma}\label{cl:from-merge-to-construction}
    For every $n$-vertex graph $G$ and merge sequence $(\PP_1,R_1),\ldots,(\PP_\ell,R_\ell)$ of $G$, there is a construction sequence $G_1,\ldots,G_\lambda$ of $G$ such that the following properties hold for all $r\ge 1$.
    Let $w_r$ be the radius-$r$ width of the merge sequence. 
    \begin{enumerate}
        \item The obtained construction sequence has radius-$r$ width at most $w_r$, and has length $\lambda\le O(nw_1)$.
        \item Suppose that for every $i\in[\ell]$, the relations $\equiv_{\PP_i}$ and $R_i$ are definable in some expansion of $G$ with~$2$ unary predicates, using two fixed first-order formulas of quantifier rank at most~$5$.

        Then, for every $t\in[\lambda-1]$ and  $v\in V(G)$, the induced partial graph
        $G_t[\Ball^{G_{t+1}}_r(v)]$ is definable in some expansion of $G$
        with $2\ell+1$ unary predicates, using formulas of quantifier rank~$7$ with $w_r$
        parameters.
        
        \item  The construction sequence can be
computed in time $O(\ell n^2)$, given an $n$-vertex graph $G$ and a merge sequence of length $\ell$.    \end{enumerate}
    \end{lemma}

\begin{proof}
We convert the merge sequence to a construction sequence by following the proof of~\cite[Lemma~3.3]{mw-arxiv}. Set $V\coloneqq V(G)$, $R_{\ell+1}\coloneqq \binom{V}{2}$, and $\PP_{\ell+1}=\set{V}$.

Start with the singleton partition $\PP$ of $V$ and empty set $R$ of resolved vertex pairs.
The construction proceeds in \emph{stages} $j=1,\ldots,\ell$.
The invariant that is maintained during stage $j$ is that in each step, the current partition $\PP$ and current set $R$ of resolved vertex pairs satisfy:
\begin{align}\label{invariant-partitions}
    \PP_j&\preceq \PP\preceq \PP_{j+1}\qquad\\ R&\subseteq R_{j+1}\label{invariant-resolved}\\
    XY\setminus R\subseteq E(G)\quad&\text{ or }\quad XY\setminus R\subseteq E(\overline G)\quad\text{for all $X,Y\in \PP$}.
\end{align}

We start the process with $\PP \coloneqq \PP_1$ being the partition into singletons and $R \coloneqq \emptyset$.
Stage
$j$ proceeds as follows. First consider all unordered pairs of parts
$X,Y\in\PP_{j}$, allowing $X=Y$, for which
$XY\subseteq R_{j+1}$.
If there are unresolved vertex pairs in $XY$ (i.e.\ $XY\nsubseteq R$), resolve the pair $X,Y$
with the common sign -- positively if $XY\setminus R\subseteq E(G)$ and negatively if $XY\setminus R\subseteq E(\overline G)$ -- and update the set of resolved vertex pairs $R\coloneqq R\cup XY$.
Then repeatedly merge pairs $A,B$ of parts of the current partition for which $A\cup B$ is contained in one part of $\PP_{j+1}$, until the current partition $\PP$ becomes $\PP_{j+1}$; if $j=\ell$, terminate. This results in a construction sequence of $G$.

Fix $r\geq 1$.
We now upper bound the radius-$r$ width of the obtained construction sequence by the radius-$r$ width $w_r$ of the merge sequence.
Fix a stage $j$ in the merge sequence and a step 
of the construction above in that phase, with current partition $\PP$ and current set of resolved pairs $R\subseteq \binom{V}{2}$. By \eqref{invariant-resolved}, for every $v\in V$ we have:
\begin{align}\label{ball-inc}
    \Ball_r^{R}(v)&\subseteq \Ball_r^{R_{j+1}}(v).
\end{align}
As $\Ball_r^{R_{j+1}}(v)$ intersects at most $w_r$ parts of $\PP_j$ and $\PP$ is coarser than $\PP_j$ by \eqref{invariant-partitions},
also $\Ball_r^{R}(v)$ intersects at most $w_r$ parts of $\PP$, proving the required bound.

The obtained construction sequence is \emph{effective}: each resolve operation introduces at least one new resolved vertex pair. By~\cite[Lemma~3.4]{mw-arxiv} (and the comment before it), the length $\lambda$ of the construction sequence is at most $n(2w_1+1) \leq O(nw_1)$.

A straightforward implementation processes each stage in time $O(n^2)$.
Namely, for each pair $X,Y$ of parts of $\PP_j$,
test whether $XY\subseteq R_{j+1}$ by scanning 
all vertex pairs $uv\in XY$, and record the sign with which $X,Y$ should be resolved. As this processes each pair $uv\in \binom{V}{2}$ once, this takes time $O(n^2)$ in total. Similarly, the merge phase can be implemented in time $O(n)$.
Altogether, the  construction sequence is computable in time $O(\ell n^2)$.
This proves the first and last conditions of the lemma.

\medskip

We now prove the second condition of the lemma, regarding definability. 
Fix a step $G_t\to G_{t+1}$ of the obtained construction sequence, and let $j$ be the stage
in which this step is performed.
Let $\PP$ and $R$ denote the partition and the set of resolved vertex pairs in $G_t$, and let $R_{+}$ denote the set of resolved vertex pairs in $G_{t+1}$; then $R\subseteq R_{+}\subseteq R_{j+1}$.

Let $\phi_\equiv^i(x,y)$ and $\phi_R^i(x,y)$ be formulas defining 
$\equiv_{\PP_i}$ and $R_i$, respectively, as given by the assumption.
For $1\le j<\ell$, the set $$\bigcup\setof{XY}{X,Y\in\PP_j, XY\subseteq R_{j+1}}$$ is defined 
by the formula
\[
 \gamma_j(x,y) \,\coloneqq\, x\ne y\ \land\forall a\forall b\,\bigl(
       (\phi_\equiv^{j}(x,a)\land\phi_\equiv^{j}(y,b)\land a\ne b)\,\to\,\phi_R^{j+1}(a,b)\bigr),
\]
which has quantifier rank at
most~$7$.  Consequently, at the start of stage~$j$, the resolved
relation, which we denote $R_j'$\,,
is defined by
\[
 \Gamma_{j}(x,y)\coloneqq \bigvee_{i=1}^{j-1}\gamma_i(x,y).
\]
This formula uses $2j$ unary predicates and has quantifier rank at most $7$.

Choose
one parameter $p_X\in X$ for every part $X\in \PP_j$.
The part $X$ is then defined by $\phi_\equiv^{j}(x,p_X)$, which 
uses $2$ unary predicates, one parameter, and has quantifier rank at most $5$.

\medskip
Fix a vertex $v\in V$.
Denote
\[
 B\coloneqq \Ball^{R_{+}}_r(v)=\Ball^{G_{t+1}}_r(v)\subseteq\Ball^{R_{j+1}}_r(v).
\]
Mark the set $B$ with a unary predicate $U_B$.

We will show how to define in $G$ the relation $R\cap \binom{B}{2}$ and the sets $X\cap B$, for each $X\in \PP$,
using Boolean combinations of the formulas defined above, 
using the $2j+1$ distinguished unary predicates, and at most $w_r$ parameters.
Once $R\cap \binom{B}{2}$ is defined, the set of  edges and non-edges of $G_t$, restricted to $\binom{B}{2}$, can be recovered by further intersecting with $E(G)$ and $E(\overline G)$, respectively. This will therefore define the induced substructure $G_t[B]$, as required in the second condition of the lemma.

\medskip
We first focus on defining $R\cap \binom{B}{2}$.
Note that $R$ 
is the union of $R_j'$
and all already processed rectangles $XY$ with $X,Y\in\PP_j, XY\subseteq R_{j+1}$.
Therefore, $R$ is defined by the disjunction 
of $\Gamma_j(x,y)$ and disjuncts of the form
\[
 (x\neq y)\land \left((\phi_\equiv^{j}(x,p_X)\land\phi_\equiv^{j}(y,p_Y))
 \lor
 (\phi_\equiv^{j}(x,p_Y)\land\phi_\equiv^{j}(y,p_X))\right).
\]
To define the intersection of $R$ with $\binom{B}{2}$,
add a conjunct $U_B(x)\land U_B(y)$,
and keep only those disjuncts which involve parts $X,Y$ that intersect $B$. As there are at most $w_r$ many such parts, this results in a formula
 using $w_r$ parameters.

Each current part $X\in\PP$ is a union of parts of $\PP_j$, so its intersection with $B$ is defined by 
the formula
$$U_B(x)\land \bigvee_{Y}\phi_\equiv^{j}(x,p_Y)$$ where $Y$ ranges over all parts $Y\in\PP_j$ with $Y\cap B\neq\emptyset$ and $Y\subseteq X$, with empty disjunction interpreted as \emph{false}.
The resulting formula uses only $w_r$ parameters.
This completes the proof of the lemma.
\end{proof}

We now prove \Cref{lem:better-definable-mw-from-sets}.
\lembetterdefinablemwfromsets*

\begin{proof}
    The empty graph is immediate, so assume $G$ is nonempty.
    Apply \Cref{lem:better-mw} to the given chain to obtain a merge sequence with $\ell$ stages, each definable by two unary predicates and fixed formulas of quantifier rank at most $5$.
    Fix $r\ge 1$ and let $k_r$ be the radius-$(5r+1)$ width of the chain.
    Then the radius-$r$ width $w_r$ of this merge sequence is at most $k_r\cdot \pi_G(k_r)$. Apply \Cref{cl:from-merge-to-construction} to obtain a construction sequence of length $\lambda\le O(nw_1)$ and radius-$r$ width at most $w_r$, with local definitions using at most $w_r$ parameters, $2\ell+1$ unary predicates, and formulas of quantifier rank at most $7$.
    Both algorithms take $O(\ell n^3)$ time and use only $G$ and the given chain.
\end{proof}

\clearpage
 \part{Model checking with type trees}
 \label{part:mc}

 In this part we prove:
{\renewcommand{\mcdepth}{d}\mcthm*}

\section{Preliminaries}\label{sec:mc-prelims}

In \Cref{part:mc},
unless stated otherwise, $G$ denotes a partial graph.
We slightly extend the notions of partial graphs and of construction sequences.

\paragraph{Partial graphs.}
To track parts over time, we give them names.
Fix a countably infinite set of unary symbols, called \emph{colors}.
We regard a partial graph $G$ as a relational structure with the two symmetric binary relations $E$ and $N$ and a unary relation $Y$ for every color $Y$.
The vertices of color $Y$ form its color class $Y^G$, and the nonempty color classes are the parts of $G$.
We write $|G|\coloneqq|V(G)|$ and $\size G\coloneqq|V(G)|+|R(G)|$ for the size of the representation of a partial graph $G$.
The \emph{distance matrix} of $G$ is the $|G|\times|G|$ matrix $D_G$ with entries $D_G(u,v)\coloneqq\dist^G(u,v)\in\N\cup\set\infty$.
Passing it as an additional input to the algorithms below saves time, as distances need not be recomputed repeatedly.

\paragraph{Construction sequences.}
For colors $P,Q$ and a partial graph $G$, we define three partial graphs
$\merge PQ(G)$ (for $P\neq Q$), $\resolve+PQ(G)$, and $\resolve-PQ(G)$, as the partial graph $H$ obtained from $G$ by leaving all relations of $G$ unchanged, apart from redefining the following:
\begin{align*}
\merge PQ(G)\colon\quad&
P^{H}\coloneqq P^G\cup Q^G,
\qquad
Q^{H}\coloneqq\emptyset
&&\text{for $P\ne Q$},\\
\resolve+PQ(G)\colon\quad&
E(H)\coloneqq E(G)\cup\bigl(P^GQ^G\setminus N(G)\bigr),\\
\resolve-PQ(G)\colon\quad&
N(H)\coloneqq N(G)\cup\bigl(P^GQ^G\setminus E(G)\bigr).
\end{align*}
After either resolve, every pair in $P^GQ^G$ is resolved.
In this representation, every step of a construction sequence $G_1,\ldots,G_\lambda$ (\Cref{sec:tech-overview}) has the form $G_{t+1}=\merge PQ(G_t)$, $\resolve+PQ(G_t)$, or $\resolve-PQ(G_t)$ for some colors $P,Q$.

\section{Distance logic}\label{sec:dist-rank}

Following \cite{mw-arxiv,mw-stoc} (and earlier work by \cite{gks}),
we define a logic called \distFO, extending first-order logic by distance atoms, and a suitable notion of rank. We then define compact representations of types tailored to this logic.

\subsection{Distance logic and distance rank}

\paragraph{First-order distance logic.}
The logic \distFO extends first-order logic by additionally allowing the following two types of atoms:
\begin{quote}
     For each radius \(r \in \N\), \distFO introduces
        a \emph{binary distance atom} \(\dist(x,y) \le r\) expressing that the distance between \(x\) and \(y\) is at most \(r\).

     For each radius \(r \in \N\) and color \(Y\), \distFO introduces
        a \emph{unary distance atom} \({\dist(x,Y) < r}\) expressing that the distance from \(x\) to \(Y\) is smaller than \(r\),
        equivalent to \({\exists y~Y(y) \land \dist(x,y) < r}\).
\end{quote}
In both cases, we call \(r\) the \emph{radius} of the distance atom.
Distance atoms are expressible in first-order logic, but only with quantifiers.
Treating them as atoms lets us measure distances without increasing the quantifier rank, as in~\cite{gks}.

\paragraph{Horizon functions.}
We fix sufficiently fast-growing horizon functions as in \cite{locality-note}:
\begin{equation}\label{eq:horizon-def}
\rhom(k,q)\coloneqq 9^{(k+q+1)q}
\qquad\text{and}\qquad
\rhop(k,q)\coloneqq 9^{(k+q)(q+1)}.
\end{equation}
They satisfy the two identities
\begin{equation}\label{eq:horizon}
\rhop(k,q)=9^k\rhom(k,q)\quad\text{for all $k,q\in\N$},
\qquad
\rhom(k,q)=9^q\rhop(k+1,q-1)\quad\text{for $q\ge1$}.
\end{equation}
In particular, as $9^q\ge2$,
\begin{equation}\label{eq:horizon-cons}
\rhop(k,q)\ge\rhom(k,q)\ge2\rhop(k+1,q-1)\quad\text{for all $k\in\N$ and $q\ge1$}.
\end{equation}

\paragraph{Distance rank.}
Let $k,q\in\N$.
A \distFO formula has \emph{distance rank \((k,q)\)} if it is a Boolean combination of
\begin{itemize}
\item first-order atoms: relational symbols in the signature or equality, applied to variables,
\item binary distance atoms with radius $r\le\rhom(k,q)$,
\item unary distance atoms with radius $r\le\rhop(k,q)$,
\item
formulas
\[
\exists y~\Bigl(\dist(x_i,y)\le \rhop(k+1,q-1)\land\phi(\tup x y)\Bigr),
\]
where \(q \ge 1\), $x_i$ is a variable in $\tup x$, and $\phi(\tup x y)$ has distance rank $(k+1,q-1)$.
\end{itemize}

A formula with distance rank \((k,q)\) has quantifier rank at most \(q\).
Observe by \eqref{eq:horizon-cons} that
the distance atoms have progressively shorter radii as one traverses the quantifiers
inwards. 

Formulas of distance rank $(k,q)$ are called \emph{local} formulas of distance rank $(k,q)$ in 
\cite{mw-arxiv,locality-note}, as opposed to the more general formulas which also allow the unrestricted quantification $\exists y\, \psi(\tup x y)$. Since we only consider local formulas in this paper, we omit this adjective.


\subsection{Type trees}
\label{sec:typetrees}
 In the context of usual first-order logic, the \emph{type} of quantifier rank $q$ of a tuple $\tup a$ in a structure $A$ is defined as the set of all formulas of quantifier rank $q$ that are satisfied by $\tup a$ in $A$ (see \Cref{sec:prelims}). It is well known~\cite{Scott1979}
  that for structures over a finite relational signature, such types can be identified with \emph{trees}, also called \emph{Hintikka types},
defined recursively as follows:
$$\tree_q^A(\tup a)=\begin{cases}
\atp^A(\tup a)&\text{if $q=0$,}\\
\setof{\tree_{q-1}^A(\tup a b)}{b\in V(A)}&\text{if $q>0$}.
\end{cases}$$
In the setting of \distFO formulas of distance rank $(k,q)$, the (local) \emph{type of distance rank $(k,q)$} of a tuple $\tup a$
is defined  \cite{mw-arxiv} as the set of (local) \distFO formulas of distance rank $(k,q)$ satisfied by $\tup a$.
In this section, we adapt the construction of trees to this setting. This is useful in the algorithmic context, as trees are much smaller objects than types.

\paragraph{Type trees.}
In the two definitions below, let $G$ be a partial graph, $k,q\in\N$, and $\tup a\in V(G)^{\le k}$, and put $\ell\coloneqq|\tup a|$.
The \emph{atomic $(k,q)$-type}, denoted $\atp_{k,q}(\tup a)$, is the set of all atomic formulas $\phi$
of distance rank \((k,q)\)
with free variables among $x_1,\ldots,x_\ell$, such that $G\models\phi(\tup a)$. Note that this set is finite, as an atom mentioning a color that does not occur in $G$ is not satisfiable in~$G$.

Define the \emph{$(k,q)$-type tree} of $\tup a$ as
\[
\tree_{k,q}\bigl(\tup a\bigr)
\coloneqq
\bigl(\atp_{k,q}(\tup a),\mathcal E_{k,q}(\tup a)\bigr) 
\]
where $\mathcal E_{k,q}(\tup a)=\emptyset$ if $q=0$, and otherwise
\[
\mathcal E_{k,q}\bigl(\tup a\bigr)
\coloneqq
\setof{\bigl(i,\tree_{k+1,q-1}(\tup a b)\bigr)}{i\in[\ell],\ b\in V(G),\ \dist(a_i,b)\le\rhop(k+1,q-1)}.
\]

Note that type trees are ``local'' in the sense that new vertices are always quantified close to previous ones (see \Cref{lem:locality}).
The members of $\mathcal E_{k,q}(\tup a)$ are the \emph{extensions} of $\tree_{k,q}(\tup a)$. The position $i$ of an extension $(i,\mu)$ is called the \emph{guard}, and we say that $i$ \emph{admits} $b$ if $\dist(a_i,b)\le\rhop(k+1,q-1)$.
We view $\tree_{k,q}(\tup a)$ as a labeled, rooted tree
 of depth $q$,
with the root labeled $\atp_{k,q}(\tup a)$ and the root of each type tree $\mu$ of an extension $(i,\mu)\in \mathcal E_{k,q}(\tup a)$ attached as a child of the root via an edge labeled $i$.
We call $\ell$ the \emph{arity} of the type tree, and $(k,q)$ its \emph{rank}.

\paragraph{Representation.}
Type trees are represented in the standard RAM model in a canonical (say, sorted,
duplicate-free) form, so that equality of two type trees $\tau_1,\tau_2$ can be
tested in time $O(\size{\tau_1}+\size{\tau_2})$.
By $\size\tau$ we denote the size of this representation (which includes $k,q$, and $\ell$).

\begin{observation}[canonical form]\label{obs:canonical}
A type tree given by a representation whose sets need not be sorted or duplicate-free can be brought into canonical form in polynomial time.
\end{observation}

The next lemma states that the type trees indeed represent the types. It is the reason for computing type trees in the first place.

\begin{lemma}[evaluation]\label{lem:eval}
Let $\phi(\tup x)$ be a \distFO formula of distance rank $(k,q)$, \(G\) be a partial graph and $\tup a\in V(G)^{\ell}$ with $\ell=|\tup x|\le k$.
Then whether $G\models\phi(\tup a)$ holds is decidable in polynomial time, given $\tree^G_{k,q}(\tup a)$ and $\phi$.
\end{lemma}

\textbf{Notation.} For a $(k,q)$-type tree $\tau=\tree^G_{k,q}(\tup a)$
and a \distFO formula $\phi(\tup x)$ of distance rank $(k,q)$,
we write $\tau \models\phi$, if $G\models\phi(\tup a)$.
The lemma implies that this does not depend on $G$ and $\tup a$, only on the type tree $\tau$.

\begin{proof}[Proof of \Cref{lem:eval}]
By induction on $q$.
Write $\tau\coloneqq\tree_{k,q}(\tup a)$.
By definition of distance rank, $\phi$ is a Boolean combination of atoms and of existential components, and we evaluate these components separately.
An atom holds of $\tup a$ if and only if it occurs in the root label $\atp_{k,q}(\tup a)$ of $\tau$.
Now let $q\ge1$ and put $H\coloneqq\rhop(k+1,q-1)$.
Every existential component $\chi$ of $\phi$ can be written as $$\exists y~(\dist(x_i,y)\le H\land\psi(\tup x y))$$ with $i\in[\ell]$ and $\psi$ of distance rank $(k+1,q-1)$.
By the inductive assumption, whether $G\models\psi(\tup c)$ holds is determined by $\tree_{k+1,q-1}(\tup c)$ and $\psi$, and by the notation above, this holds if $\tree_{k+1,q-1}(\tup c)\models \psi$.
Unfolding the semantics, then applying the inductive assumption, and finally the definition of $\mathcal E_{k,q}(\tup a)$, we obtain
\begin{align*}
G\models\chi(\tup a)
&\iff\exists b\in V(G)\colon\ \dist(a_i,b)\le H\ \text{and}\ G\models\psi(\tup ab)\\
&\iff\exists b\in V(G)\colon\ \dist(a_i,b)\le H\ \text{and}\ \tree_{k+1,q-1}(\tup ab)\models\psi\\
&\iff\exists\mu\colon\ (i,\mu)\in\mathcal E_{k,q}(\tup a)\ \text{and}\ \mu\models\psi.
\end{align*}
So $\chi$ is decided by evaluating $\psi$, via the inductive assumption, on the extensions of $\tau$ with guard~$i$.

This proof yields a polynomial-time algorithm.
\end{proof}

\begin{corollary}\label{cor:tp-tree}
   Fix $\ell\le k$ and $q$,  a partial graph $G$, and  $\tup a,\tup b\in V(G)^\ell$. Then $\tree^G_{k,q}(\tup a)=\tree^G_{k,q}(\tup b)$ if and only if $G\models \phi(\tup a)\Leftrightarrow \phi(\tup b)$ for every \distFO formula $\phi(\tup x)$  of distance rank $(k,q)$ with $|\tup x|=\ell$.
\end{corollary}
\begin{proof}
   The rightwards implication follows from \Cref{lem:eval}. The converse is proved by an easy induction on $q$.
\end{proof}

\begin{lemma}[locality]\label{lem:locality}
Let $k\ge1$, $q\in\N$, let $G$ be a partial graph, and let $a\in V(G)$.
Every node at depth $j$ of $\tree_{k,q}(a)$ is $\tree_{k+j,q-j}(\tup v)$ for some tuple $\tup v$ from $\Ball_{\rhop(k,q)}(a)$, and every color mentioned in the root label intersects this ball.
\end{lemma}
\begin{proof}
By the definition of extensions and induction on $j$, a node at depth $j$ is $\tree_{k+j,q-j}(\tup v)$ for a tuple $\tup v=(v_0,\ldots,v_j)$ with $v_0=a$ in which every $v_i$ with $i\ge1$ is within distance $\rhop(k+i,q-i)$ of an earlier entry.
By induction on $i$ and the triangle inequality, $\dist(a,v_i)\le\sum_{i'=1}^{i}\rhop(k+i',q-i')$.
The root label mentions colors only in atoms $Y(x_i)$ and $\dist(x_i,Y)<r$ with $r\le\rhop(k+j,q-j)$, and such an atom holds of $\tup v$ only if a vertex of $Y$ is within distance $r$ of $v_i$.
Hence every entry of $\tup v$, and some vertex of every color in an atom in the root label, is within distance
\[
\sum_{i=1}^{j}\rhop(k+i,q-i)+\rhop(k+j,q-j)\le\rhop(k,q)
\]
of $a$, where the inequality uses $\rhop(k+i,q-i)\le2^{-i}\rhop(k,q)$, which follows by applying \eqref{eq:horizon-cons} $i$ times.
\end{proof}

\subsection{Operations on type trees}

\begin{lemma}[ordered coordinate restriction]\label{lem:restrict}
There is an operation $\restriction$ on type trees
such that for all \(k,q\in\N\), partial graphs \(G\),
tuples $\tup w\in V(G)^{\ell}$ for $\ell\le k$ and ordered lists $J$ of distinct positions in $[\ell]$,
\[
\tree_{k,q}(\tup w)\restriction J
=
\tree_{k,q}(\tup w\restriction J).
\]
Given $\tau$ and $J$ one can compute $\tau\restriction J$ in time polynomial in $\size\tau$.
\end{lemma}
\begin{proof}
By induction on $q$.
Let $J=(j_1,\ldots,j_p)$, and $H\coloneqq\rhop(k+1,q-1)$.
Renaming variables gives
\[
\atp_{k,q}(\tup w\restriction J)=\setof{\psi(x_1,\ldots,x_p)}{\psi(x_{j_1},\ldots,x_{j_p})\in\atp_{k,q}(\tup w)},
\]
which settles the root label.
Let $q\ge1$, and put $J^+\coloneqq(j_1,\ldots,j_p,\ell+1)$, so that $(\tup w\restriction J)\,b=(\tup w b)\restriction J^+$ for every vertex $b$.
Unfolding the definition of $\mathcal E_{k,q}(\tup w\restriction J)$, then applying the inductive assumption, and finally the definition of $\mathcal E_{k,q}(\tup w)$, we obtain
\begin{align*}
\mathcal E_{k,q}(\tup w\restriction J)
&=\setof{\bigl(i,\ \tree_{k+1,q-1}((\tup w b)\restriction J^+)\bigr)}{i\in[p],\ b\in V(G),\ \dist(w_{j_i},b)\le H}\\
&=\setof{\bigl(i,\ \tree_{k+1,q-1}(\tup w b)\restriction J^+\bigr)}{i\in[p],\ b\in V(G),\ \dist(w_{j_i},b)\le H}\\
&=\setof{(i,\ \mu\restriction J^+)}{i\in[p],\ (j_i,\mu)\in\mathcal E_{k,q}(\tup w)}.
\end{align*}
So the guard $j_i$ is renumbered to $i$, as the variable $x_{j_i}$ is renamed to $x_i$ in the root.
The right-hand side depends only on $\tree_{k,q}(\tup w)$ and $J$ and we define $\tau\restriction J$ accordingly.
It is computed by one pass over $\tau$, followed by restoring canonical form. As no subtree grows, this takes polynomial time by \Cref{obs:canonical}.
\end{proof}

\begin{lemma}[rank shift]\label{lem:shift}
There is a polynomial-time algorithm which maps
$\tree_{k,q}(\tup a)$ to $\tree_{k+1,q-1}(\tup a)$ for all \(k\in\N\), \(q\ge1\),
partial graphs \(G\) and nonempty $\tup a\in V(G)^{\le k}$.
\end{lemma}
\begin{proof}
Since $\dist(a_1,a_1)=0$, the guard $1$ admits $a_1$, so $\tree_{k,q}(\tup a)$ has an extension $(1,\mu)$.
Every such $\mu$ is $\tree_{k+1,q-1}(\tup ab)$ for some vertex $b$, and applying the algorithm of \Cref{lem:restrict} yields
$\mu\restriction(1,\ldots,|\tup a|)=\tree_{k+1,q-1}(\tup a)$.
\end{proof}

The following is an adaptation of \cite[Lemma~3.1]{locality-note}.
\begin{lemma}[far composition]\label{lem:compose}
There is an operation $\ast$ on type trees
such that for all \(k,q\in\N\), partial graphs \(G\) and
nonempty tuples $\tup a_1,\tup a_2$ with $|\tup a_1\tup a_2|\le k$ and $\dist(\tup a_1,\tup a_2)>\rhom(k,q)$,
\[
\tree_{k,q}(\tup a_1)\ast\tree_{k,q}(\tup a_2)
=
\tree_{k,q}(\tup a_1\tup a_2).
\]
Given $\tau_1,\tau_2$ one can compute $\tau_1\ast\tau_2$ in time $\poly_{k,q}(\size{\tau_1}+\size{\tau_2})$.
\end{lemma}
\begin{proof}
By induction on $q$.
Write $\ell_i\coloneqq|\tup a_i|$ and $\tau_i\coloneqq\tree_{k,q}(\tup a_i)$. Note that \(\ell_i\) can be deduced from~$\tau_i$.

Consider an atom of distance rank $(k,q)$ that mentions variables of both $\tup a_1$ and $\tup a_2$.
If the atom was true, we would have $\dist(\tup a_1,\tup a_2)\le1$ for an equality, edge, or non-edge atom, and $\dist(\tup a_1,\tup a_2)\le\rhom(k,q)$ for a binary distance atom, each contradicting $\dist(\tup a_1,\tup a_2)>\rhom(k,q)$.
Every other atom mentions only variables of $\tup a_1$ or only variables of $\tup a_2$, and whether it holds is recorded in $\atp_{k,q}(\tup a_1)$ or $\atp_{k,q}(\tup a_2)$.
Hence $\atp_{k,q}(\tup a_1\tup a_2)$ is determined by $\atp_{k,q}(\tup a_1)$ and $\atp_{k,q}(\tup a_2)$.
This settles the root label.

Let now \(q \ge 1\).
Put $H\coloneqq\rhop(k+1,q-1)$, and let $\ast$ also denote the operation given by the inductive assumption for rank $(k+1,q-1)$.
Write $\tup a_1\tup a_2=(c_1,\ldots,c_{\ell_1+\ell_2})$, and split $\mathcal E_{k,q}(\tup a_1\tup a_2)$ by guards into $\mathcal E_1\cup\mathcal E_2$, where
\begin{align*}
\mathcal E_1&\coloneqq\setof{\bigl(i,\ \tree_{k+1,q-1}(\tup a_1\tup a_2 b)\bigr)}{i\in[\ell_1],\ b\in V(G),\ \dist(c_i,b)\le H},\\
\mathcal E_2&\coloneqq\setof{\bigl(i,\ \tree_{k+1,q-1}(\tup a_1\tup a_2 b)\bigr)}{\ell_1<i\le\ell_1+\ell_2,\ b\in V(G),\ \dist(c_i,b)\le H}.
\end{align*}

Consider $\mathcal E_1$, and let $i,b$ satisfy its conditions.
Then $\dist(\tup a_1,b)\le H$, and the triangle inequality, the hypothesis, \eqref{eq:horizon-cons} and \eqref{eq:horizon} give
\[
\dist(\tup a_1b,\tup a_2)
\;\ge\;\dist(\tup a_1,\tup a_2)-\dist(\tup a_1,b)
\;>\;\rhom(k,q)-H
\;\ge\;H
\;\ge\;\rhom(k+1,q-1).
\]
Hence, we may apply the inductive assumption for rank $(k+1,q-1)$ to the tuples $\tup a_1b$ and $\tup a_2$.
Moreover, let $J$ be the list of positions with $\tup a_1\tup a_2 b=(\tup a_1 b\tup a_2)\restriction J$.
Applying in this order the property of \(J\), \Cref{lem:restrict} and the inductive assumption, and finally the definition of $\mathcal E_{k,q}(\tup a_1)$, we obtain
\begin{align*}
\mathcal E_1
&=\setof{\bigl(i,\ \tree_{k+1,q-1}(\tup a_1 b\tup a_2)\restriction J\bigr)}{i\in[\ell_1],\ b\in V(G),\ \dist(c_i,b)\le H}\\
&=\setof{\bigl(i,\ \bigl(\tree_{k+1,q-1}(\tup a_1b)\ast\tree_{k+1,q-1}(\tup a_2)\bigr)\restriction J\bigr)}{i\in[\ell_1],\ b\in V(G),\ \dist(c_i,b)\le H}\\
&=\setof{\bigl(i,\ \bigl(\mu\ast\tree_{k+1,q-1}(\tup a_2)\bigr)\restriction J\bigr)}{(i,\mu)\in\mathcal E_{k,q}(\tup a_1)},
\end{align*}
where $\tree_{k+1,q-1}(\tup a_2)$ is obtained from $\tau_2$ by \Cref{lem:shift}.
Symmetrically, with the guard $i$ of $\tup a_2$ renumbered to $\ell_1+i$ as the variables of its root are shifted by $\ell_1$, and without reordering since $\tup a_1\tup a_2 b$ already ends with $b$,
\[
\mathcal E_2
=\setof{\bigl(\ell_1+i,\ \tree_{k+1,q-1}(\tup a_1)\ast\mu\bigr)}{(i,\mu)\in\mathcal E_{k,q}(\tup a_2)}.
\]
The right-hand sides depend only on $\tau_1$ and $\tau_2$ and we define $\tau_1\ast\tau_2$ accordingly.
By \Cref{lem:restrict,lem:shift} and the inductive assumption, the work per input extension is $\poly_{k,q}(\size {\tau_1}+\size{\tau_2})$, and the recursion has fixed depth $q$. Restoring canonical form by \Cref{obs:canonical} stays within this bound, which proves the stated running time.
\end{proof}

\section{Scatter types and far trees}\label{sec:localglobal}

The type tree of a tuple $\tup a$ records extensions only by vertices near $\tup a$.
A resolve, however, can bring vertices far from $\tup a$ within guard range, so lifting type trees through a resolve (\Cref{sec:merge-resolve}) also requires the type trees of such far away vertices.
For a vertex $b$ far from $\tup a$, the type tree of $\tup ab$ is obtained from those of $\tup a$ and $b$ by far composition (\Cref{lem:compose}).
As for quantifiers over far witnesses in the proof of Gaifman's locality theorem, it therefore suffices to know which type trees are realized far from $\tup a$.
We collect them in the \emph{far tree} of $\tup a$.

The far tree is not determined by the type tree of $\tup a$ alone. It also depends on global information about $G$.
As in Gaifman's proof, this global information is captured by an analogue of the basic local sentences, which we call the \emph{scatter type} of $G$.
In this section we show that the far tree of $\tup a$ is computable from the type tree of $\tup a$ and the scatter type of $G$.

\Cref{sec:stp} defines scatter types, and \Cref{sec:ftree} uses them to compute the far tree.
The model checking algorithm maintains the type trees of all vertices, and \Cref{lem:compute-stp} shows that the scatter type can be computed from these.

\subsection{Scatter types}\label{sec:stp}

Let $G$ be a partial graph.
A set $S$ of vertices of $G$ is \emph{$r$-scattered} if all vertices in $S$ have pairwise distance larger than $r$ in $G$.
Fix, once and for all, a linear order on $V(G)$. The algorithms below use it without further mention.
For a set $X\subseteq V(G)$ and $r\in\N$, let $S_r(X)$ denote the inclusion-wise maximal $r$-scattered subset of $X$ constructed by the greedy process that repeatedly adds the least available vertex.
If $\nu$ is a type tree of distance rank $(k,q)$ and arity one, write
\[
[\nu]^G
\coloneqq
\setof{b\in V(G)}{\tree_{k,q}(b)=\nu}
\]
for the set of its realizations.

\begin{definition}[scatter type]\label{def:stp}
Let $G$ be a partial graph and $k,q\in\N$. The \emph{$(k,q)$-scatter type} $\stp_{k,q}^G$ is the function  with domain
\begin{align*}
\bigl\{(r,\tree_{k+i,q-i}(b))\ \big|\
i\in[q],\ b\in V(G),\ r\in[9^{k+i}\rhom(k+i,q-i)]\bigr\}
\end{align*}
defined by
\[
\stp^G_{k,q}(r,\nu)
\coloneqq
\min\bigl(|S_r([\nu]^G)|,\ k+q\bigr).
\]
For $q=0$ the domain is empty and $\stp_{k,0}$ is the empty function. We write $\stp_{k,q}=\stp^{G}_{k,q}$.
\end{definition}

We represent a scatter type canonically as a sorted list of its triples
$(r,\nu,\stp_{k,q}(r,\nu))$, using the canonical representation of $\nu$.
Its representation size, denoted by $\size{\stp_{k,q}}$, is the total size of
this list, including the encodings of the radii and values.

\begin{lemma}\label{obs:stp-restrict}
For every partial graph $G$ and $q\ge 1$, the function $\stp_{k+1,q-1}$ is the restriction of $\stp_{k,q}$ to $\operatorname{domain}(\stp_{k+1,q-1})$.
\end{lemma}
\begin{proof}
Level $i$ of $\operatorname{domain}(\stp_{k+1,q-1})$ equals level $i+1$ of $\operatorname{domain}(\stp_{k,q})$: the ranks $\bigl((k+1)+i,(q-1)-i\bigr)=\bigl(k+(i+1),q-(i+1)\bigr)$ and the radius bounds coincide, the value caps $(k+1)+(q-1)=k+q$ coincide, and both functions use the same canonical set $S_r([\nu]^G)$.
\end{proof}

In the model checking application, the type trees of all vertices are maintained by a dynamic-programming algorithm, and the scatter type is computed from them.
Recall that $|G|=|V(G)|$ and $\size G=|V(G)|+|R(G)|$.

\begin{lemma}\label{lem:compute-stp}
There is an algorithm which, given a nonempty partial graph $G$, numbers
$k\ge1$ and $q\in\N$, its distance matrix $D_G$, and the family
$(\tree_{k,q}(a):a\in V(G))$, computes $\stp^G_{k,q}$ in time
\[
|G|^2\cdot
\poly_{k,q}\left(\max_{a\in V(G)}\size{\tree_{k,q}(a)}\right).
\]
\end{lemma}
\begin{proof}
Put $M\coloneqq\max_{a\in V(G)}\size{\tree_{k,q}(a)}$.
For every $a\in V(G)$ and $i\in[q]$, applying \Cref{lem:shift} $i$ times to $\tree_{k,q}(a)$ yields $\tree_{k+i,q-i}(a)$, of size $\poly_{k,q}(M)$.
This takes $|G|\cdot\poly_{k,q}(M)$ time and gives $\operatorname{domain}(\stp_{k,q})$, as each radius is $O_{k,q}(1)$; in particular, $|\operatorname{domain}(\stp_{k,q})|=O_{k,q}(|G|)$.

For each $(r,\nu)\in \operatorname{domain}(\stp_{k,q})$, run the greedy process defining $S_r([\nu]^G)$, stopping after $k+q$ vertices were discovered.
It scans $V(G)$ in order, testing membership in $[\nu]^G$ by comparing the trees computed above with $\nu$. After adding a vertex $v$, scan the row $D_G(v,\cdot)$ and mark all vertices $u$ with $D_G(v,u)\le r$ as unavailable.
At most $k+q$ rows are scanned, so one run takes $|G|\cdot\poly_{k,q}(M)$ time, and the $|\operatorname{domain}(\stp_{k,q})|=O_{k,q}(|G|)$ runs together take the stated time.
\end{proof}

\subsection{The far tree}\label{sec:ftree}

\begin{definition}\label{def:ftree}
For a partial graph $G$, $k\in\N$, $q\ge1$, and a nonempty tuple $\tup a\in V(G)^{\le k}$, define
\[
\ftree_{k,q}(\tup a)\coloneqq
\setof{\tree_{k+1,q-1}(b)}{b\in V(G),\ \dist(\tup a,b)>\rhom(k+1,q-1)}.
\]
\end{definition}

The goal of this subsection is to compute $\ftree_{k,q}(\tup a)$ from $\tree_{k,q}(\tup a)$ and $\stp_{k,q}$ (\Cref{lem:to-ftree}).
Our main tool is the following clustering lemma~\cite[Corollary~3.4 and Lemma~3.5]{locality-note}.

\begin{lemma}[{\cite{locality-note}}]\label{lem:clustering}
Let $r\in\N$, let $\ell\ge1$, let $G$ be a graph, and let $\tup a=(a_1,\ldots,a_\ell)\in V(G)^\ell$.
There exist a radius $\crad$ with $r\le \crad\le9^{\ell-1}r$ and a partition $\mathcal B$ of $\set{a_1,\ldots,a_\ell}$ with the following property.

For every $X\subseteq V(G)$, inclusion-wise maximal $4\crad$-scattered subset $S \subseteq X$, and $H\ge4\crad+r$,
\[
\exists x\in X\colon \dist^G(x,\tup a)>r
\quad\Longleftrightarrow\quad
\begin{aligned}[t]
&\exists x\in X\colon r<\dist^G(x,\tup a)\le H,\\[-2pt]
&\text{or}\quad \bigl|\setoff{B\in\mathcal B}{\exists x\in X\colon \dist^G(x,B)\le r}\bigr|< |S|.
\end{aligned}
\]
Moreover, the radius $\crad$ and the partition $\mathcal B$ can be determined by tests of the form $\dist^G(a_i,a_j)\le s$, where $i,j\in[\ell]$ and $s\le8\cdot9^{\ell-1}r$.
\end{lemma}

\begin{lemma}\label{lem:to-ftree}
Fix $k\in\N$ and $q\ge1$, let $G$ be a partial graph, and let $\tup a\in V(G)^{\le k}$ be nonempty.
Then $\ftree_{k,q}(\tup a)$ is computable from $\tree_{k,q}(\tup a)$ and $\stp_{k,q}$ in time
\[
\size{\stp_{k,q}}\cdot
\poly_{k,q}\bigl(\size{\tree_{k,q}(\tup a)}\bigr).
\]
\end{lemma}
\begin{proof}
Let
$\tau\coloneqq\tree_{k,q}(\tup a)$,
$\sigma\coloneqq\stp_{k,q}$,
$r\coloneqq\rhom(k+1,q-1)$,
$H\coloneqq\rhop(k+1,q-1)$, and
$\ell\coloneqq|\tup a|$.
The entries with $i=1$ in the domain of $\sigma$ let us compute the set
\[
\mathcal T
\coloneqq
\setoff{\tree_{k+1,q-1}(b)}{b\in V(G)}.
\]
Then
\[
\ftree_{k,q}(\tup a) =\setoff{\nu\in\mathcal T}{\exists b\in[\nu]^G\colon\dist(\tup a,b)>r}.
\]

\paragraph{Rewriting the target.}
The tree $\tau$ only records vertices within distance $H$ of $\tup a$, while a vertex $b\in[\nu]^G$ with $\dist(\tup a,b)>r$ may be farther away than $H$.
We will now use \Cref{lem:clustering} to replace the existence of such a vertex by two conditions that we can check.
Apply the lemma with radius $r$ to the Gaifman graph of $G$ (with vertex set $V(G)$ and edge set $R(G)$) and to $\tup a$, and obtain a radius $\crad$ and a partition $\mathcal B$ of $\set{a_1,\ldots,a_\ell}$.
Since $\ell\le k$,
\[
4\crad\le4\cdot9^{\ell-1}r\le9^{k+1}r,
\qquad
4\crad+r\le(4\cdot9^{\ell-1}+1)r\le9^{k}r\le H.
\]
The first chain shows that $(4\crad,\nu)$ lies in the domain of $\sigma$ for every $\nu\in\mathcal T$. The second is the hypothesis $H\ge4\crad+r$ of \Cref{lem:clustering}, so the lemma applies to every $X\subseteq V(G)$ together with $S_{4\crad}(X)$.
Define
\begin{align*}
\mathcal N&\coloneqq\setof{\tree_{k+1,q-1}(b)}{b\in V(G),\ r<\dist(\tup a,b)\le H},\\
\mathcal N_B&\coloneqq\setof{\tree_{k+1,q-1}(b)}{b\in V(G),\ \dist(b,B)\le r}
\qquad\text{for $B\in\mathcal B$.}
\end{align*}
For $\nu\in\mathcal T$, \Cref{lem:clustering} applied to $X\coloneqq[\nu]^G$ and $S\coloneqq S_{4\crad}([\nu]^G)$ states that
\[
\exists b\in[\nu]^G\colon\dist(\tup a,b)>r
\quad\Longleftrightarrow\quad
\nu\in\mathcal N\ \text{ or }\ \bigl|\setoff{B\in\mathcal B}{\nu\in\mathcal N_B}\bigr|<|S_{4\crad}([\nu]^G)|.
\]
As $\sigma(4\crad,\nu)=\min(|S_{4\crad}([\nu]^G)|,k+q)$ by \Cref{def:stp}, and $\bigl|\setoff{B\in\mathcal B}{\nu\in\mathcal N_B}\bigr|\le|\mathcal B|\le\ell\le k<k+q$, we can equivalently replace the bound $|S_{4\crad}([\nu]^G)|$ above by $\sigma(4\crad,\nu)$.
Therefore,
\[
\ftree_{k,q}(\tup a)
=\setoff{\nu\in\mathcal T}{\nu\in\mathcal N\ \text{ or }\ \bigl|\setoff{B\in\mathcal B}{\nu\in\mathcal N_B}\bigr|<\sigma(4\crad,\nu)}.
\]
In this expression, $\mathcal T$ and the values $\sigma(4\crad,\nu)$ can be read from $\sigma$.
It remains to compute $\crad$, $\mathcal B$, $\mathcal N$, and the sets $\mathcal N_B$ from $\tau$.

\paragraph{Reading $\crad$ and $\mathcal B$ off $\tau$.}
By \Cref{lem:clustering}, $\crad$ and $\mathcal B$ are determined by tests $\dist(a_i,a_j)\le s$ with $i,j\in[\ell]$ and
\[
s\le8\cdot9^{\ell-1}r\le9^kr\le H\le\rhom(k,q).
\]
These are binary distance atoms of radius at most $\rhom(k,q)$ and hence determined by the label $\atp_{k,q}(\tup a)$ of the root of $\tau$.
We represent every set $B\in\mathcal B$ by its set of positions $I_B\coloneqq\setof{i\in[\ell]}{a_i\in B}$, which is determined by the equality atoms in the root atomic type of $\tau$.

\paragraph{Reading $\mathcal N$ and $\mathcal N_B$ off $\tau$.}
The trees attached to the root of $\tau$ are exactly 
the trees $\tree_{k+1,q-1}(\tup ab)$ with $b\in V(G)$ and $\dist(\tup a,b)\le H$.
Such a tree $\mu$ determines two things about $b$: by \Cref{lem:restrict}, $\mu\restriction{(\ell+1)}=\tree_{k+1,q-1}(b)$, and since the root atomic type $\atp_{k+1,q-1}(\tup ab)$ of $\mu$ contains the binary distance atoms of radius at most $\rhom(k+1,q-1)=r$, we have $\mu\models\dist(x_{j},x_{\ell+1})\le r$ if and only if $\dist(a_{j},b)\le r$, for every $j\in[\ell]$.
Therefore,
\[
\mathcal N
=\setof{\mu\restriction{(\ell+1)}}{(i,\mu)\in\mathcal E_{k,q}(\tup a),\ \mu\models\textstyle\bigwedge_{j\in[\ell]}\dist(x_{j},x_{\ell+1})>r}.
\]
Likewise, since $r\le H$, the condition $\dist(b,B)\le r$ implies $\dist(\tup a,b)\le H$, and
\[
\mathcal N_B
=\setof{\mu\restriction{(\ell+1)}}{(i,\mu)\in\mathcal E_{k,q}(\tup a),\ \mu\models\textstyle\bigvee_{j\in I_B}\dist(x_{j},x_{\ell+1})\le r}.
\]
The right-hand sides depend only on $\tau$.
Substituting them into the expression for $\ftree_{k,q}(\tup a)$ above gives a description in terms of $\tau$ and $\sigma$ only.

\paragraph{Running time.}
The radius $\crad$ and the sets $I_B$ are read off the root label of $\tau$.
For all extensions $(i,\mu)$ of $\tau$, the trees $\mu\restriction({\ell+1})$ are obtained by \Cref{lem:restrict} and the atoms $\dist(x_{j},x_{\ell+1})\le r$ are read off the root of $\mu$, in time $\poly_{k,q}(\size\tau)$ in total.
Collecting them yields $\mathcal N$ and the at most $k$ sets $\mathcal N_B$ within the same bound.
The expression for $\ftree_{k,q}(\tup a)$ is then evaluated by one scan of $\sigma$: for every entry $(4\crad,\nu,\sigma(4\crad,\nu))$ with $\nu$ of rank $(k+1,q-1)$, membership of $\nu$ in $\mathcal N$ and in each $\mathcal N_B$ is tested by comparing $\nu$ with every element of these at most $k+1$ sets, in time $O(k\cdot\size\tau\cdot(\size\nu+\size\tau))$.
Summed over all entries of $\sigma$ this is $O(\size\sigma\cdot\poly_{k,q}(\size\tau))$, which gives the stated running time.
\end{proof}

\section{Lifting type trees through merges and resolves}
\label{sec:merge-resolve}
The model checking algorithm will process 
the construction sequence step by step and maintain the type trees of all vertices.
We first show how to lift type trees through a merge, and then through a resolve.

\begin{lemma}[merge lift]\label{lem:merge-lift}
Fix $\ell\le k$ and $q\in\N$.
Let $G$ be a partial graph and $\tup a\in V(G)^\ell$.
Given
$\tree^G_{k,q}(\tup a)$ and two distinct colors $P,Q$, one can compute
 $\tree^{H}_{k,q}(\tup a)$, where $H=\merge PQ(G)$,
in time
\[
\poly_{k,q}(\size{\tree^G_{k,q}(\tup a)}).
\]
\end{lemma}
\begin{proof}
We proceed by induction on $q$.

\paragraph{Root labels.}
The partial graphs $G$ and $H$ have the same resolved pairs, so all distances agree, and only the colors $P$ and $Q$ change.
Hence, for every vertex $u$ and $s\in\N$,
\begin{align*}
H\models P(u)&\quad\Longleftrightarrow\quad G\models P(u)\lor Q(u),\\
H\models Q(u)&\quad\Longleftrightarrow\quad\textit{false},\\
H\models\dist(u,P)<s&\quad\Longleftrightarrow\quad G\models(\dist(u,P)<s)\lor(\dist(u,Q)<s),\\
H\models\dist(u,Q)<s&\quad\Longleftrightarrow\quad\textit{false},
\end{align*}
and every other atom has the same interpretation in $G$ and $H$.
This allows us to rewrite every atom \(\phi\) 
into a quantifier-free formula \(\phi'\)
of the same distance rank, so that 
$H\models\phi(\tup a)$ if and only if $G\models\phi'(\tup a)$.
Since whether or not $G\models\phi'(\tup a)$ is determined by $\atp^G_{k,q}(\tup a)$, it follows that the new root label $\atp^{H}_{k,q}(\tup a)$ can be computed  from $\atp^G_{k,q}(\tup a)$.
Only atoms mentioning $P$ or $Q$ change, so the new root label is obtained by one pass over the old root label and is no larger.
This settles the root label.

\paragraph{Extensions.}
Let $q\ge1$.
Since $\dist^{H}=\dist^G$, every guard admits the same vertices in $G$ and $H$, so the inductive assumption gives $\mathcal E^{H}_{k,q}(\tup a)=\setof{(i,F(\mu))}{(i,\mu)\in\mathcal E^G_{k,q}(\tup a)}$,
where $F(\mu)$ is obtained inductively.
Together with the root label, this gives $\tree^{H}_{k,q}(\tup a)$.

\paragraph{Running time.}
The transformation replaces atoms in the label of every node and keeps the guards, so it visits each node of the input tree once, and no tree grows.
By \Cref{obs:canonical}, restoring canonical form takes time polynomial  in the size of the tree.
\end{proof}


A resolve can bring vertices far from a tuple within guard range.
Their type trees are not recorded in the old type tree of the tuple, and we obtain them from the far tree instead.

\begin{lemma}[resolve lift]\label{lem:resolve-lift}
Let $k\ge1$ and $q\in\N$.
Given a partial graph $G$, its distance matrix $D_G$, colors $P,Q$,
a sign $\pm$, and
$(\tree^G_{k,q}(a):a\in V(G))$, one can compute
$H=\resolve\pm PQ(G)$, its distance matrix $D_H$, and
$(\tree^{H}_{k,q}(a):a\in V(G))$ in time
\[
|G|^2\cdot\poly_{k,q}(M+T),
\]
where
\begin{align*}
M&\coloneqq\max_{a\in V(G)}\size{\tree^G_{k,q}(a)},\\
T&\coloneqq\max_{a\in V(G)}\bigl|\bigl\{\tree^G_{k+i,q-i}(\tup v)\colon 0\le i\le q,\ \ell\in[k+i],\ \tup v\in\bigl(\Ball^{H}_{\rhop(k,q)}(a)\bigr)^{\ell}\bigr\}\bigr|.
\end{align*}
Each output tree has size $\poly_{k,q}(T)$.
\end{lemma}
\begin{proof}
Assume the resolve is positive, as the negative case is symmetric.
By induction on $q$, 
we construct an auxiliary map $F$ on trees of arbitrary nonempty tuples, so that $\tree^{H}_{k,q}(\tup a)=F(\tree^G_{k,q}(\tup a))$ for all $\ell\le k$ and $\tup a\in V(G)^\ell$.
Fix $\tup a \in V(G)^\ell$ and 
put $\tau\coloneqq\tree^G_{k,q}(\tup a)$ and $\sigma\coloneqq\stp^G_{k,q}$.

\paragraph{Root labels.}
We may assume that $P^G$ and $Q^G$ are nonempty, as otherwise $H=G$.
We rewrite every atom $\phi$ in $H$ to an equivalent formula $\phi'$ in $G$,
so that $\phi'$ has the same distance rank as $\phi$,
and $H\models\phi(\tup a)\iff G\models\phi'(\tup a)$.
Below, $u,v$ are vertices, $Y$ is a color, and $s\in\N$.

\emph{First-order atoms.}
Equality, color, and non-edge atoms $N(x,y)$ have the same interpretation in $G$ and~$H$.
By definition of the positive resolve,
\[
H\models E(u,v)
\quad\Longleftrightarrow\quad
G\models E(u,v)\lor\Bigl(\bigl((P(u)\land Q(v))\lor(Q(u)\land P(v))\bigr)\land u\ne v\land\lnot N(u,v)\Bigr).
\]

\emph{Distances.}
We have $R(H)=R(G)\cup P^GQ^G$.
A shortest path from $u$ to $v$ in $H$ is a path in $G$, or runs in $G$ from $u$ to $P^G$ or $Q^G$, then crosses to the other part or returns to the same part through the other one, and then runs in $G$ to $v$:
otherwise, the segment between its first and last vertex in $P^G\cup Q^G$ could be replaced by one pair from $P^GQ^G$, or by two through the other part if both ends lie in the same part.
Hence
\begin{equation}\label{eq:resolve-dist}
\dist^H(u,v)=\min\left\{
\begin{aligned}
&\dist^G(u,v),\\
&\dist^G(u,P)+1+\dist^G(Q,v),\quad \dist^G(u,Q)+1+\dist^G(P,v),\\
&\dist^G(u,P)+2+\dist^G(P,v),\quad \dist^G(u,Q)+2+\dist^G(Q,v)
\end{aligned}\right\}.
\end{equation}

\emph{Unary distance atoms.}
Taking the minimum of \eqref{eq:resolve-dist} over $v\in Y^G$ gives
\begin{align*}
&H\models\dist(u,Y)<s
\quad\Longleftrightarrow\quad
G\models\dist(u,Y)<s\\
&\qquad\lor\dist(u,P)<(s-1-\dist^G(Q,Y))\lor\dist(u,Q)<(s-1-\dist^G(P,Y))\\
&\qquad\lor\dist(u,P)<(s-2-\dist^G(P,Y))\lor\dist(u,Q)<(s-2-\dist^G(Q,Y)).
\end{align*}
The values $\dist^G(P,Y)$ and $\dist^G(Q,Y)$ do not depend on $u$, and are computed once. A unary-distance comparison whose upper bound is nonpositive, or from which an infinite distance is subtracted, is interpreted as \emph{false}.

\emph{Binary distance atoms.}
Define
\begin{align*}
&\delta^s_{P,Q}(x,y)\coloneqq\\
&\bigvee_{s_1+1+s_2=s}\bigl(\dist(x,P)<s_1{+}1\land\dist(y,Q)<s_2{+}1\bigr)\lor\bigl(\dist(x,Q)<s_1{+}1\land\dist(y,P)<s_2{+}1\bigr)\\
\lor{}&\bigvee_{s_1+2+s_2=s}\bigl(\dist(x,P)<s_1{+}1\land\dist(y,P)<s_2{+}1\bigr)\lor\bigl(\dist(x,Q)<s_1{+}1\land\dist(y,Q)<s_2{+}1\bigr).
\end{align*}
By \eqref{eq:resolve-dist},
\begin{align}\label{eq:bin-dist}
H\models\dist(u,v)\le s
\quad\Longleftrightarrow\quad
G\models(\dist(u,v)\le s)\lor\delta^s_{P,Q}(u,v).
\end{align}

\emph{Root label.}
For an atom \(\phi\), let $\phi'$ be the corresponding formula as obtained above.
Then $H\models\phi(\tup a)$ if and only if $G\models\phi'(\tup a)$ and $\phi'$ has the same distance rank as $\phi$.
Since whether or not $G\models\phi'(\tup a)$ is determined by $\atp^G_{k,q}(\tup a)$, it follows that the new root label $\atp^{H}_{k,q}(\tup a)$ can be computed  from $\atp^G_{k,q}(\tup a)$.
Algorithmically, there are $O_{k,q}(1)$ atoms without colors, and each nonempty color contributes $O_{k,q}(1)$ further atoms.

\paragraph{Extensions.}
Let $q\ge1$, and put $r\coloneqq\rhom(k+1,q-1)$ and $R\coloneqq\rhop(k+1,q-1)$.
On trees of rank $(k+1,q-1)$, $F$ denotes the map given by the inductive assumption, using the restriction of $\sigma$ to $\operatorname{domain}(\stp_{k+1,q-1})$ from \Cref{obs:stp-restrict} and the same precomputed distances.
By the inductive assumption,
\begin{align}\label{neqE}
\mathcal E^{H}_{k,q}(\tup a)=\setof{\bigl(i,F(\tree^G_{k+1,q-1}(\tup a b))\bigr)}{i\in[\ell], \dist^H(a_i,b)\le R}.
\end{align}

For $i\in[\ell]$, put
\[
\gamma_i(x_1,\ldots,x_{\ell+1})\coloneqq\delta^R_{P,Q}(x_i,x_{\ell+1}).
\]
By \eqref{eq:bin-dist},
\begin{align}\label{eq:Disteq}
H\models\dist(a_i,b)\le R
\quad\Longleftrightarrow\quad
G\models\dist(a_i,b)\le R\ \lor\ \gamma_i(\tup a b).
   \end{align}
The right-hand side of \eqref{eq:Disteq}
is equivalent to the disjunction of three conditions:
\begin{enumerate}
   \item  $\dist^G(a_i,b)\le R$,
   \item  $\dist^G(\tup a,b)\le R$ and $G\models\gamma_i(\tup a b)$,
    \item  $\dist^G(\tup a,b)>r $ and $G\models\gamma_i(\tup a b)$.
\end{enumerate}
This holds since $r\le R$ by \eqref{eq:horizon}.
Consequently,
\[
\mathcal E^{H}_{k,q}(\tup a)=\mathcal E_{\mathrm{old}}\cup\mathcal E_{\mathrm{near}}\cup\mathcal E_{\mathrm{far}},
\]
where
\begin{align*}
\mathcal E_{\mathrm{old}}&\coloneqq\setof{\bigl(i,\ F(\tree^{G}_{k+1,q-1}(\tup a b))\bigr)}{i\in[\ell],\ b\in V(G),\ \dist^G(a_i,b)\le R},\\
\mathcal E_{\mathrm{near}}&\coloneqq\setof{\bigl(i,\ F(\tree^{G}_{k+1,q-1}(\tup a b))\bigr)}{i\in[\ell],\ b\in V(G),\ \dist^G(\tup a,b)\le R,\ G\models\gamma_i(\tup a b)},\\
\mathcal E_{\mathrm{far}}&\coloneqq\setof{\bigl(i,\ F(\tree^{G}_{k+1,q-1}(\tup a b))\bigr)}{i\in[\ell],\ b\in V(G),\ \dist^G(\tup a,b)>r,\ G\models\gamma_i(\tup a b)}.
\end{align*}

The first set is obtained via
\[
\mathcal E_{\mathrm{old}}
=
\setof{(i,F(\mu))}{(i,\mu)\in\mathcal E^G_{k,q}(\tup a)}.
\]
Every witness $b$ for $\mathcal E_{\mathrm{near}}$ is admitted by some guard $j$ in $G$, so $\tree^G_{k+1,q-1}(\tup a b)$ occurs as an extension $(j,\mu)$ of $\tau$ with $\mu\models\gamma_i$.
Conversely, every such $\mu$ equals $\tree^G_{k+1,q-1}(\tup a b)$ for some $b$ with $\dist^G(a_j,b)\le R$ and $G\models\gamma_i(\tup a b)$.
Hence, using the notation $\mu\models\gamma$ introduced after \Cref{lem:eval},
\[
\mathcal E_{\mathrm{near}}
=
\setof{(i,F(\mu))}{i\in[\ell],\ (j,\mu)\in\mathcal E^G_{k,q}(\tup a),\ \mu\models\gamma_i}.
\]
For $\mathcal E_{\mathrm{far}}$, consider $b$ with $\dist^G(\tup a,b)>r$.
Since $r=\rhom(k+1,q-1)$ is exactly the separation required by \Cref{lem:compose} at rank $(k+1,q-1)$, that lemma 
gives
\[
\tree^G_{k+1,q-1}(\tup a b)
=
\tree^G_{k+1,q-1}(\tup a)\ast\tree^G_{k+1,q-1}(b),
\]
where $\tree^G_{k+1,q-1}(\tup a)$ is obtained from $\tau=\tree^G_{k,q}(\tup a)$ by \Cref{lem:shift}.
As $b$ ranges over all such vertices, $\tree^G_{k+1,q-1}(b)$ ranges exactly over $\ftree^G_{k,q}(\tup a)$, computed from $\tau$ and $\sigma$ by \Cref{lem:to-ftree}.
Therefore
\[
\mathcal E_{\mathrm{far}}
=
\setof{\bigl(i,\ F(\tree^G_{k+1,q-1}(\tup a)\ast\nu)\bigr)}{i\in[\ell],\ \nu\in\ftree^G_{k,q}(\tup a),\ \tree^G_{k+1,q-1}(\tup a)\ast\nu\models\gamma_i}.
\]
So the new root label, $\mathcal E_{\mathrm{old}}$, $\mathcal E_{\mathrm{near}}$, and $\mathcal E_{\mathrm{far}}$ are computed from $\tau$ together with $\sigma$ and the values $\dist^G(P,Y)$ and $\dist^G(Q,Y)$, which do not depend on $\tup a$. This defines $F(\tau)=\tree^{H}_{k,q}(\tup a)$.

\paragraph{Running time.}
Computing $H$ takes time $O(\size{H})=O(|G|^2)$.
Scanning $D_G$ gives $\dist^G(u,P)$ and $\dist^G(u,Q)$ for all vertices $u$, and thus $\dist^G(P,Y)$ and $\dist^G(Q,Y)$ for all nonempty colors $Y$, in time $O(|G|^2)$.
Then \eqref{eq:resolve-dist} gives $D_H$ in time $O(|G|^2)$, and \Cref{lem:compute-stp} computes $\sigma\coloneqq\stp^G_{k,q}$ from $D_G$ in time $|G|^2\cdot\poly_{k,q}(M)$.
Every tree $\tree^G_{k+i,q-i}(b)$ is obtained from $\tree^G_{k,q}(b)$ by at most $q$ applications of \Cref{lem:shift}, so it has size $\poly_{k,q}(M)$, and $\size\sigma\le |G|\cdot\poly_{k,q}(M)$, since the domain of $\sigma$ has $O_{k,q}(1)$ entries per vertex.

\emph{Number of calls.}
Fix $a\in V(G)$ and consider the recursion computing $F(\tree^G_{k,q}(a))$.
We collect the distinct old extension trees admitted by at least one guard in $H$ before recursing, retaining the list of guards for each tree.
The calls at depth $j$ are on trees $\tree^G_{k+j,q-j}(\tup v)$ for tuples $\tup v$ reached by the guards in $H$.
By \Cref{lem:locality} applied to $H$, every such $\tup v$ is a tuple from $\Ball^{H}_{\rhop(k,q)}(a)$.
So by definition of $T$, a call makes at most $T$ recursive calls, and as the depth is $q$, there are $\poly_{k,q}(T)$ calls for $a$.

\emph{Cost per call and output size.}
A recursive input is a child of the previous input, or is obtained by rank shift and far composition with a  tree of arity one recorded in $\sigma$.
These trees have size $\poly_{k,q}(M)$, so by \Cref{lem:shift,lem:compose} and the depth bound $q$, every recursive input and every candidate extension tree has size $\poly_{k,q}(M)$.
For a call with positive remaining depth, restricting $\sigma$ and applying \Cref{lem:to-ftree} takes time $O(\size\sigma\cdot\poly_{k,q}(M))=O(|G|\cdot\poly_{k,q}(M))$.
The far tree has at most $|G|$ members, so composing them and testing the formulas $\gamma_i$ 
(which are quantifier-free formulas of distance rank $(k+1,q-1)$, so are determined by the root label of a tree of rank $(k+1,q-1)$) takes the same time.
We scan the candidates, test their guards, and compare each admitted candidate with the list of at most $T$ distinct admitted trees collected so far.
Including the candidates from the old extensions, this takes $O(|G|\cdot\poly_{k,q}(M+T))$ time.

The colors recorded in each new root label meet $\Ball^{H}_{\rhop(k,q)}(a)$ by \Cref{lem:locality} applied to $H$, and as vertices of distinct colors have distinct old trees, there are at most $T$ of them.
So each new root label has size $O_{k,q}(T)$ and is computed in time $O(|G|\cdot\size\tau)$.
Together with the branching and depth bounds, this gives output size $\poly_{k,q}(T)$; restoring canonical form has the same polynomial bound by \Cref{obs:canonical}.

Altogether the recursion for $a$ takes time $O(|G|\cdot\poly_{k,q}(M+T))$, and for all vertices $O(|G|^2\cdot\poly_{k,q}(M+T))$.
Adding the construction of $H$ and $D_H$, the distance preprocessing, and the scatter type gives the claimed bound.
\end{proof}

\section{The model checking algorithm}\label{sec:model-checking}
We now give the model checking algorithm,
proving \Cref{thm:mc}.
Recall the width notion used in the statement.
\defrqwidth*

The width bounds numbers of ordinary types, and the following lemma converts such bounds into bounds on type trees, as required by \Cref{lem:resolve-lift}.

\begin{lemma}[type tree sizes]\label{lem:ball-tree-bound}
Let $k\ge1$ and $q\in\N$, and put $r\coloneqq\rhop(k,q)$ and $d\coloneqq r+k+q$.
The following holds for every partial graph $G$.
\begin{enumerate}[label=(\roman*)]
\item\label{lem:ball-count}
$\bigl|\setoff{\tree^G_{k+i,q-i}(\tup v)}{\tup v\in X^{\ell}}\bigr|
\le|\Types^G_{d,d}(X)|$ for all $i\in\set{0,\ldots,q}$, $\ell\in [k+i]$, and $X\subseteq V(G)$.
\item\label{lem:ball-size}
$\size{\tree^G_{k,q}(a)}
\le
\poly_{k,q}\bigl(|\Types^G_{d,d}(\Ball^G_r(a))|\bigr)$ for every $a\in V(G)$.
\end{enumerate}
\end{lemma}
\begin{proof}
We first show that ordinary types determine type trees: for all $k',q'\in\N$, every partial graph $G$, and all nonempty tuples $\tup a,\tup b\in V(G)^{\le k'}$ of the same length,
\begin{equation}\label{eq:tp-to-tree}
\tp^G_{\rhop(k',q')+q'}(\tup a)=\tp^G_{\rhop(k',q')+q'}(\tup b)
\quad\Longrightarrow\quad
\tree^G_{k',q'}(\tup a)=\tree^G_{k',q'}(\tup b).
\end{equation}
By expressing distance atoms by paths along $E\lor N$, every \distFO formula of distance rank $(k',q')$ is equivalent to a first-order formula of quantifier rank at most $\rhop(k',q')+q'$. Thus, the left-hand side of \eqref{eq:tp-to-tree} implies that $\tup a$ and $\tup b$ satisfy 
the same \distFO formulas of distance rank $(k',q')$.
By \Cref{cor:tp-tree}, this yields the right-hand side.

For \ref{lem:ball-count}, extend every tuple $\tup v\in X^\ell$ to a tuple $\tup v'\in X^d$ by repeating its last coordinate.
Then $\tp_d(\tup v')$ determines $\tp_d(\tup v)$, which determines $\tree_{k+i,q-i}(\tup v)$ by \eqref{eq:tp-to-tree}, as $\rhop(k+i,q-i)+q-i\le \rhop(k,q)+q\le d$ by \eqref{eq:horizon-cons}.
So distinct type trees yield distinct elements of $\Types^G_{d,d}(X)$.

For \ref{lem:ball-size}, put $X\coloneqq\Ball_r(a)$ and $K\coloneqq|\Types^G_{d,d}(X)|$.
As every vertex has exactly one color, vertices of distinct colors have distinct type trees, so at most $K$ colors meet $X$ by \ref{lem:ball-count}.
By \Cref{lem:locality}, every node of $\tree_{k,q}(a)$ is $\tree_{k+j,q-j}(\tup v)$ for a tuple $\tup v$ from $X$, and every color in a true atom of its root meets $X$.
So every root has size $O_{k,q}(K)$, and by \ref{lem:ball-count}, every node has at most $K$ children per guard, and thus at most $(k+q)K$ children.
As the depth is $q$, this gives $\size{\tree_{k,q}(a)}\le\poly_{k,q}(K)$.
\end{proof}

{\renewcommand{\mcdepth}{d}\mcthm*}
\begin{proof}
Let $q$ be the quantifier rank of $\phi$, and let $r,d$ be as in
\Cref{lem:ball-tree-bound} for $k=1$ and this $q$.  We prove the statement
with radius $r$ and depth $d$.
If $n\le1$, evaluate $\phi$ directly, so assume $n\ge2$.

Let $G_1,\ldots,G_\lambda$ be the given
construction sequence of $G$, and let $K$ be its radius-$r$ depth-$d$
width.  We iteratively compute $(\tree^{G_t}_{1,q}(a):a\in V(G))$ for $t=1,\ldots,\lambda$. To save time we also maintain the distance matrix $D_{G_t}$.
Initially, for \(t=1\), there are no resolved pairs and each nonempty part is a singleton.
Hence, every guard admits only vertices of the guarded tuple itself, so
each \(\tree^{G_1}_{1,q}(a)\) is determined by the part containing \(a\), and can therefore be computed in time \(O_q(1)\).
Initialize also $D_{G_1}$ in $O(n^2)$ time by putting $0$ on its diagonal and $\infty$ elsewhere.

Now let \(t<\lambda\), and assume the trees in $G_t$ have been computed.
All these trees have size $\poly_{q}(K)$: for every $t'\in[\lambda-1]$
and $a\in V(G)$, \Cref{lem:ball-tree-bound}\ref{lem:ball-size} bounds
$\size{\tree^{G_{t'}}_{1,q}(a)}$ by a polynomial in
$|\Types^{G_{t'}}_{d,d}(\Ball^{G_{t'}}_{r}(a))|$.  Using $R(G_{t'})\subseteq R(G_{t'+1})$
and therefore $\Ball^{G_{t'}}_r(a)\subseteq \Ball^{G_{t'+1}}_r(a)$, this is at most
$|\Types^{G_{t'}}_{d,d}(\Ball^{G_{t'+1}}_{r}(a))|\le K$ by the definition of
the width.

For a merge, construct $G_{t+1}$ in time $O(n)$, keep $D_{G_{t+1}}=D_{G_t}$ unchanged, and apply \Cref{lem:merge-lift} to each tree of arity one, taking $O(n\cdot\poly_q(K))$ time in total.
For a resolve, apply \Cref{lem:resolve-lift} to $G_t$, the supplied matrix $D_{G_t}$, and all its trees of arity one. The lemma returns both $D_{G_{t+1}}$ and the new trees.
Its parameter $M$ is $\poly_q(K)$ by the bound above.
Its parameter $T$ is
$$T\coloneqq \max_{a\in V(G)}\left|\setof{\tree_{1+i,q-i}(\tup v)}{\tup v\in \left(\Ball^{G_{t+1}}_{\rho^+(1,q)}(a)\right)^\ell,\,  0\le i\le q,\, \ell\in [i+1]}\right|.$$
The tuples $\tup v$ above lie in $X\coloneqq\Ball^{G_{t+1}}_r(a)$ since $\rhop(1,q)\le r$.
By \Cref{lem:ball-tree-bound}\ref{lem:ball-count}, there are at most $|\Types^{G_t}_{d,d}(X)|\le K$ trees for each $i$ and arity $\ell$.
There are at most $(q+1)^2$ choices of $i$ and $\ell$, so $T\le(q+1)^2K$.

As each step takes $O(n^2\cdot\poly_q(K))$ time, the total time is $O(\lambda n^2\cdot\poly_\phi(K))$.
After $\lambda-1$ steps we have all the type trees in $G_\lambda$.
These final trees also have size $\poly_q(K)$: if the last step is a merge, no tree grows; if it is a resolve, use the output-size bound $\poly_q(T)$ in \Cref{lem:resolve-lift} and $T\le(q+1)^2K$.

As all pairs of $G_\lambda$ are resolved,
we can turn each quantifier of $\phi$ into a local one without changing its truth value.
Construct \(\phi^{\mathrm{loc}}(x)\) from \(\phi\) by
introducing a fresh variable $x$, and replacing every subformula
$\exists y\,\psi$ of $\phi$ that lies in the scope of quantifiers
$\exists z_1,\ldots,\exists z_j$ by
\[
\exists y~\Bigl(
\dist(x,y)\le\rhop(j+2,q-j-1)
\land\psi
\Bigr),
\]
where $\psi$ is transformed in the same way.  This yields a \distFO formula
$\phi^{\mathrm{loc}}(x)$ such that, for
every $a\in V(G)$,
\[
G_\lambda\models\phi
\quad\Longleftrightarrow\quad
G_\lambda\models\phi^{\mathrm{loc}}(a).
\]
Note that $\phi^{\mathrm{loc}}$ has distance rank $(1,q)$.

Now choose any $a\in V(G)$,
and use \Cref{lem:eval} for $k=1$ to compute whether $G_\lambda\models\phi^{\mathrm{loc}}(a)$,
and hence whether $G_\lambda\models\phi$, from
$\tree^{G_\lambda}_{1,q}(a)$ in time
$\poly\bigl(|\phi^{\mathrm{loc}}|+\size{\tree^{G_\lambda}_{1,q}(a)}\bigr)\le\poly_\phi(K)$.
As $E(G_\lambda)=E$ and $\phi$ is a sentence over graphs, $G_\lambda\models\phi$ if and only if $G\models\phi$.
This fits the preceding time bound and completes the proof.
\end{proof}

\addtocontents{toc}{\protect\vspace{1em}}
\section*{AI methodology}
\addcontentsline{toc}{section}{AI methodology}
We developed the approach to the main theorem in summer 2025, without AI assistance.
By summer 2026, we had a draft of the present paper, and the companion paper \cite{md-abmw} was nearly complete.
The draft already had essentially the current proof structure.
It contained a complete proof of the result of \Cref{part:types}, and definitions and lemma statements for the other parts, though some of their proofs were still sketches.
Believing that these sketches could be completed, we used GPT-6 to help check the argument.

In \Cref{part:definable-construction-sequences}, we had proposed a proof of a stronger version of \Cref{cl:from-merge-to-construction}, in which the number of unary predicates is bounded by a constant independent of the length of the input merge sequence.
This would have let us apply the results of \cite{md-abmw} without controlling the sequence length.
The model identified a gap in this proof.
We expected the gap to be easy to repair with the methods already in the paper, but the model did not find a repair.

We therefore asked a narrower question: are there construction sequences of width $n^{o(1)}$ that are locally definable using a constant number of unary predicates?
The model observed that a slightly weaker width bound of $n^\eps$, for any fixed $\eps>0$, can be achieved by tuning the parameter $q$ in Theorem 4.3 in \cite{md-abmw}, so that it produces chains of length $O(r/\eps)$.
Since the number of unary predicates depends only on the length of the input sequence, this gives a bound independent of the size of the graph.
We immediately recognized that this was enough to close the gap within our original proof strategy.

The text of \Cref{part:types,part:definable-construction-sequences} was written by the authors.
We used LLMs to help write some proofs in \Cref{part:mc} that are mechanical adaptations of proofs from \cite{locality-note}.
GPT-6 also proposed a shorter proof of \Cref{lem:ball-types} in \Cref{part:types}, which we verified and adopted.
Finally, we used LLMs to streamline the presentation.
The models used were GPT-6, Claude Opus~5.5 and Claude Fable~5.1.

The authors verified the correctness and originality of all content, including references.

\printbibliography[heading=bibintoc]

\part*{Appendix}
\addcontentsline{toc}{part}{Appendix}
\appendix

\section{Forbidden patterns}\label{sec:forbidden-patterns}
\label{app:patterns}

In this appendix we give a self-contained proof of \Cref{lem:phi-psi-patterns}, which is a corollary of~\cite[Lemma~C.1]{ordered-tww}. The proofs in this appendix apply to arbitrary classes of relational structures, not just classes of finite graphs.
Recall the notion of $(\phi,\psi)$-patterns.
\defphipsi*

\noindent We call a $(\phi,\psi)$-pattern \emph{clean} if the above conditions hold with $\neq$ in place of $>$ in both conditions.

We define dp-minimal classes  \cite{DolichGoodrickLippel2011} as follows.
\begin{definition}
A class $\CC$ of structures is \emph{dp-minimal} 
if for every first-order formulas $\phi(\bar x,z),\psi(\bar y,z)$ in the signature of $\CC$ there is a constant $c\in\N$ such that no structure $G\in\CC$ contains a clean $(\phi,\psi)$-pattern of order $c\times c$.   
\end{definition}

We will prove the following two statements.
\begin{lemma}\label{lem:mondep-dpmin}
    Every monadically dependent class $\CC$ of structures is dp-minimal.
\end{lemma}

\begin{lemma}\label{lem:dpmin-phipsi-patterns}
    Let $\CC$ be a dp-minimal class of structures, and let $\phi(\bar x,z)$ and $\psi(\bar y,z)$ be first-order formulas in the signature of $\CC$. Then there is a constant $t\in\N$ such that no structure $G\in\CC$ contains a $(\phi,\psi)$-pattern of order $t\times t$.
\end{lemma}

Clearly, \Cref{lem:mondep-dpmin} and \Cref{lem:dpmin-phipsi-patterns} imply \Cref{lem:phi-psi-patterns}, repeated below.
\lemphipsi*

We first prove \Cref{lem:mondep-dpmin}. This follows easily 
from the following characterization of monadic dependence in terms of the independence property.
A formula $\phi(\bar x,\bar y)$ has the \emph{independence property} on a class $\CC$ of structures if for every $n\in\N$ there is some $G\in \CC$ and tuples $\bar a_1,\ldots,\bar a_n\in V(G)^{|\bar x|}$ and $(\bar b_J)_{J\subseteq[n]}$ in $V(G)^{|\bar y|}$ such that for all $i\in[n]$ and $J\subseteq[n]$,
\[
    G\models\phi(\bar a_i,\bar b_J)
    \quad\Longleftrightarrow\quad
    i\in J.
\]

We use the following result of Baldwin and Shelah~\cite[Theorem~8.1.8]{BS1985monadic} (see also~\cite[Theorem~2.58 and Proposition~2.65]{wojtekthesis} for a discussion).
\begin{theorem}\label{thm:bs}
A class $\CC$ is monadically dependent if and only if 
for all $k\in\N$, no first-order formula $\phi(\bar x,\bar y)$ has the independence property on the class $\CC^{(k)}$ of all $k$-colorings of graphs in $\CC$.    
\end{theorem}

\begin{proof}[Proof of \Cref{lem:mondep-dpmin}]
    Suppose that $\CC$ is monadically dependent, and let $\phi(\bar x,z)$ and $\psi(\bar y,z)$ be first-order formulas in the signature of $\CC$.
     Let $G\in\CC$ and suppose that $\bar a_1,\ldots,\bar a_m\in V(G)^{|\bar x|}$, $\bar b_1,\ldots,\bar b_m\in V(G)^{|\bar y|}$, and $\bar c_{i,j}\in V(G)^{|z|}$ for $i,j\in[m]$
       form a clean $(\phi,\psi)$-pattern of order $m\times m$ in~$G$.
    
    Fix an arbitrary binary relation $B\subseteq [m]\times [m]$,
and let $G^+$ be the expansion of $G$ with a unary relation $U$ 
interpreted as $\setof{c_{i,j}}{(i,j)\in B}$.
 Consider the following formula:
 $$\gamma(\bar x,\bar y)\coloneqq \exists z\, U(z)\land\phi(\bar x,z)\land \psi(\bar y,z).$$
 By the definition of a clean pattern, the above imply that
 \begin{align}\label{eq:encode-binary}
 G^+\models \gamma(\bar a_i,\bar b_j)\text{ \quad if and only if\quad} (i,j)\in B.   
 \end{align}

 Indeed, if $(i,j)\in B$ then the existential quantifier is witnessed by $c_{i,j}$.
 Conversely, a witness to the existential quantifier must be of the form $c_{i',j'}$ for some $(i',j')\in B$, and then the definition of a clean pattern implies that
 $i=i'$ and $j=j'$, thus $(i,j)\in B$.

Let $k$ be the largest integer such that $2^k\le m$. 
Fix an arbitrary injection $h\from P([k])\to [m]$ from the powerset of $[k]$ into $[m]$.
As $B\subseteq[m]\times[m]$ may be chosen arbitrarily,
we may set $$B\coloneqq \setof{(i,h(J))}{i\in[k],J\subseteq [k],i\in J}.$$
By \eqref{eq:encode-binary},
 $G^+\models\gamma(\bar a_i,\bar b_{h(J)})$ holds if and only if $i\in J$, for all $i\in[k]$ and $J\subseteq [k]$.
 As $\CC$ is monadically dependent, 
 \Cref{thm:bs} implies that $k\le O_{\CC,\gamma}(1)$.
Since the formula $\gamma$ depends only on the original formulas $\phi$ and~$\psi$, this implies $k\le O_{\CC,\phi,\psi}(1)$. As $m< 2^{k+1}$, we conclude that $m\le O_{\CC,\phi,\psi}(1)$, as required.
\end{proof}

To prove \Cref{lem:dpmin-phipsi-patterns},
we use the following version of Ramsey's theorem (see~\cite[full version, Lemma~4.4]{flip-breakability}).
\newcommand{\otp}{\mathrm{otp}}
For a tuple $\bar a=(a_1,\ldots,a_\ell)\in[n]^\ell$,
the \emph{order type} $\otp(\bar a)$ of 
$\bar a$ is the atomic type of $\bar a$ in the ordered set $([n],\le)$.
More explicitly, it stores the information, for each $i,j\in[\ell]$,
whether $a_i<a_j$, $a_i=a_j$, or $a_i>a_j$.

\begin{lemma}\label{lem:bip-ramsey}
    For every $k,\ell\in\N$
    there is an unbounded, non-decreasing function $U_{k,\ell}\from\N\to\N$ with the following property.
For every $t\in\N$ and coloring 
    $$c\from [t]^{\ell}\times [t]^{\ell}\to [k]$$
    there 
    are subsets $I,J\subseteq [t]$ of size $U_{k,\ell}(t)$ such that $c(\bar a,\bar b)$ only depends on $\otp(\bar a)$ and $\otp(\bar b)$, for all $\bar a\in I^\ell$ and $\bar b\in J^{\ell}$.
\end{lemma}

\begin{proof}[Proof of \Cref{lem:dpmin-phipsi-patterns}]
    Let $\CC$ be a dp-minimal class of structures, and let $\phi(\bar x,z)$ and $\psi(\bar y,z)$ be first-order formulas in the signature of $\CC$.
Suppose that
$(\bar a_i:i\in [t])$,
$(\bar b_j:j\in [t])$ and $(c_{i,j}:i,j\in[t])$ form a $(\phi,\psi)$-pattern of order $t\times t$ in some $G\in\CC$. We prove that $t$ is bounded by some constant depending on $\CC,\phi$, and $\psi$.

The proof proceeds in two steps, restricting the considered profiles depicted in the right-hand side part of \Cref{fig:phi-psi-pattern}.
In the first step, we arrive at a $(\phi,\psi)$-pattern where all of the $?$'s in those profiles are equal to some fixed boolean value $0$ or $1$. In the second step, we 
arrive at an $(\alpha,\beta)$-pattern, for some related formulas $\alpha,\beta$,
for which the $?$'s are further restricted to $0$.

\paragraph{Step 1.}
As a first step we show that
we can find sets $I,J\subseteq [t]$ of large size $m$, to ensure the following property:

\begin{quote}
There are $e_1,e_2\in\set{\textit{true},\textit{false}}$
 such that for all $i,i'\in I$ and $j,j'\in J$,

\begin{enumerate}[start=3]
    \item $G\models\phi(\bar a_{i'},c_{i,j})\Leftrightarrow e_1$ holds if $i'<i$,
    \item $G\models\psi(\bar b_{j'},c_{i,j})\Leftrightarrow e_2$ holds if $j'<j$,
\end{enumerate}
\end{quote}

\noindent Indeed, 
define a coloring 
$$c\from [t]^2\times [t]^2\to \set{\textit{true},\textit{false}}^2$$
where $c((i,i'),(j,j'))$ 
stores the truth values of
$G\models\phi(\bar a_{i'},c_{i,j})$
and 
$G\models\psi(\bar b_{j'},c_{i,j})$.
By \Cref{lem:bip-ramsey}, there are subsets $I,J\subseteq [t]$ 
of size $U(t)$, for some unbounded function $U=U_{4,2}$,
such that 
$c((i,i'),(j,j'))$ only depends on $\otp(i,i')$ and $\otp(j,j')$
for all $i,i'\in I$ and $j,j'\in J$.

Let $e_1$ be the first component of $c((i,i'),(j,j'))$ for  $i'<i\in I$ and $j=j'\in J$,
and let $e_2$ be the second component of $c((i,i'),(j,j'))$
for $i=i'\in I$ and $j'<j\in J$. This choice satisfies the conditions 3. and 4. above.

\paragraph{Step 2.}
By reindexing and removing the last element of each of the sequences if necessary, 
assume that $I=J=[2m]$ for some $m\in\N$ with $m\ge U(t)/2-1$ for some unbounded function~$U$, and that all conditions 1.--4. above hold.
As a next step, we show that we can modify the formulas and sequences for which conditions 1.--4. hold with $e_1=e_2=\textit{false}$.

Fix $i,j\in [2m]$
and for $i'\in [2m]$ let $f_{i'}\in\set{\textit{true}, \textit{false}}$ denote the truth value 
of $G\models \phi(\bar a_{i'},c_{i,j})$.
Then $f_1,\ldots,f_{2m}$ 
consists of a sequence of $e_1$'s, followed by a single \emph{true} at position $i$, which is followed by a sequence of \emph{false}'s.
Grouping the sequence into consecutive pairs,
and assuming that $i$ is odd,
we get a sequence of the form:
$$(e_1,e_1),\ldots,(e_1,e_1),(\textit{true},\textit{false}),(\textit{false},\textit{false}),\ldots, (\textit{false},\textit{false}).$$
There is exactly one pair $(p,q)$ in this sequence, namely the $\frac{i+1}2$th one, with distinct truth values,
that is, for which $p\mathbin{\mathrm{XOR}}q$ is \emph{true}. A similar reasoning applies to $\psi$ and $(\bar b_{j'}:j'\in [2m])$.
This motivates us to consider the following formulas and sequences. Define
\begin{align*}
\alpha(\bar x_1\bar x_2,z)&\coloneqq \phi(\bar x_1,z)\mathbin{\mathrm{XOR}} \phi(\bar x_2,z),\\
\beta(\bar y_1\bar y_2,z)&\coloneqq \psi(\bar y_1,z)\mathbin{\mathrm{XOR}} \psi(\bar y_2,z).
\end{align*}
For $i\in[m]$, 
let $\bar u_i$ be the $2|\bar x|$-tuple 
 $\bar a_{2i-1}\bar a_{2i}$,
 and for $j\in[m]$, let $\bar w_j$
 be the $2|\bar y|$-tuple $\bar b_{2j-1}\bar b_{2j}$.
 Finally, let $d_{i,j}=c_{2i-1,2j-1}$.
 By the observation above,
 for all $i,i',j,j'\in[m]$ we have that 
 \begin{enumerate}[label=(\textit{\alph*}),ref=(\textit{\alph*})]
    \item\label{item:pattern-alpha} $G\models \alpha(\bar u_{i'},d_{i,j})$ if and only if $i'=i$,
    \item\label{item:pattern-beta} $G\models \beta(\bar w_{j'},d_{i,j})$ if and only if $j'=j$.
 \end{enumerate}

Thus, $(\bar u_i:i\in[m])$, $(\bar w_j:j\in[m])$, and $(d_{i,j}:i,j\in[m])$ form a clean $(\alpha,\beta)$-pattern of order $m\times m$.
Dp-minimality therefore bounds $m$ by a constant depending only on $\CC,\alpha$, and $\beta$, hence only on $\CC,\phi$, and $\psi$.
Since $m\ge U(t)/2-1$ and $U$ is unbounded and non-decreasing, this also bounds $t$ by a constant depending only on $\CC,\phi$, and $\psi$, as required.
\end{proof}

\section{Corollary of Gaifman's locality theorem}\label{app:gaifman}
In this appendix, we prove \Cref{lem:local-to-global} and \Cref{lem:sep-types-tuple}.

\newcommand{\adj}{\mathrm{adj}}

We will use Gaifman's Locality Theorem~\cite{gaifman}.
We recommend~\cite[Section~4.1]{logic-graphs-algorithms} for an introduction.
First some definitions.

Fix a finite relational signature $\sigma$ of arity $\alpha$.
Two $\sigma$-formulas $\phi(\bar x)$ and $\psi(\bar x)$ are \emph{equivalent} if for every $\sigma$-structure $A$ and every tuple $\bar a \in V(A)^{|\bar x|}$
    \[
    A\models \phi(\bar a) \Leftrightarrow A \models \psi(\bar a).
    \]

Let $\adj_\sigma(x_1,x_2)$ be the formula that tests whether $x_1$ and $x_2$ are adjacent in the Gaifman graph.
More precisely, $\adj_\sigma$ tests for every  relation $R \in \sigma$ of arity $\beta$ with $2\le\beta \leq \alpha$, whether $x_1\neq x_2$ and there exist elements $x_3, \ldots, x_\beta$ such that some permutation of $x_1, \ldots, x_\beta$ satisfies $R$.
Thus, the quantifier rank of $\adj_\sigma$ can be bounded by $\alpha - 2$.
We can naturally derive a formula $\dist(x_1,x_2) \leq r$ checking whether $x_1$ and $x_2$ have distance at most $r$ in the Gaifman graph.
We can (generously) bound the quantifier rank of $(\dist(x_1,x_2) \leq r)$ by $r + \alpha$ (quantifying candidate vertices for the path and then checking the required adjacencies ``in parallel'').

A $\sigma$-formula $\phi(\bar x)$ is \emph{$r$-local}
if every quantifier appearing in $\phi(\bar x)$ is of the form \mbox{$\forall y \in \Ball_s(\bar x)$} or $\exists y \in \Ball_s(\bar x)$ for some $s \leq r$, where the ball checks are implemented using the $\dist$ formulas (which we exempt from this condition).
A sentence $\psi$ is \emph{basic $r$-local} if there is some $r' \leq r$ such that $\psi$ is of the form
\[
  \exists x_1, \ldots, x_m \bigwedge_{1 \leq i < j \leq m} \dist(x_i,x_j) > 2r'
  \wedge
  \bigwedge_{1 \leq i \leq m} \phi(x_i)
\]
for an $r'$-local formula $\phi$.

Intuitively: whether an $r$-local formula holds on a tuple depends only on the $r$-ball of that tuple, as made precise in the following lemma.

\begin{lemma}\label{lem:local-substructure}
Let $\phi(\bar x)$ be an $r$-local $\sigma$-formula, let $A$ be a $\sigma$-structure, let $\bar a \in V(A)^{|\bar x|}$, and let $\Ball_r^A(\bar a)\subseteq B\subseteq V(A)$.
Then $A \models \phi(\bar a) \Leftrightarrow A[B] \models \phi(\bar a)$.
\end{lemma}
\begin{proof}
When $\phi$ is evaluated in $A[B]$, its bounded quantifiers range over balls in the Gaifman graph of $A[B]$, which for arity above two can have fewer edges than the Gaifman graph of $A$ restricted to $B$.
Still, $\Ball_s^{A[B]}(\bar a)=\Ball_s^A(\bar a)$ for all $s\le r$, since a shortest path from $\bar a$ of length at most $r$ survives in $A[B]$:
every tuple witnessing its $j$-th edge has all entries at distance at most $j\le r$ from $\bar a$, hence in $B$.
So the bounded quantifiers of $\phi$ range over the same subsets of $B$ in $A$ and in $A[B]$, and since $A[B]$ is an induced substructure of $A$, the claim follows.
\end{proof}

The following theorem is one of the key tools for limiting the expressiveness of first-order logic.

\newcommand{\loc}{\mathrm{loc}}
\begin{theorem}[Gaifman's Locality Theorem \cite{gaifman}]
    \label{fact:gaifman}
    For all $\alpha,q,k \in \N$ there is a $p = p(\alpha,q,k)$ computable from $\alpha,q,k$  with the following property.
    Let $\sigma$ be a finite relational signature of arity at most $\alpha$.
    Every $\sigma$-formula $\phi(\bar x)$ of quantifier rank at most $q$ and with $|\bar x| \leq k$ free variables is equivalent to a $\sigma$-formula $\phi^\loc(\bar x)$, computable from $\phi$, that is a Boolean combination of basic $7^q$-local sentences and $7^q$-local formulas, \textbf{\textup{each of quantifier rank at most $\textit{p}$}}.
\end{theorem}

The above theorem was proven in \cite{gaifman}, but the paper does not explicitly state the highlighted bound $p$ on the quantifier rank.
Analyzing the proof in \cite{gaifman} shows that, indeed, there exists a bound $p$ that only depends on $\alpha$, $q$, and $k$, but not on the specific choice of the signature $\sigma$.
This observation is made explicit in the recent rank-preserving locality theorems presented in~\cite{locality-note} and~\cite{vanbergerem2026rank}.
In all three papers~\cite{gaifman,locality-note,vanbergerem2026rank}, the dependence of $p$ on the arity $\alpha$ is subtle.
Each of them produces a formula in some extension of first-order logic (FO).
While these extensions of FO are equivalent in expressive power to plain FO, they each contain predicates measuring distances in the Gaifman graph of a structure.
As discussed above, converting such distance formulas to plain FO introduces an additional nesting of quantifiers depending on the arity $\alpha$.

\lemLocalToGlobal*

\begin{proof}
Let $h\coloneqq 7^q$, and let $q'\coloneqq p$ be given by \Cref{fact:gaifman} for $\alpha,q,k$.
Let $\bar a,\bar b\in V(A)^k$, $B_a\supseteq \Ball_h^A(\bar a)$ and $B_b\supseteq \Ball_h^A(\bar b)$ with $\tp_{q'}^{A[B_a]}(\bar a)=\tp_{q'}^{A[B_b]}(\bar b)$.
By \Cref{fact:gaifman}, every formula $\phi(\bar x)$ of quantifier rank at most $q$ is equivalent to a Boolean combination of sentences and $h$-local formulas of quantifier rank at most $q'$.
The sentences have a fixed truth value in $A$ independent of $\bar a$ and $\bar b$.
For an $h$-local formula $\psi(\bar x)$ of quantifier rank at most $q'$,
\[
  A\models\psi(\bar a)
  \iff
  A[B_a]\models\psi(\bar a)
  \iff
  A[B_b]\models\psi(\bar b)
  \iff
  A\models\psi(\bar b),
\]
where the middle equivalence follows from the assumed equality of the $q'$-types.
The first and third equivalences follow from \Cref{lem:local-substructure}.
Hence, $A\models\phi(\bar a)\iff A\models\phi(\bar b)$ for all formulas of quantifier rank at most $q$, so $\tp_q^A(\bar a)=\tp_q^A(\bar b)$.
\end{proof}

\begin{lemma}\label{lem:sep-types}
    For every $q,k,c\in\N$ there are numbers $r=r(q)$ and $T=T(q,k,c)$ with the following property. Let $H$ be a graph expanded by at most $c$ unary predicates, and let $U\subseteq V(H)$ be finite.

    Consider the set $F \coloneqq \{a\in V(H) \,:\, \dist^H(a,U) > r\}$ of vertices that are far from $U$. We can bound
    \[
      |\Types^H_{k,q}(F/U)|\leq T(q,k,c).
    \]
\end{lemma}

\begin{proof}
  Let $h$ and $q'$ be given by \Cref{lem:local-to-global} for $k$, $q$, and $\alpha=2$.
  Since $h=h(q)$, the radius $r\coloneqq h+1$ depends only on $q$.

  Let $H^+$ be the expansion of $H$ by the unary predicates $P_u\coloneqq\{u\}$ and $N_u\coloneqq N^H(u)$ for all $u\in U$.
  Then $H^+$ has a finite relational signature of arity two, and its Gaifman graph is the underlying graph of $H$.
  Take a formula with parameters from $U$.
  Replace every atom $x=u$ or $u=x$ by $P_u(x)$, every atom $E(x,u)$ or $E(u,x)$ by $N_u(x)$, and every atom without variables by its truth value in $H$.
  This gives a parameter-free formula in the signature of $H^+$ that has the same quantifier rank and is satisfied by the same tuples.
  Hence $\tp^{H^+}_q(\bar a)$ determines $\tp^H_q(\bar a/U)$.

  Let $\bar a\in F^k$ and $B\coloneqq \Ball^{H^+}_h(\bar a)$.
  Every $w\in B$ satisfies $\dist^H(w,U)>r-h=1$, so $B$ avoids $U$ and all neighbors of vertices in $U$.
  Thus every $P_u$ and $N_u$ is empty in $H^+[B]$.
  So $H^+[B]$ is the induced subgraph $H[B]$ expanded by empty predicates.
  Then $\tp^{H[B]}_{q'}(\bar a)$ determines $\tp^{H^+[B]}_{q'}(\bar a)$.
  By \Cref{lem:local-to-global}, $\tp^{H^+[B]}_{q'}(\bar a)$ determines $\tp^{H^+}_q(\bar a)$, and hence $\tp^H_q(\bar a/U)$.
  Here we crucially use that $h$ and $q'$ do not depend on the signature of $H^+$, which grows with $|U|$.
  So $|\Types^H_{k,q}(F/U)|$ is at most the number of parameter-free rank-$q'$ types of $k$-tuples in graphs with $c$ unary predicates.
  This number is finite and depends only on $q'$, $k$, and $c$, hence only on $q$, $k$, and $c$.
  Call it $T(q,k,c)$.
\end{proof}

The following is an adaptation of~\cite[Corollary~16]{number-of-types}. 
\septypestuples*
\begin{proof}
  Let $r=r(q)$ and $T=T(q,k,c)$ be given by \Cref{lem:sep-types},
  yielding the desired bound
  \[
  \left|
    \{
    \tp_q^H(\bar a/U) \,:\, \bar a \in F^k  
    \}
  \right|
  \leq
  T(q,k,c)
  \]
  in case no tuple $\bar u$ is involved.
  To lift that result, it remains to show that for a fixed tuple $\bar u$ with elements from $U$, the type 
  $\tp^H_q(\bar a\bar u/U)$ depends only on $\tp^H_q(\bar a/U)$.
  More precisely, it suffices to show that there exists a map $\phi \mapsto \phi_{\bar u}$ such that every formula
  $\phi(\bar x \bar y)$ with 

  \begin{center}
    quantifier rank ${\leq}q$,
    \qquad
    $|\bar x| = k$ and $|\bar y| = |\bar u|$, 
    \qquad
    and parameters from $U$
  \end{center}
  gets mapped to a formula $\phi_{\bar u}(\bar x)$ of the same quantifier rank and again with parameters from $U$ such that 
  for every $\bar a \in F^k$
  \[
    \phi\in\tp^H_q(\bar a\bar u/U)
    \iff H\models\phi(\bar a\bar u)
    \iff H\models\phi_{\bar u}(\bar a)
    \iff \phi_{\bar u}\in\tp^H_q(\bar a/U).
  \]
  We can obtain a suitable $\phi_{\bar u}(\bar x)$ from $\phi(\bar x \bar y)$ by substituting in $\phi$ each free variable $y_i \in \bar y$ with the corresponding parameter $u_i \in \bar u$ from $U$.
  This translation depends only on $\bar u$ and not on $\bar a$. 
  It does not change the quantifier rank, and still all parameters of $\phi_{\bar u}$ are contained in $U$.
\end{proof}
\begin{remark}
  In \Cref{lem:sep-types-tuple},
  the bound crucially relies on $\bar u$ being fixed, as the translation $\phi\mapsto\phi_{\bar u}$ differs for different $\bar u$.
  Without fixing $\bar u$ there is no bound independent of $|U|$:
  for $k=1$, $a\in F$, and $\bar u=(u)$ ranging over $U$, the formulas $\psi_u(xy) \coloneqq (y=u)$ (each using a different parameter $u$) show that the tuples $au$ realize $|U|$ different types over $U$.
\end{remark}

\section{Locally definable construction sequences in nowhere dense classes}\label{sec:nd}
We prove \Cref{thm:nd-def-seq}, that in nowhere dense classes, we can obtain definable construction sequences of length $O(n^{1+\eps})$ for any $\eps>0$, with $O(n^\eps)$ parameters and quantifier rank $0$.

\thmnddefseq*

\begin{proof}[Proof sketch]
Let $G\in\CC$ have $n$ vertices.
We use the almost bounded \emph{weak coloring numbers} of nowhere dense classes~\cite{NesetrilM11a,gks}, through the following consequence~\cite{drabik2026flips}.
For $v\notin U$, let $S_s(v,U)$ be the set of vertices in $U$ reachable from $v$ by a path of length at most $s$ whose internal vertices avoid $U$.
There is a total order on $V(G)$ such that $|S_{r+1}(v,U)|\le O_{\CC,r,\eps}(n^\eps)$ for every prefix $U$ of this order and every $v\notin U$.
Such an order can be computed in time $O_{\CC,r,\eps}(n^{1+\eps})$ by the algorithm of Grohe, Kreutzer, and Siebertz~\cite{gks} for weak coloring orders: every vertex of $S_{r+1}(v,U)$ is weakly $(r+1)$-reachable from $v$, since it is the smallest vertex on the witnessing path.

For each prefix $U$, consider the partial graph $H_U$ on $V(G)$ whose positive resolved pairs are the edges of $G[V(G)\setminus U]$, whose negative relation is empty, and whose parts are the atomic types in $G$ over $U$.
Thus the vertices of $U$ are singleton parts, and the remaining vertices are grouped by their neighborhoods in $U$.
Starting with $U=V(G)$, remove the last vertex $u$ of $U$ one at a time.
To pass from $H_U$ to $H_{U'}$, where $U'=U\setminus\{u\}$, first resolve positively between $\{u\}$ and each part outside $U$ adjacent to $u$ in $G$, and then merge parts with the same atomic type over $U'$.
Finally, when $U$ is empty, one negative resolve within the unique part completes the sequence.
There are at most $|E(G)|$ positive resolves and $n-1$ merges, giving length $O_{\CC,r,\eps}(n^{1+\eps})$ by the sparsity of $\CC$.

To implement the construction, first compute the atomic-type partitions over the prefixes in forward order by partition refinement: when adding $u$, split off $\{u\}$ and refine the remaining parts by adjacency to $u$.
Using adjacency lists, this takes $O(n+|E(G)|)$ time in total.
Record these splits and the resulting adjacent parts outside the prefix.
Processing the prefixes in reverse order, resolve positively against the recorded adjacent parts and undo the recorded splits by merges.
Thus the sequence can be computed in time $O_{\CC,r,\eps}(n^{1+\eps})$, including the computation of the order.

To see local definability, consider two consecutive partial graphs $H,H'$ during the transition from $H_U$ to $H_{U'}$, and put $B=\Ball^{H'}_r(v)$.
Vertices of $U'$ are still isolated, so their balls are singletons and are easy to define.
Otherwise $B\subseteq V(G)\setminus U'$. Every resolved pair of $H'$ is an edge of $G$, so every vertex of $U'$ adjacent in $G$ to a vertex of $B$ belongs to $S_{r+1}(v,U')$.
All vertices of $U'$ outside this separator are therefore nonadjacent to every vertex of $B$.
Thus, on $B$, the atomic types in $G$ over $U$ are determined by adjacency and equality tests against the parameters
\[
S_{r+1}(v,U')\cup\{u\}.
\]
This set has size $O_{\CC,r,\eps}(n^\eps)$.
Each atomic-type part is therefore quantifier-free definable, and merging parts corresponds to taking disjunctions of their definitions.
The resolved edges are the original edges between vertices outside $U$, together with edges from $u$ to the parts already resolved against it, so they too have quantifier-free definitions using these parameters.
One unary predicate naming $B$ restricts these definitions to the required domain.
The final negative resolve is immediate, since just before it the positive relation is $E(G)$ and there is only one part.
Thus the same bounds apply to $H[\Ball^{H'}_r(v)]$, including the intermediate merges and resolves.
\end{proof}

\end{document}